\documentclass[preprint,11pt]{elsarticle}

\usepackage[a4paper,top=2cm,bottom=2cm,left=1.5cm,right=1.5cm]{geometry}

\usepackage{soul,color}

\usepackage{subcaption}

\usepackage{algorithm}  
\usepackage{algpseudocode}

\usepackage{amsmath,amssymb}
\usepackage{graphicx,booktabs}
\usepackage{hyperref}
\hypersetup{colorlinks=true,
	linkcolor=blue,
	filecolor=magenta,      
	urlcolor=cyan,
	citecolor=blue}

\newcommand{\bb}{ {\boldsymbol b} }

\newcommand{\bc}{ {\boldsymbol c} }
\newcommand{\bC}{ {\boldsymbol C} }

\newcommand{\bD}{ {\boldsymbol D} }
\newcommand{\be}{ {\boldsymbol e} }

\newcommand{\boldf}{ {\boldsymbol f} }
\newcommand{\bF}{ {\boldsymbol F} }
\newcommand{\bg}{ {\boldsymbol g} }

\newcommand{\bh}{ {\boldsymbol h} }

\newcommand{\bI}{ {\boldsymbol I} }

\newcommand{\br}{ {\boldsymbol r} }

\newcommand{\bS}{ {\boldsymbol S} }

\newcommand{\bu}{ {\boldsymbol u} }

\newcommand{\bv}{ {\boldsymbol v} }
\newcommand{\bV}{ {\boldsymbol V} }

\newcommand{\bx}{ {\boldsymbol x} }

\newcommand{\by}{ {\boldsymbol y} }

\newcommand{\bz}{ {\boldsymbol z} }

\newcommand{\bzero}{\mbox{\boldmath{$0$}}}

\newcommand{\bepsilon}{ {\boldsymbol \epsilon} }

\newcommand{\bLambda}{ {\boldsymbol \Lambda} }
\newcommand{\bmu}{ {\boldsymbol \mu} }

\newcommand{\bomega}{ {\boldsymbol \omega} }
\newcommand{\bOmega}{ {\boldsymbol \Omega} }

\newcommand{\bsigma}{ {\boldsymbol \sigma} }

\newcommand{\bSigma}{ {\boldsymbol \Sigma} }

\newcommand{\bxi}{ {\boldsymbol \xi} }

\usepackage{natbib}

\allowdisplaybreaks

\begin{document}

\begin{frontmatter}	
	
	\title{\bf A practical and efficient approach for Bayesian reservoir inversion: 
		Insights from the Alvheim field data}
	
	\author{Karen S. Auestad}
	\author{The Tien Mai\corref{cor1}}\ead{the.t.mai@ntnu.no}
	\author{Mina Spremic}
	\author{Jo Eidsvik\corref{cor2}}\ead{ jo.eidsvik@ntnu.no}	

	\affiliation{organization={		Department of Mathematical Sciences,
			Norwegian University of Science and Technology},
		city={Trondheim},
		postcode={7034}, 
		country={Norway}}

	\begin{abstract}
		Stochastic reservoir characterization, a critical aspect of subsurface exploration for oil and gas reservoirs, relies on stochastic methods to model and understand subsurface properties using seismic data. This paper addresses the computational challenges associated with Bayesian reservoir inversion methods, focusing on two key obstacles: the demanding forward model and the high dimensionality of Gaussian random fields. Leveraging the generalized Bayesian approach, we replace the intricate forward function with a computationally efficient multivariate adaptive regression splines method, resulting in a 34 acceleration in computational efficiency. For handling high-dimensional Gaussian random fields, we employ a fast Fourier transform (FFT) technique. Additionally, we explore the preconditioned Crank–Nicolson method for sampling, providing a more efficient exploration of high-dimensional parameter spaces. The practicality and efficacy of our approach are tested extensively in simulations and  its validity is demonstrated in application to the Alvheim field data. 
	\end{abstract}

\end{frontmatter}

	\section{Introduction}
	
	Stochastic reservoir characterization from seismic data involves using stochastic methods and techniques to model and characterize subsurface reservoirs based on seismic data, see e.g. \cite{bosch2010seismic} and \cite{grana2021seismic,grana2022probabilistic} for an overview
	of the problem. Seismic data are measurements of the reflections of sound waves or vibrations sent into the subsurface layers of the Earth. These reflections provide information about the geological structure and properties of the subsurface, including the presence and characteristics of hydrocarbon reservoirs. The goal of stochastic reservoir characterization is to provide a range of plausible subsurface models that honor the available data and capture the uncertainty associated with the reservoir properties. This information is crucial for making informed decisions in the exploration and development of oil and gas reservoirs, as it helps in assessing risks, optimizing well placement, and designing efficient production strategies.
	
	Stochastic reservoir modeling entails the representation of reservoir properties, such as oil, gas, and clay, as random fields. This involves assigning probability distributions to these properties to capture the spatial variability within the reservoir. Bayesian inversion has emerged as a well-suited method for reservoir characterization challenges, as it enables the integration of prior knowledge with observational data. In many cases, there is a need for a priori knowledge of Earth parameters since the information obtained from measurements alone may be insufficient. This a priori knowledge is encapsulated by a prior distribution, allowing for the incorporation of trends and uncertainties in our initial beliefs \citep{malinverno2004expanded,spremic2023bayesian,forberg2021bayesian,grana2022bayesian}. In Bayesian inversion, a likelihood function, grounded in geophysical principles, informs us about which Earth parameter values align with the observed data. By amalgamating the prior distribution and the likelihood function, a posterior distribution of Earth parameters can be formed. Although, examining posterior densities in Bayesian inversion can be analytically challenging, various methods are available for approximating the posterior distribution \citep{mosegaard1995monte,khoshkholgh2021informed,khoshkholgh2022full,grana2022markov}.	
	
	The practical implementation of Bayesian reservoir inversion methods often grapples with computational challenges, primarily stemming from two key sources. Firstly, the forward model or function integral to the inversion process is frequently demanding in terms of computational resources. Secondly, in order to faithfully represent the spatial variability within a reservoir, a random field, typically a Gaussian random field, is utilized. However, the utilization of such random fields results in a high-dimensional distribution that necessitates extensive sampling and evaluation efforts.
	
	To illustrate geophysical inversion, we examine seismic amplitude data from the Alvheim field. In this specific study, a Gaussian random field is employed to capture reservoir variability extending into a dimension exceeding 44,000. This considerable dimensionality adds to the computational complexity, intensifying the challenges associated with sampling and evaluating the distribution. Consequently, it presents a noteworthy hurdle in the effective application of Bayesian reservoir inversion methods.
	
	Another significant challenge is the design of an efficient Markov Chain Monte Carlo (MCMC) method in high-dimensional settings. This challenge is particularly crucial as Bayesian methods face a daunting aspect: the slow convergence of the sampling of the posterior distribution, which becomes increasingly sluggish as the dimensionality escalates. This poses a notable obstacle in the application of Bayesian reservoir inversion methods, emphasizing the need for innovative solutions to address the intricacies of high-dimensional parameter spaces.

	In this study, our primary objective is to address and mitigate the computational challenges associated with Bayesian reservoir inversion methods. The initial computational complexity arising from the evaluation of the forward model is effectively tackled by employing a cutting-edge statistical approach known as the generalized Bayesian approach. This approach, as detailed in abundant literature \citep{bissiri2016general, grunwald2017inconsistency,mai2017pseudo,alquier2020concentration,hong2020model,Knoblauch, Matsubara,Mai2023ARA,hammer2023approximate,alquier2024user}, represents a generalization of classical Bayesian methods. Instead of relying on the true likelihood, this approach allows the use of a fractional likelihood or a loss-based function. The generalized Bayesian approach has demonstrated efficacy both theoretically and in practical applications. To leverage this novel insight, we address the first computational challenge by substituting the intricate forward function with an estimated function derived from training data. Specifically, we employ the multivariate adaptive regression splines (MARS),  \citep{friedman1991multivariate} for this purpose. Notably, our implementation of this approach yields a noteworthy acceleration in computational efficiency. In our case study of the Alvheim field data, the reported speedup is approximately 34 times faster when compared to the use of the true forward function. This substantial improvement in running time underscores the practical utility of the generalized Bayesian approach in enhancing the computational efficiency of Bayesian reservoir inversion methods.
	
	To address the computational challenges posed by the substantial dimensionality of the Gaussian random field, we adopt a strategy that leverages the efficiency of the Fast Fourier transform (FFT) for simulating and evaluating the Gaussian random field, as outlined by the works \cite{chan1997algorithm, davies2013circulant, rue2005gaussian,abrahamsen2018simulation}. This technique involves the circulant embedding of a structured covariance matrix onto a torus. The implementation of this method not only significantly reduces the required memory but also yields remarkably fast computations. Notably, this method proves to be highly efficient, enabling simulations and evaluations on standard computing hardware. In our case study focused on the Alvheim data, which involves a Gaussian random field of dimensionality exceeding 44,000, the computations are easily managed on a laptop with 16GB RAM. Furthermore, the simulation time for a vector within this high-dimensional space is impressively less than a second. This underscores the practical feasibility and efficiency of employing FFT-based methods to handle and streamline the simulation and evaluation of Gaussian random fields in large-scale applications.

	The impetus behind the present study is rooted in the recognition that a more tailored sampling methodology, specifically designed for the high dimensionality, has the potential to yield substantial computational speed enhancements. Notably, \cite{cotter2013mcmc} introduced a groundbreaking approach to Markov Chain Monte Carlo (MCMC) sampling that addresses the notorious 'curse of dimensionality' associated with Gaussian priors, given appropriate assumptions on the likelihood function. This innovative method, termed ``preconditioned Crank–Nicolson" (pCN), represents a modification of the standard random-walk Metropolis sampling technique. The key feature of pCN involves scaling the position from the previous iteration before adding a random shift and generating the proposal. This adjustment effectively mitigates the computational challenges associated with high-dimensional spaces. By doing so, pCN provides a more efficient and effective means of exploring the parameter space, particularly in scenarios where traditional methods may encounter difficulties due to the curse of dimensionality \citep{hairer2014spectral}. The introduction of pCN thus opens up new avenues for achieving significant computational speedups in the context of MCMC sampling, offering a promising solution to challenges posed by increased system dimensionality. Utilizing the pCN method within our framework, where the proposed approach involves incorporating a physically informed prior, enables effective exploration of various high-probability regions within the intricate posterior distribution. This aspect holds particular significance in the context of Bayesian inverse problems, as emphasized in previous works such as \cite{mosegaard1995monte,khoshkholgh2021informed,khoshkholgh2022full}.

In application to the Alvheim field data, our study reveals varying saturation levels around wells, with more gas near gas-producing wells and more oil near oil-producing ones. Uncertainty is low near these wells but higher in transition zones. Compared to \cite{spremic2023bayesian}, our study shows similar high saturation levels near wells but differs in saturation distributions and uncertainty levels, with our MCMC method generally yielding lower uncertainty. We also note higher saturation levels in clay content. Ternary plots highlight differences, especially in shallow areas, where our results suggest higher oil saturation and less uncertainty. While \cite{spremic2023bayesian} use local ensemble-based techniques, their effectiveness in reducing uncertainty remains unclear.

	The subsequent sections of the paper are organized as follows: Section \ref{sc_background} provides an introduction to the Bayesian rock physics problem and offers insights into the Alvheim field data. Our proposed procedure is outlined in Section \ref{sc_proposedmethod}. The findings from the Alvheim field real data are presented and deliberated upon in Section \ref{sc_result_realdata}. Further, Section \ref{sc_result_simulation} encompasses the presentation of numerical studies. For additional information on the fast Fourier transform sampling method and more insights into the simulation studies, refer to \ref{sc_detail_FFT}, \ref{sec:npkr} and \ref{sc_extra_simulation}. We discuss and conclude our work in Section \ref{sc_discuss_conclusion}.

	\section{Background}
	\label{sc_background}

	\subsection{The Alvheim field}
	\label{sc_Alvheim_data}
\begin{figure}[htb]
		\includegraphics[width=0.4\textwidth]{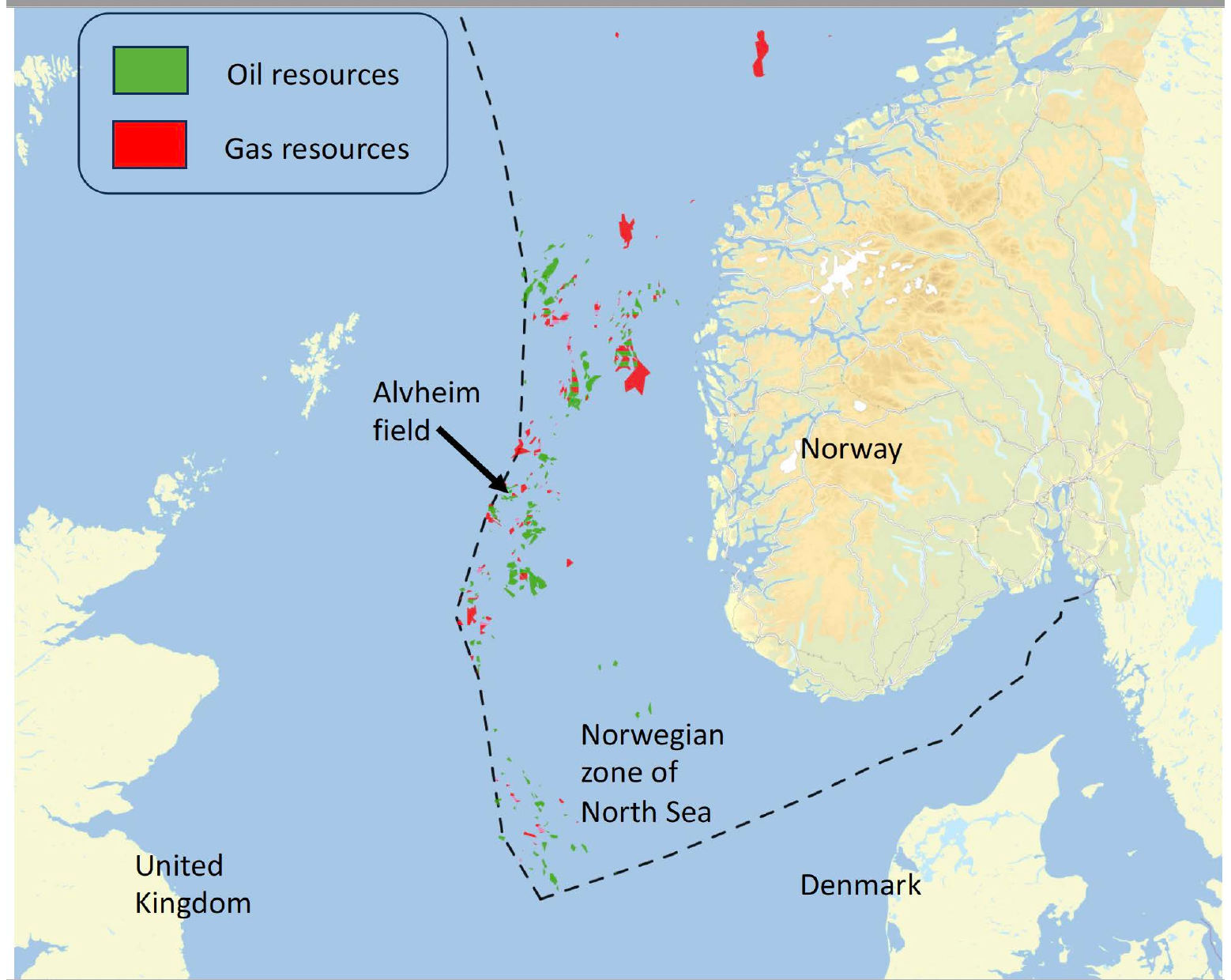}
		\caption{Location of the Alvheim field in the North Sea. Map from: \url{https://www.norskpetroleum.no/en/interactive-map-quick-downloads/interactive-map/}}
		\label{fig:traveltimes}
	\end{figure}

	The Alvheim field, situated in the North Sea,  Figure~\ref{fig:traveltimes}, off the coast of Norway and dating back to the Paleocene era, is a turbiditic oil and gas reservoir. The field's characteristics include a complex sand distribution and hydrocarbon trapping system, featuring multiple submarine fan lobes and significant variations in depositional facies, net-to-gross ratios, and sand textures. The sands range from massive, thick-bedded formations to more diverse, inter-bedded sand-shale intervals. The overall depositional pattern is influenced by structural topography linked to deeper faults and salt tectonics.
	
	This field poses unique challenges for seismic reservoir characterization due to its complex geological features. The reservoir sands, attributed to the Heimdal Formation and capped by Lista Formation shales, exhibit intricate rock physics properties influenced by a compaction history related to burial depths of approximately 2 kilometers below the seafloor and temperatures around 70°C. Diagenetic processes, such as the transformation of smectite-rich shales to illite and cementation of quartzose sands, contribute to significant elastic stiffening of the rock frame. The presence of both unconsolidated sands and cemented sandstones further complicates the fluid sensitivity of seismic signals, see e.g \cite{avseth2008shale}. Additionally, the distribution of hydrocarbons, encompassing both gas and oil, is complex and poorly understood, likely influenced by regional tectonic events. These challenges underscore the difficulty in accurately characterizing the Alvheim field using seismic methods, requiring careful consideration of geological intricacies and fluid dynamics.

\begin{figure}[htb]
	\begin{subfigure}[b]{0.3\textwidth}
		\includegraphics[width=\textwidth]{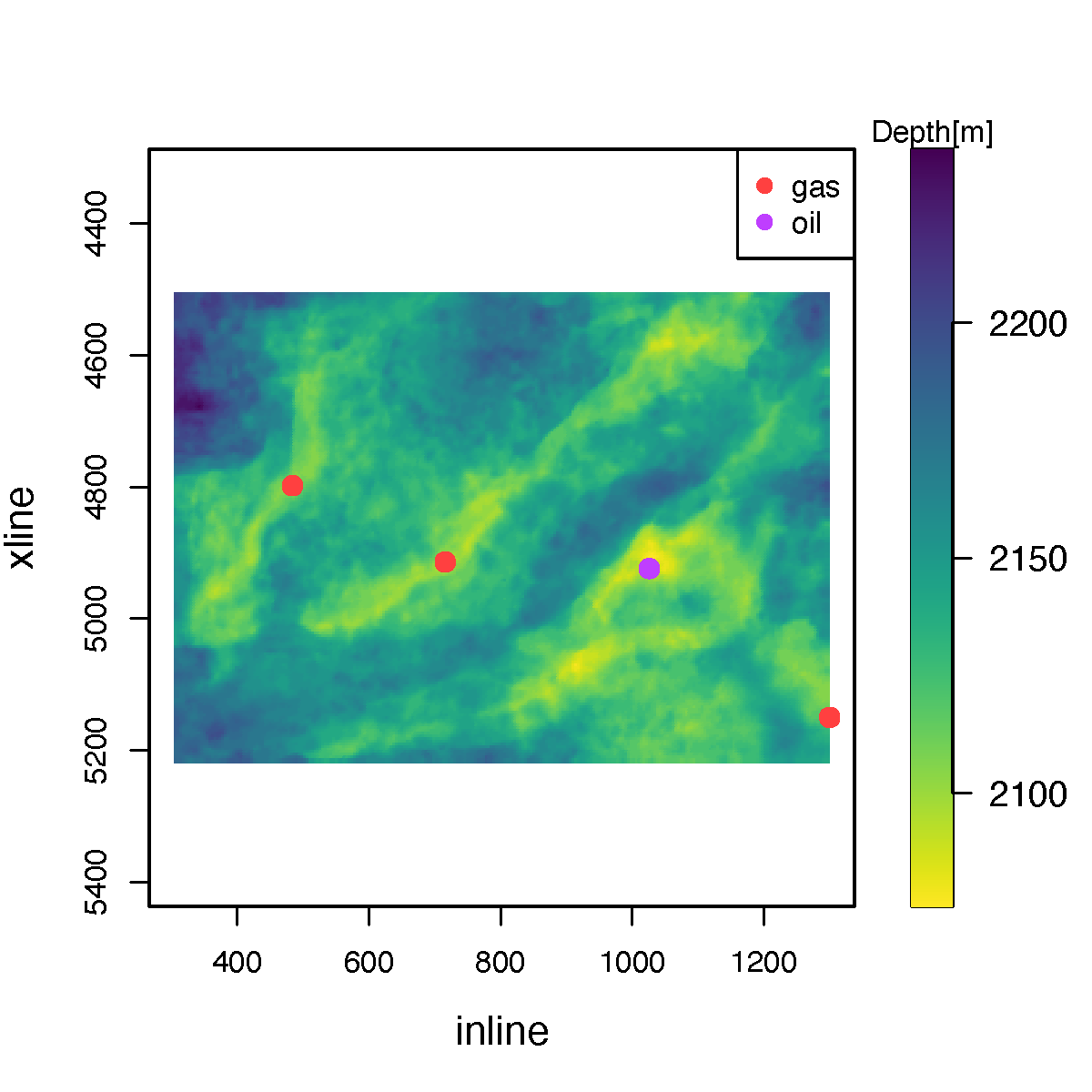}
		\caption{The depth at the Alvheim field. }
		\label{fig:depthfield}
	\end{subfigure}
	\begin{subfigure}[b]{0.3\textwidth}
		\centering
		\includegraphics[width=\textwidth]{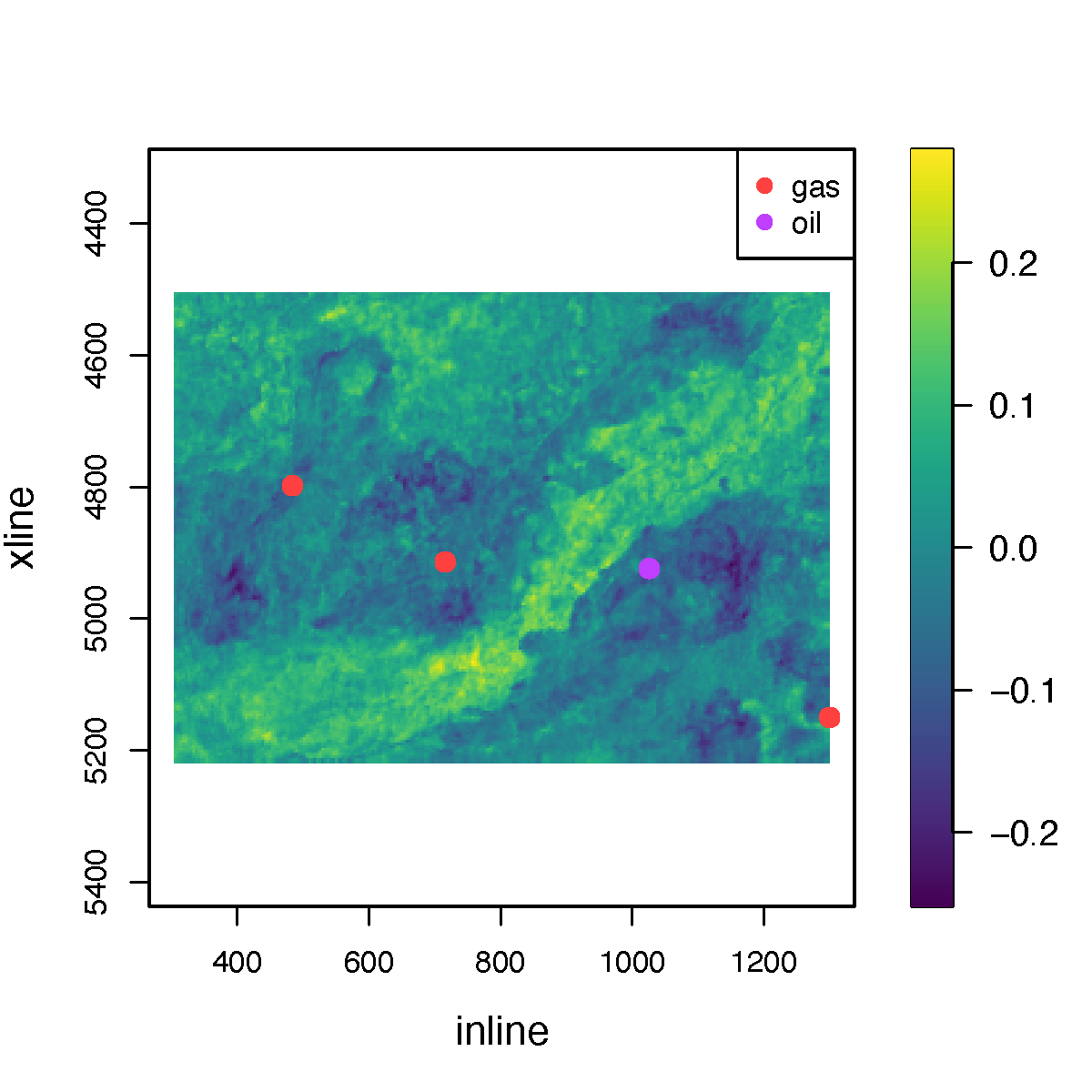}
		\caption{Zero offset data, $R_0$.}
		\label{fig:zeroOffset}
	\end{subfigure}
	\begin{subfigure}[b]{0.3\textwidth}
		\centering
		\includegraphics[width=\textwidth]{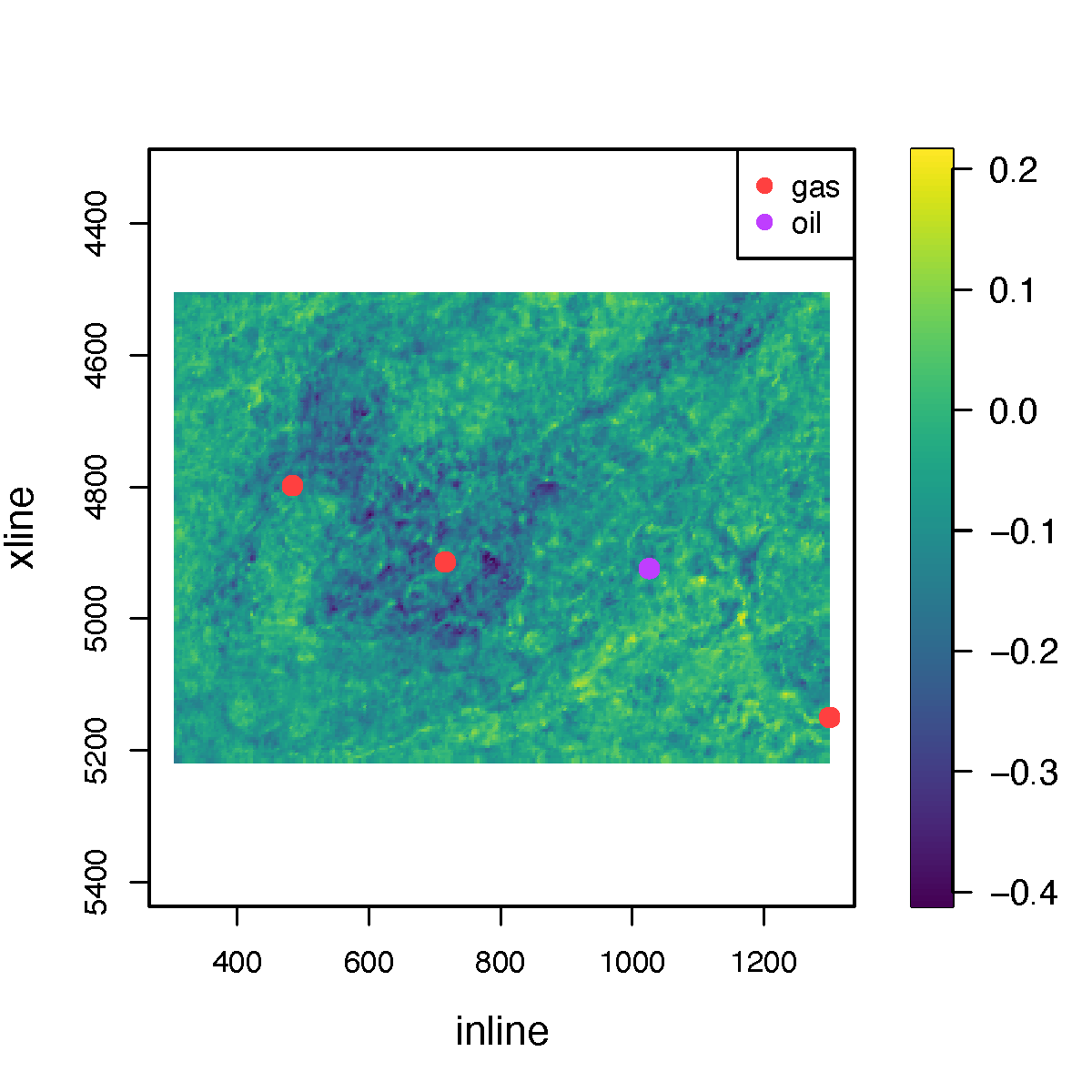}
		\caption{Gradient data, $G$.}
		\label{fig:gradient}
	\end{subfigure}
	\caption{AVO data presented in the inline-crossline coordinate system for the Alvheim field. Four wells are marked by coloured circles, where the colour indicates the majority type of hydrocarbons found at that well. The wells are located at $($inline$, $crossline$)$-locations $(484, 4798)$, $(716,4914)$, $(1026, 4924)$ and $(1300, 5150)$.}
	\label{fig:avodata}
\end{figure}

	\subsubsection{Seismic AVO data}
	
	This investigation employs seismic amplitude-versus-offset (AVO) data acquired from the Alvheim field, covering angle stacks within near ($12^{\circ}$), mid ($22^{\circ}$), and far ($31^{\circ}$) angle ranges. The data have undergone processing and preconditioning to ensure their appropriateness for quantitative analysis, following the methodology outlined by \cite{rimstad}. AVO attributes, such as zero-offset reflectivity ($R_0$) and AVO gradient ($G$), have been extracted and calibrated using scalars derived from upscaled well log data \citep{avseth2016combining}. Each inline and crossline not only includes two data points ($R_0$ and $G$) but also incorporates associated traveltimes, as shown in Figure~\ref{fig:depthfield}, subsequently converted to depth.
	
	We focuses on the top-reservoir horizon, where, in addition to spatial variations in the inline-crossline domain, depth information is available for the 2D slice at each inline-crossline location. Referred to as a 2.5D inverse problem, the AVO data for the Alvheim field, depicted in Figure~\ref{fig:avodata}, encompass $R_0$ and $G$ for each inline and crossline of the top-reservoir. Covering an area of approximately {12.35}{km}$\times${8.85}{km}, the dataset is subsampled by selecting measurements from every 4 inlines and crosslines, equivalent to {50}{m} intervals in both crossline and inline directions. The spatial grid representation in Figure~\ref{fig:depthfield} is utilized for modeling, maintaining the original inline and crossline units without conversion to distance metrics.

	\subsubsection{Well log data}
	
	This study utilizes data from four wells, namely 24/6--2, 24/6--4, 25/4--7, and 25/4--8, whose locations and outcomes at the top-reservoir are illustrated in Figure~\ref{fig:depthfield}. These well details will be referred to in figures throughout the paper, although the legends will be omitted. All these wells are situated within the seismic inline-crossline ranges and drilled on structural highs in the shallower sections of the top-reservoir. The well log data encompass sonic velocities ($V_{\text{p}}$ and $V_{\text{s}}$), densities, and petrophysical logs, including fluid saturations (oil, gas, and brine) and clay content, as outlined in \cite{rimstad} and \cite{spremic2023bayesian}. 
	
	In the case of well 24/6--2, commercial gas is found in the Heimdal Formation, with a gas column of {52}{m} down to the gas-oil contact at {2151}{m}. Additionally, there is a thin oil column of {17}{m} down to the oil-water contact at {2168}{m}. Well 25/4--7 encountered commercial oil saturation, with an oil column of {48}{m} down to the oil-water contact at {2133}{m}. Hydrocarbon saturations generally fall within commercial ranges (0.6--0.9), although lower values are observed in heterolithic zones where the presence of clay influences reservoir transport properties. Thin clay laminations may also induce patchy saturations at the log scale. For Wells 24/6--4 and 25/4--8, constant interpretations are made for different zones due to missing saturation logs estimated from resistivity logs.

	\subsection{Bayesian model}
	
	\subsubsection{Probabilistic geophysical inversion}
	
	The inverse problem arises when there is the ability to collect data, denoted as $\by$, which are associated with an underlying variable of significance referred to as $\bx$. The relationship between the variable of interest and the observed data is expressed as:
	\begin{equation}
	\by = h(\bx) + \be,
	\label{inverseProblem}
	\end{equation}
	where $h$ defines a recognized functional relationship and $\be$ represents stochastic perturbations.
	
The computational challenge of inversion is multifaceted, primarily stemming from the intricate interplay of high-dimensional, nonlinear, and uncertain data. In many real-world scenarios, the variables of interest and observed data inhabit high-dimensional spaces, necessitating sophisticated algorithms to explore parameter space efficiently. Nonlinear function $ h $ further compounds the challenge, requiring iterative methods to converge on optimal solutions. Moreover, uncertainty pervades the inversion process, demanding robust techniques such as Bayesian inference to quantify and incorporate uncertainty into parameter estimates. \citep{InverseProblem}.
	
Bayesian inversion involves combining a likelihood model for observed data with prior knowledge about the variable of interest to generate a posterior distribution $f(\bx|\by)$. Prior knowledge is expressed through a prior distribution, denoted $f(\bx)$. The likelihood probability density function, $f(\by | \bx)$, indicates the likelihood of obtaining the observed data given a specific value of $\bx$. According to Bayes' rule, the posterior distribution function can be written as:
$ f(\bx|\by)  \propto f(\by | \bx) f(\bx), $
where the posterior $f(\bx | \by)$ is referred to as $\pi(\bx)$.

	\subsubsection{Observation model}
	
	Before delving into the above mentioned models, we provide essential notation, adopting conventions akin to those in \cite{spremic2023bayesian}. Reservoir variables are denoted at locations $\bu\in \mathcal{D}$, with $\mathcal{D}$ representing the inline and crossline grid used to portray seismic data at the top reservoir. The grid dimensions are $248 \times 178$, yielding  $N=44,144 $ grid cells with inline-crossline locations $\bu_1,\ldots,\bu_N$. Seismic AVO data at locations $\bu\in \mathcal{D}$ are 
	$$
	\by(\bu)
	=
	[R_0(\bu),G(\bu)]
	.
	$$
	The main interest lies in the oil and gas saturation, and clay content at each of the inline-crossline locations. Results will be displayed for the reservoir variables of interest $\br(\bu)$, which we denote by $\br(\bu) = [\bS_{\text{g}}(\bu),\bS_{\text{o}}(\bu), \bV_{\text{clay}}(\bu)]$. 
		
	For the purpose of efficient computations, we work with transformed saturations and clay content. Variables are transformed back and forth by applying logistic functions at all locations $\bu \in \mathcal{D}$,
	\begin{eqnarray*}
	S_{l}(\bu) &=& \frac{\exp\left[x_{l} (\bu)\right]}{1+\exp\left[x_{\text{g}} (\bu)\right]+\exp\left[x_{\text{o}} (\bu)\right]}, \, l=\text{o,g} \mbox{(oil,gas)}  
	\\
	S_{\text{b}}(\bu)&=& \frac{1}{1+\exp\left[x_{\text{g}}(\bu)\right]+\exp\left[x_{\text{o}}(\bu)\right]},
	\end{eqnarray*}
meaning that $S_{\text{g}}(\bu)+S_{\text{o}}(\bu)+S_{\text{b}}(\bu)=1$ for every location $\bu$,
	where $x_{\text{o}}(\bu) $ and $x_{\text{g}}(\bu) \in \mathbb{R} $. In doing so, we keep brine as a reference or background saturation.
	Similar for clay content $V_{\text{clay}}(\bu)$,
	\begin{equation*}
	V_{\text{clay}}(\bu) = \frac{\exp \left[x_{\text{clay}}(\bu) \right]}{1+\exp\left[x_{\text{clay}}(\bu)\right]},
	\end{equation*}
	where $x_{\text{clay}}(\bu) \in \mathbb{R} $.
	Altogether, we then get the reservoir variables of interest from a logistic function, compactly denoted by
	\begin{equation}
	\label{eq:r_x_s}
	\br(\bu) = \boldf_0 \left[\bx(\bu)\right], \hspace{3mm} \bu \in \mathcal{D}.
	\end{equation}
	
	It is  note that our Bayesian model adopts fixed hyperparameters in both the prior and likelihood functions. An alternative approach in a fully Bayesian formulation involves placing prior distributions on these parameters, as exemplified in \cite{malinverno2004expanded}. However, in our specific setting, we opt for a sensitivity analysis approach, focusing on the examination of the most relevant input parameters.

	At a location $\bu$, we let $\by(\bu)$ denote the zero-offset coefficient $R_0$ and gradient $G$ at that location. The model can be written in an additive manner,
	\begin{equation}\label{forw}
	\by(\bu) = \bh_0\left[\bx(\bu)\right] + \bepsilon(\bu), \quad \bepsilon(\bu) \sim N(\bzero, \bOmega_0),
	\end{equation}
	where $\bh_0\left[\bx(\bu)\right]=\bg_0\left\{\boldf_0 \left[\bx(\bu)\right]\right\}$, and $\bg_0[\br(\bu)]$ represents the rock physics forward model function applied to the reservoir variables $\br(\bu)$ of primary interest. The $2 \times 2$ matrix $\bOmega_0$ characterizes the variances in $R_0$ and $G$ and the correlation between the two. Through the utilization of seismic reflection data in proximity to the four wells and the corresponding well log data in the reservoir zones, we define the parameters $\mbox{Var}(R_0)=0.003$, $\mbox{Var}(G)=0.03$, and $\mbox{Corr}(R_0,G)=-0.6$. Notably, these parameter values closely align with those employed in \cite{eidsvik2004}.
	
	Therefore, the likelihood model is characterized by a Gaussian distribution, with the mean derived from the nonlinear geophysical forward model and a spatially independent variance. The applicability of geophysical forward model relationships is assumed for each inline and crossline, considering the known depth of the top reservoir at each grid location. The likelihood model assumes conditional independence, implying that data at different locations are independent given the rock properties. Combining data from all inline-crossline locations into a length $2N$ vector $\by$, the probability density function $p(\by|\bx)$ follows a Gaussian distribution with mean $ \bh_0(\bx) $ and covariance matrix $ \bOmega $. Assuming conditional independence between AVO data at different inline-crossline grid locations, the $N \times N$ covariance matrix $\bOmega$ is block diagonal, with the $2 \times 2$ matrix $\bOmega_0$ repeating along the block diagonal elements. The function $\bh(\bx)$ is constructed based on the same rock physics relations applied to various spatially varying input reservoir variables.
	
	The realistic forward model, denoted as $\bh_0(\cdot)$, is constructed by integrating rock physics and geophysical models with geological properties. This forward model takes as input random fields representing clay content and fluid saturation. To address the cementation effect presumed to occur at a specific depth, we incorporate rock physics models based on mechanical and chemical compaction. It is also noted that the forward model is considered to be independent of locations and depth. Depending on the depth of the input reservoir variables, we employ either an unconsolidated sand model or a contact cement model, as outlined in \cite{avseth2010}. Additionally, Gassmann fluid substitution is implemented, followed by the Shuey approximation, to derive the amplitude-versus-offset (AVO) attributes. Further details regarding the forward model can be found in \cite{spremic2023bayesian}.
	
	\subsubsection{Physical-informed priors}
	\label{sc_prior_distribu}
	
	The prior distributions are formulated based on spatial random field models for the reservoir variables $\br(\bu)$ as defined in equation~{\eqref{eq:r_x_s}}. Independent Gaussian random fields are employed to represent the spatial structures within the transformed saturation variables $\bx_{\text{g}}=[x_{\text{g}}(\bu_1),\ldots,x_{\text{g}}(\bu_N)]$, $\bx_{\text{o}}=[x_{\text{o}}(\bu_1),\ldots,x_{\text{o}}(\bu_N)]$, and $\bx_{\text{clay}}=[x_{\text{clay}}(\bu_1),\ldots,x_{\text{clay}}(\bu_N)]$. To be more specific, given the spatial discretization of locations, we obtain Gaussian multivariate distributions that,
	\begin{equation}
	\bx_{\text{g}} \sim N(\bmu_{\text{g}}, \bSigma_{\text{g}}),\hspace{3mm}
	\bx_{\text{o}} \sim N(\bmu_{\text{o}}, \bSigma_{\text{o}}), \hspace{3mm}
	\bx_{\text{clay}} \sim N(\bmu_{\text{clay}}, \bSigma_{\text{clay}}).
	\label{eq_priors}
	\end{equation}
	In this context, the mean vectors $\bmu_{\text{g}}$, $\bmu_{\text{o}}$, and $\bmu_{\text{clay}}$ have dimensions of $N$, while the covariance matrices $\bSigma_{\text{g}}$, $\bSigma_{\text{o}}$, and $\bSigma_{\text{clay}}$ are sized $N \times N$. These parameters are derived from physical data, specifically well log information gathered at various depths.

\begin{figure}[htb]
	\centering
	\begin{subfigure}[b]{0.32\textwidth}
		\centering
		\includegraphics[width=\textwidth]{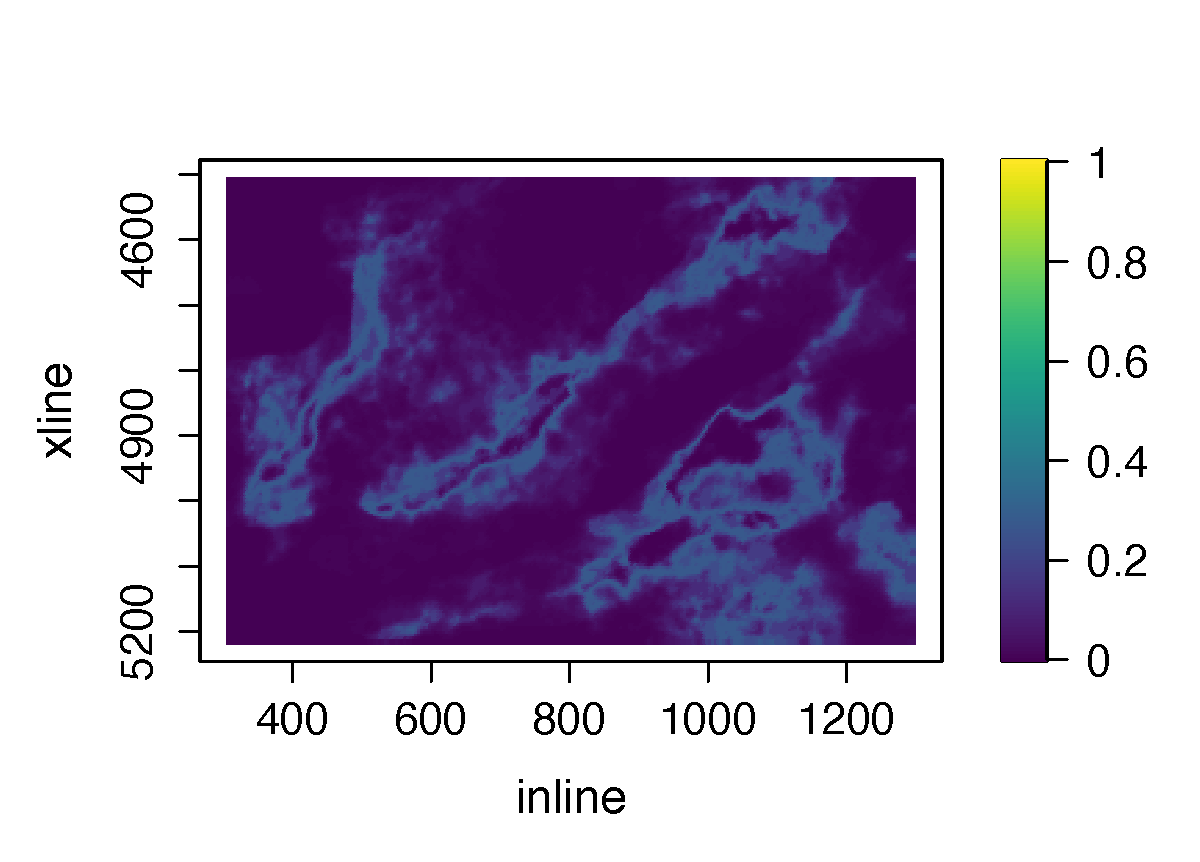}
	\end{subfigure}
	\begin{subfigure}[b]{0.32\textwidth}
		\centering
		\includegraphics[width=\textwidth]{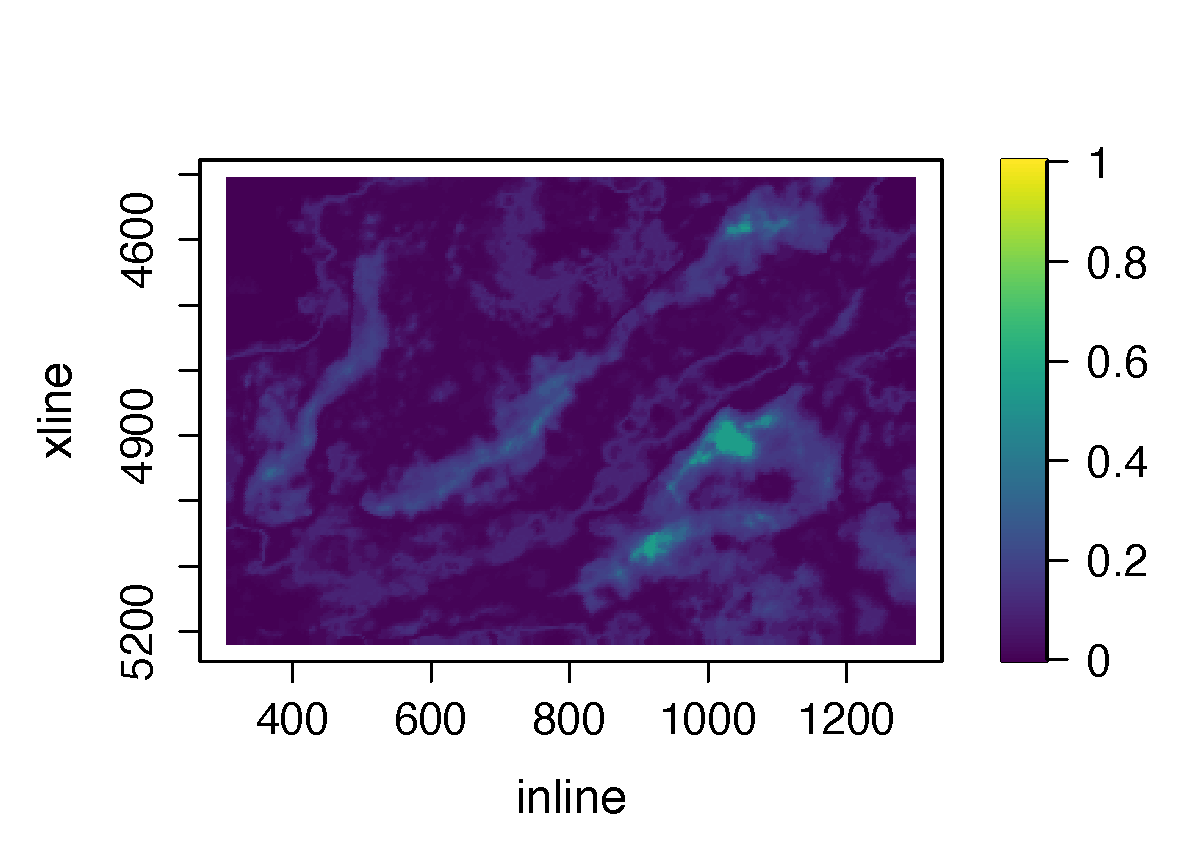}
	\end{subfigure}
	\begin{subfigure}[b]{0.32\textwidth}
		\centering
		\includegraphics[width=\textwidth]{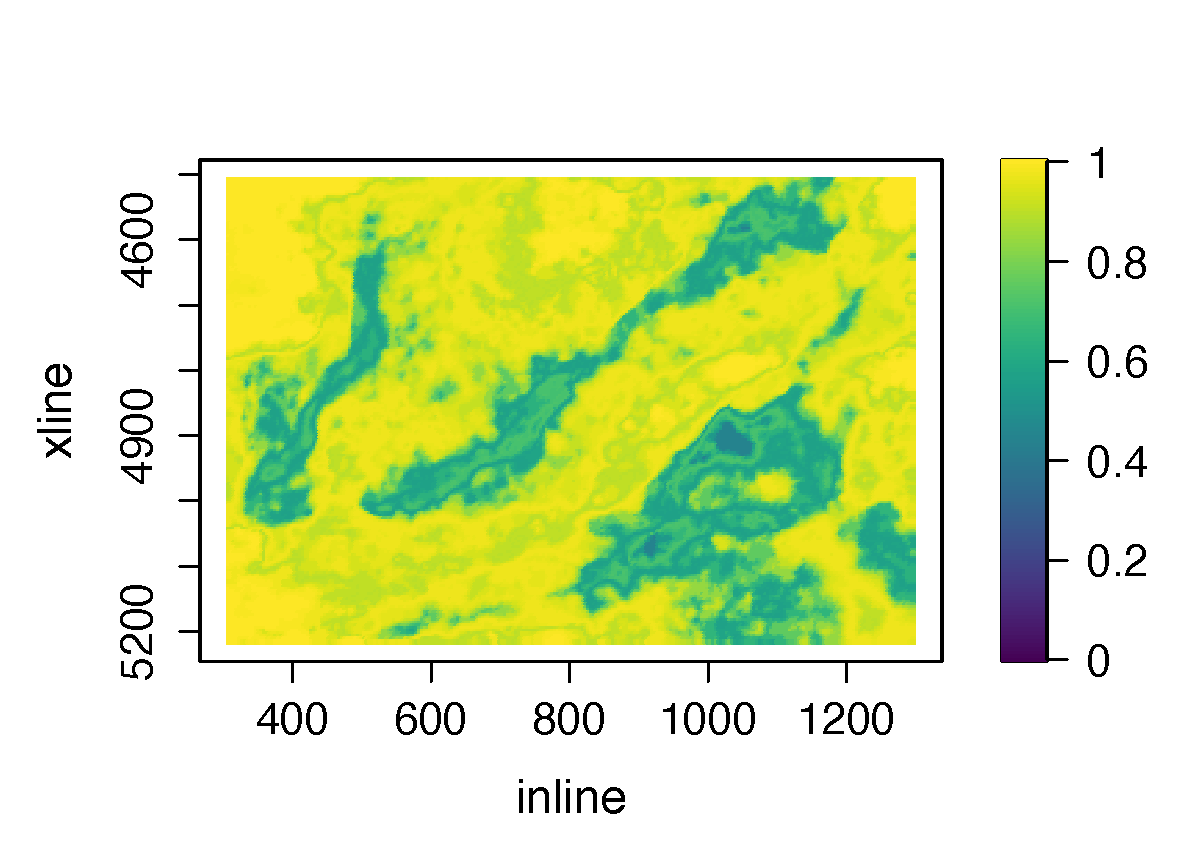}
	\end{subfigure}
	\caption{Prior mean saturations of gas, oil and brine (from left to right).}
	\label{fig:PriorAlvheim}
\end{figure}

	The mean values contribute to establishing a smoothly varying background model with depth. Predominantly high gas saturation is anticipated in the very shallow inline-crossline locations, elevated oil saturation is expected at relatively shallow positions, and brine is highly probable to be predominant at deeper locations. The standard deviation for each variable is assumed to remain constant with depth. 
	
	Regarding the covariance matrix, we incorporate spatial correlation using a Gaussian correlation function with an effective correlation length of $15$ grid cells. This correlation function is applied to $x_{\text{o}}$, $x_{\text{g}}$, and $x_{\text{clay}}$ and is specified based on prior assumptions about the geological depositional environment, particularly considering geographically separated lobe structures.
		
	The prior mean values of oil, gas and brine are depicted in Figure \ref{fig:PriorAlvheim}. Generally, the visualization illustrates that in shallow regions, the prior mean saturations of gas and oil are elevated. Specifically, the prior mean saturation of oil is relatively high in the vicinity of the well marked by a violet circle in Figure \ref{fig:depthfield}, while the prior mean saturation of gas is comparatively low in this region. In the deepest areas, such as the top-left corner, both gas and oil saturations exhibit lower values. In the intermediate terrain between the middle ridge and the elevated region containing the predominantly oil-filled well, the prior mean saturation for oil surpasses that of gas. Moreover, in the shallowest regions, gas exhibits a slightly higher prior mean saturation than oil.

	\section{Proposed method}
	\label{sc_proposedmethod}
			
	\subsection{Approximate forward function via MARS}

	Evaluating the forward model in the Alvheim case at a value $\bx$ takes a relatively long time. At each iteration in the MCMC algorithm, $h(\bx^p)$ needs to be calculated to find the likelihood at the proposed sample $\bx^p$. The goal is to find a good approximation that considerably reduces computation time.
	
	The idea here is to use some ensemble data to train an approximate model that is faster to evaluate in compared to the underlying forward model. This idea maybe similar those approaches based on Kalman filtering but we take advantage of these ensemble data for training an relatively good approximate function for the purpose of reducing the cost of calculating the acceptance rate within the MCMC algorithm.
	Let $\bx^i \sim N(\bmu, \bSigma), \by^i = \bh(\bx^i), i=1,...,n $ be some ensemble data. Based on this ensemble data, we train a function $ \hat{\bh} $ that is not only a good approximation of $ \bh $ but also faster to evaluate. For this purpose, we make use of the multivariate adaptive regression splines (MARS), introduced in \cite{friedman1991multivariate}.

MARS constitutes a non-parametric regression method utilized to elucidate intricate associations among variables. It operates by delineating the connection between predictor variables and the response variable through a sequence of basis functions, typically piecewise linear in nature, amalgamated to compose the ultimate regression model. The algorithm initiates with a constant prediction and discerns pivotal breakpoints in the predictor variables, segmenting the data into distinct sections where varied linear functions are employed. During the forward pass, MARS integrates basis functions into the model predicated on the identified breakpoints, encompassing simple linear terms, hinge functions, or interaction terms. Subsequently, a backward pass ensues wherein MARS prunes redundant basis functions to streamline the model, thus averting overfitting and enhancing interpretability. The resultant MARS model embodies a blend of these basis functions, yielding a piecewise linear regression model adept at capturing the dataset's nonlinearities. MARS proves advantageous in datasets exhibiting nonlinear or interactive relationships, offering flexibility in modeling without imposing specific functional forms. Its utility extends to scenarios where conventional linear models fall short in complexity and more intricate machine learning models risk overfitting. Particularly well-suited for regression tasks, MARS stands as a potent tool for data exploration and predictive analytics \cite{ESL}.

	The MARS approximation takes the following form:
	\[ \widehat{h}(\bx) = \beta_0 + \sum_{d=1}^{D} \beta_d g_d(\bx), \]
	where $\beta_1, \beta_2, ..., \beta_D$ represent coefficients jointly determined by minimizing the residual sum of squares. The linear basis functions $g_d(\bx)$ are in the set $\mathcal{C} = \{ (x_j-c_j)_{+}, (c_j-x_j)_{+} \}_{c_j \in \{ x_{1j}, x_{2j}, ... x_{nj} \}}$ for each covariate $x_j$ and observed value $x_{ij}$ in the dataset, where $(a)_+ = \max(a,0), \forall a \in \mathbb{R}$. Despite $g_d$ relying solely on one or multiple covariates, it functions across all covariates collectively. The model encompasses $D$ terms in total.
	
	Building a MARS model consists of two main steps, a forward model building procedure and a backward pruning procedure. In the forward step, functions $g_d(\bx)$ are chosen successively until a large, often overfitted model is obtained, which then is pruned to a smaller model. To go through these steps in detail, let $\mathcal{D}$ be the set of chosen combinations of basis functions for the model. Initially, $\mathcal{D} = \{ g_0 \}$, with $g_0 = 1$. Next, all functions in $\mathcal{D}$ are successfully multiplied by the pairs $\mathcal{C}$ and the term 
	\begin{equation*}
	\widehat{\beta}_{D+1} g_r(\bx)(x_j - c_j)_{+} + \widehat{\beta}_{D+2} g_r(\bx)(c_j - x_j)_{+}, \quad g_r(x) \in \mathcal{D}
	\end{equation*}
	that produces the largest decrease in the training error is added to the model. This greedy step is repeated until a maximum number of terms is reached or some other stopping criterion is fulfilled. 
	
	After the stopping criterion is met, the model is pruned by removing the term that leads to the smallest increase in residual square error. This produces an estimated best model $\widetilde{f} _{\lambda}$, for each number of terms $\lambda$. For each $\widetilde{f}_{\lambda}$ the generalized cross-validation 
	\begin{equation*}
	GCV_f(\lambda) = \frac{\sum_{i=1}^{n} (f_i - \widetilde{f}_{\lambda}(\bx_i))^2}{\left(1- \frac{ \mathcal{D}(\lambda)}{n} \right)^2},
	\end{equation*}
	is calculated, where $f_i$ is the response at data point $i$. $\mathcal{D}(\lambda)$ is the effective number of parameters as described in \cite{ESL}. Finally, the model that minimises $GCV_h(\lambda)$ is chosen to be $\widehat{f}$. In this work, we make use of the MARS implementation from the \texttt{R}-package ``earth" \cite{earthpackage}.

	\subsection{MCMC Algorithms}
		
	\subsubsection{Physical-informed proposal via preconditioned Crank–Nicolson algorithm}
	
	The primary challenge encountered in Bayesian methods lies in the slow convergence of posterior distribution sampling, a problem exacerbated as the dimension increases. To address this, a groundbreaking solution was presented in \cite{cotter2013mcmc}, introducing a transformative approach to Markov Chain Monte Carlo (MCMC) method that effectively eliminates the ``curse of dimensionality" for Gaussian priors, given appropriate assumptions on the likelihood. This innovative technique, known as the ``preconditioned Crank–Nicolson" (pCN) algorithm, exhibits dimension-independent sampling efficiency within its resulting Markov chain. The pCN algorithm, a modification of standard random-walk Metropolis sampling, scales the previous iteration's position before adding a random shift to generate a new proposal. Notably, the pCN proposal kernel arises from discretizing an Ornstein–Uhlenbeck process with an invariant measure, ensuring reversibility with respect to the prior. This unique feature simplifies the acceptance ratio to a straightforward ratio of the likelihood.

	The physical-informed prior distribution in our framework is a Gaussian distribution with non-zero mean $\bmu$ and covariance matrix $\bSigma$. Then the pCN proposal distribution, outlined in Algorithm \ref{alg_pCM}, at step $ k $, is as
	\begin{equation}
	q(\bx' | \bx^{(k)}  ) =  \phi \left(\bmu + \sqrt{1-s^2}(\bx^{(k)} - \bmu),  s^2 \bSigma \right), 
	\label{eq_pCB}
	\end{equation} 
	with $s>0$ being a tuning parameter. This proposal is reversible with respect to the prior distribution \citep{pinski2015algorithms}. It means that the acceptance probability reduces to the likelihood ratio
	$$
	A(\bx' , \bx^{(k)}) = \min \Bigg\{1, \frac{f(\by | \bx')}{f(\by | \bx^{(k)} )}\Bigg\}.
	$$
	This results in notable acceleration for problems discretized on an $ N $-dimensional grid. Moreover, the pCN method exhibits sampling efficiency that is independent of dimension, meaning that unlike the standard random walk, the number of steps required with pCN does not grow as the dimension $ N $ increases. Consequently, pCN efficiently explores the posterior support. The tuning of the parameter $s$ in pCN, as discussed in \cite{cotter2013mcmc}, is aimed at achieving an acceptance rate of approximately 23.4\%, \cite{gelman1997weak}.

	\begin{algorithm}[htb]
		\caption{Proposed sampling procedure}
		\label{alg_pCM}
		\begin{itemize}
			\item[1.] Choose the stepsize $ s \in (0,1) $. Set $ k = 0 $ and draw $ \bx^{(0)}  $ from the prior, see \eqref{eq_priors}.
			\item[2.] Draw $ \bxi \sim N(\mathbf{0} , \bSigma)$  (using FFT method).
			\\
			Then propose the new point as
			\[ 
			\bx' = \bmu + \sqrt{1-s^2} (\bx^{(k)} - \bmu) + s \bxi .
			\]
			\item[3.] Set $ \bx^{(k+1)} = \bx' $ with probability 
			$ A(\bx' , \bx^{(k)}) $.
			\\
			Otherwise set $ \bx^{(k+1)} = \bx^{(k)} $.
			\\
			Increment $ k $ by one and return to step 2.
		\end{itemize}
	\end{algorithm}

	Utilizing the pCN method within our framework, where the proposed approach involves incorporating a physically informed prior, enables effective exploration of various high-probability regions within the intricate posterior distribution. This aspect holds particular significance in the context of Bayesian inverse problems, as emphasized in previous works such as \cite{mosegaard1995monte,khoshkholgh2021informed,khoshkholgh2022full}.

	\subsubsection{Other MCMC samplers}
\label{sc_other_proposals}	
	Choosing an effective proposal distribution poses a significant and frequently challenging undertaking in the application of the Metropolis-Hastings (MH) algorithm. An optimal proposal distribution should adequately cover the tails of the posterior distribution while also bearing similarity to the posterior itself, as outlined by \cite{givens2012computational}. Additionally, the proposal distribution should facilitate straightforward sampling. The selection of the proposal distribution plays a crucial role in influencing the efficiency of the MH algorithm. In addressing our specific problem, we will explore various suggestions for proposal distributions, including the random walk, pCN, and Metropolis adjusted Langevin algorithm (MALA).
	
	Given the physics-informed prior as the Gaussian distribution, $ \mathcal{N}( \bmu, \bSigma) $ , in Section \ref{sc_prior_distribu}, with $ s>0 $ being a tuning parameter, the random walk proposal is as
	\begin{equation*}
	q_{1}(\bx^p | \bx_{t}) = \phi(\bx_{t},s^2 \bI),
	\end{equation*}
	random walk with covariance from the prior
	\begin{equation*}
	q_{2} (\bx^p | \bx_{t}) = \phi(\bx_{t},s^2 \bSigma),
	\end{equation*}
	the preconditioned Crank-Nicolson proposal in \eqref{eq_pCB} denoted by $ q_3 $, and the MALA
	\begin{equation*}
	q_{4} (\bx^{p} | \bx_{t}) = \phi \left( \bx_{t} + \frac{s^2}{2} \nabla \log \left( \pi (\bx_{t})\right), s^2 I \right).
	\end{equation*}

	\subsubsection{Sampling and evaluating large Gaussian random fields via FFT}

	Let us emphasize that the random variable dimensionality is \(N = 44,144\). The proposed methodology involves obtaining high-dimensional Gaussian random fields from a distribution of this size using the fast Fourier transform (FFT) algorithm, as detailed in \cite{davies2013circulant}. Similar techniques were previously employed in expedited 3D elastic inversion, as seen in \cite{buland2003rapid}. 
	
	The FFT algorithm, embedding the covariance matrix onto a torus, avoids the need to construct the full prior covariance matrix for the entire domain, enabling efficient realization generation in the Fourier domain. The general idea is to use structure of circulant matrices and perform matrix-vector multiplication in a lower-dimensional space in the Fourier domain.  Additional details can be found in \ref{sc_detail_FFT}. In the context of sampling high-dimensional Gaussian random fields, the approach overcomes memory limitations by leveraging circulant matrix structures and Fourier domain transformations, as documented in \cite{chan1997algorithm, rue2005gaussian, davies2013circulant,abrahamsen2018simulation}. This method facilitates rapid sample generation and proves computationally advantageous for estimating log-density in scenarios requiring the evaluation of probability density functions, such as calculating acceptance rates in 
	MCMC algorithms.

\section{Numerical study: small-scale data}
	\label{sc_result_simulation}
In this section, in Subsection \ref{sc_evaluate_MARS_against_true}, we initially conduct numerical investigations to assess the efficacy of MARS in approximating the true forward function. Subsequently, in Subsection \ref{sec:compare_small_area}, we scrutinize comparisons between MCMC samples derived from the true model and those from the approximated model within a limited zone of the Alvheim field. Finally, in Subsection \ref{sc_comparing_proposals}, we assess the efficiency of various proposal distributions.

\subsection{Evaluating approximation of the forward model}
\label{sc_evaluate_MARS_against_true}
To approximate the forward model in the Alvheim case, a training data set is created. The data set has $20,000$ observations of covariates $x_g$, $x_o$, $x_c$ and $x_d$. The value of data point $i$ at covariate $j$ is denoted by $x_{ij}$, where $j = g, o, c, d$ corresponds to gas, oil, clay and depth respectively.
Data point $i$ is sampled as follows: First, a value for $x_{id}$ is sampled from a uniform distribution on the interval between the minimum and maximum depth at the top reservoir of the Alvheim field shown in Figure \ref{fig:depthfield}. Next $x_{id}$ is used to find the expected value of the prior and draw $x_{ig}$, $x_{io}$ and $x_{ic}$ from the prior distribution. The test data are created in the same manner as the training data. It has $44144$ test data points, which is the size of one sample $\bx$ at the Alvheim field.

Recall that there are two types of seismic AVO data, the zero offset ($R_0$) and the seismic gradient ($G$). That is, $\bh : \mathbb{R}^4 \to \mathbb{R}^2$. To approximate $\bh$, two models are created. A model for the zero offset, $\widehat{h_{R_0}}$, and one for the seismic gradient $\widehat{h_{G}}$ such that $\widehat{\bh}$ collectively denotes the two models $\widehat{h_{R_0}}$ and $\widehat{h_{G}}$ with $\widehat{h_{R_0}} : \mathbb{R}^4 \to \mathbb{R}$ and $\widehat{h_{G}} : \mathbb{R}^4 \to \mathbb{R}$. 

As candidates for approximating $\bh$, the MARS models  $\widehat{\bh}_{\text{MARS}}$ and $\widehat{\bh}_{\text{MARS}_{\text{OF}}}$ are trained on the $20,000$ data points. The large MARS models for the two AVO properties $h_{R_0}$ and $h_{G}$ have $39$ and $40$ terms of which $26$ and $24$ are interaction terms respectively. The pruned MARS model for $h_{R_0}$ has $25$ terms of which $17$ are interaction terms, while the pruned MARS model for $h_G$ has $25$ of which $13$ are interaction terms. In addition the kernel regression models $\widehat{\bh}_{\text{NPKR}_{1000}}$ and $\widehat{\bh}_{\text{NPKR}_{4000}}$ are trained, as described in  \ref{sec:npkr}, with $1000$ and $4000$ of the training data respectively. They are trained on subsets of the training data set because of computation time. As the NPKR models are weighted averages of the training data, the number of training data affects the prediction time. The subsets of the training data are random subsets, however one could try to select these points in a more optimal way to improve the NPKR models as discussed in \ref{sec:npkr}.

The accuracy of the predictions are examined by evaluating the sample correlation 
\begin{equation}
\text{Corr}\left(f, \widehat{f}\right) = \frac{\sum_{i=1}^{n} \left(f - \bar{f}\right)\left(\widehat{f}_i- \bar{\widehat{f}}\right)}{\sqrt{\sum_{i=1}^{n} \left(f_i - \bar{f}\right)^2} \sqrt{\sum_{i=1}^{n} \left(\widehat{f}_i - \bar{\widehat{f}}\right)^2}}.
\label{sample_corr}
\end{equation}
and MSE
\begin{equation}
\text{MSE}\left(f, \widehat{f}\right) = \frac{1}{n} \sum_{i=1}^{n}\left(f_i - \widehat{f}_i \right)^2. 
\label{mse}
\end{equation}
Precision metrics, such as correlation and MSE are the mean of these metrics for the two models. That is, the correlation between a model $\widehat{\bh}$ and the response $\bh$ is the mean of $\text{Corr}\left(h_{R_0}, \widehat{h_{R_0}}\right)$ and $\text{Corr}\left(h_{G}, \widehat{h_{G}}\right)$ calculated using equation \eqref{sample_corr}. Similarly, the MSE is the mean of $\text{MSE}\left(h_{R_0}, \widehat{h_{R_0}}\right)$ and $\text{MSE}\left(h_{G}, \widehat{h_{G}}\right)$ from equation \eqref{mse}. Moving on $\widehat{\bh}_{\text{MARS}}$, $\widehat{\bh}_{\text{MARS}_{\text{OF}}}$, $\widehat{\bh}_{\text{NPKR}_{1000}}$ and $\widehat{\bh}_{\text{NPKR}_{4000}}$ are referred to as one model, although they are two models, as described in the second paragraph of this section. 

The computation time for predicting the test data, correlation and MSE are reported in Table \ref{table:h_hatt}. The computation time  is the average of predicting the test data 50 times. For comparison, the computation time of the true forward model used on average $2.059$ seconds to compute $\bh(\bx)$. The fastest model was $\widehat{\bh}_{\text{MARS}}$, which was on average $ 34 $ times proximately faster than the true forward model. The larger MARS model was also very fast, although slower than the smaller MARS model. The larger MARS model is marginally better than $\widehat{\bh}_{\text{MARS}}$ when considering the correlation and MSE, however, it is much more complex. By setting the maximum number of terms in the MARS model after pruning to a lower number, the computation time is reduced. This is at a cost of increasing MSE and decreasing correlation.

The correlation achieved by $\widehat{\bh}_{\text{NPKR}_{4000}}$ approaches that of the MARS models, the MARS models demonstrate an MSE that is half that of $\widehat{\bh}_{\text{NPKR}_{4000}}$, along with computation times that are, on average, over 200 times faster. Notably, $\widehat{\bh}_{\text{NPKR}_{1000}}$ displays the poorest performance in terms of both MSE and correlation. The results in Table \ref{table:h_hatt} reveal significantly slower computational speeds for the NPKR (kernel regression) models compared to the MARS models, including the forward function $\bh$. Moreover, the computational time increases with the augmentation of data points in the NPKR model, evident in the slower computation of $\widehat{\bh}_{\text{NPKR}_{4000}}$ compared to $\widehat{\bh}_{\text{NPKR}_{1000}}$. Beyond the reported computation times in Table \ref{table:h_hatt}, an additional test involved predicting test data using an NPKR model trained with 100 data points, which averaged 0.674 seconds for prediction. Although this time is considerably shorter than that of $\widehat{\bh}_{\text{NPKR}_{4000}}$ and $\widehat{\bh}_{\text{NPKR}_{1000}}$, it still remains significantly longer than the computation times observed for the MARS models.

{\renewcommand{\arraystretch}{1.6}%
	\begin{table}[htb]
		\centering
		\begin{tabular}{|c|c|c|c|} 
			\hline 
			$\widehat{\bh}$ & correlation & MSE$\times 10^{-5} $ & computation time [sec] \\
			\hline
			$\widehat{\bh}_{\text{MARS}}$ & 0.995 & 3.8 & 0.06
			\\ 
			$\widehat{\bh}_{\text{MARS}_{\text{OF}}}$ & 0.996 & 3.4 & 0.11
			\\ 
			$\widehat{\bh}_{\text{NPKR}_{1000}}$ & 0.983 & 14.0 & 5.69
			\\ 
			$\widehat{\bh}_{\text{NPKR}_{4000}}$ & 0.991 & 7.2 & 22.7
			\\
			\hline
		\end{tabular}
		\caption{Correlation, MSE and computation time for $\widehat{\bh}_{\text{MARS}}$, $\widehat{\bh}_{\text{MARS}_{\text{OF}}}$, $\widehat{\bh}_{\text{NPKR}_{1000}}$ and $\widehat{\bh}_{\text{NPKR}_{4000}}$.}
		\label{table:h_hatt}
\end{table}}

\begin{figure}[htb]
	\centering
	\begin{subfigure}[b]{0.17\textwidth}
		\centering
		\includegraphics[width=\textwidth]{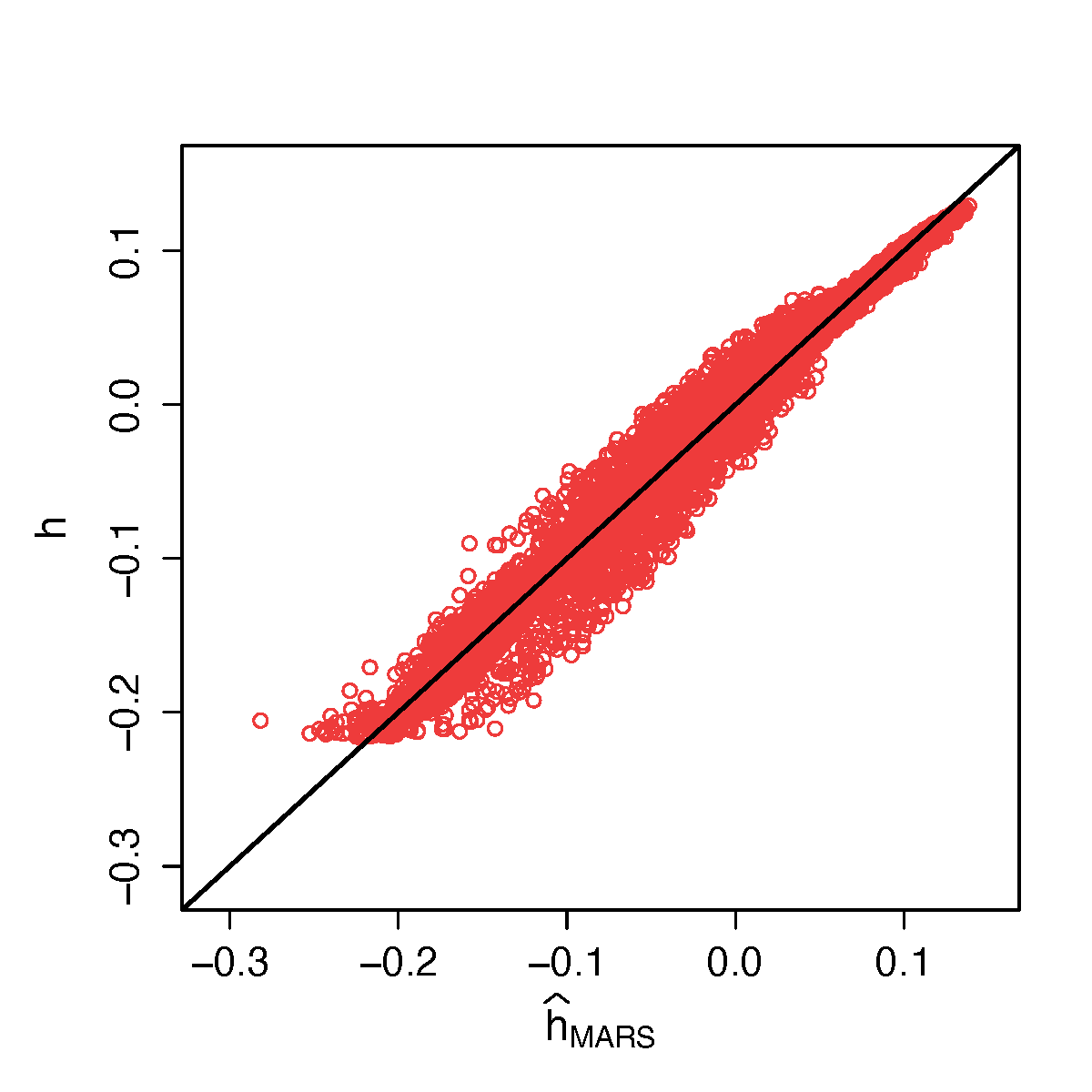}
		\caption{True forward model values for $h_{R_0}$ plotted against $\widehat{\bh}_{\text{MARS}}$.}
		\label{fig:R0_sys_err}
	\end{subfigure}	
\hfill
	\begin{subfigure}[b]{0.17\textwidth}
		\centering
		\includegraphics[width=\textwidth]{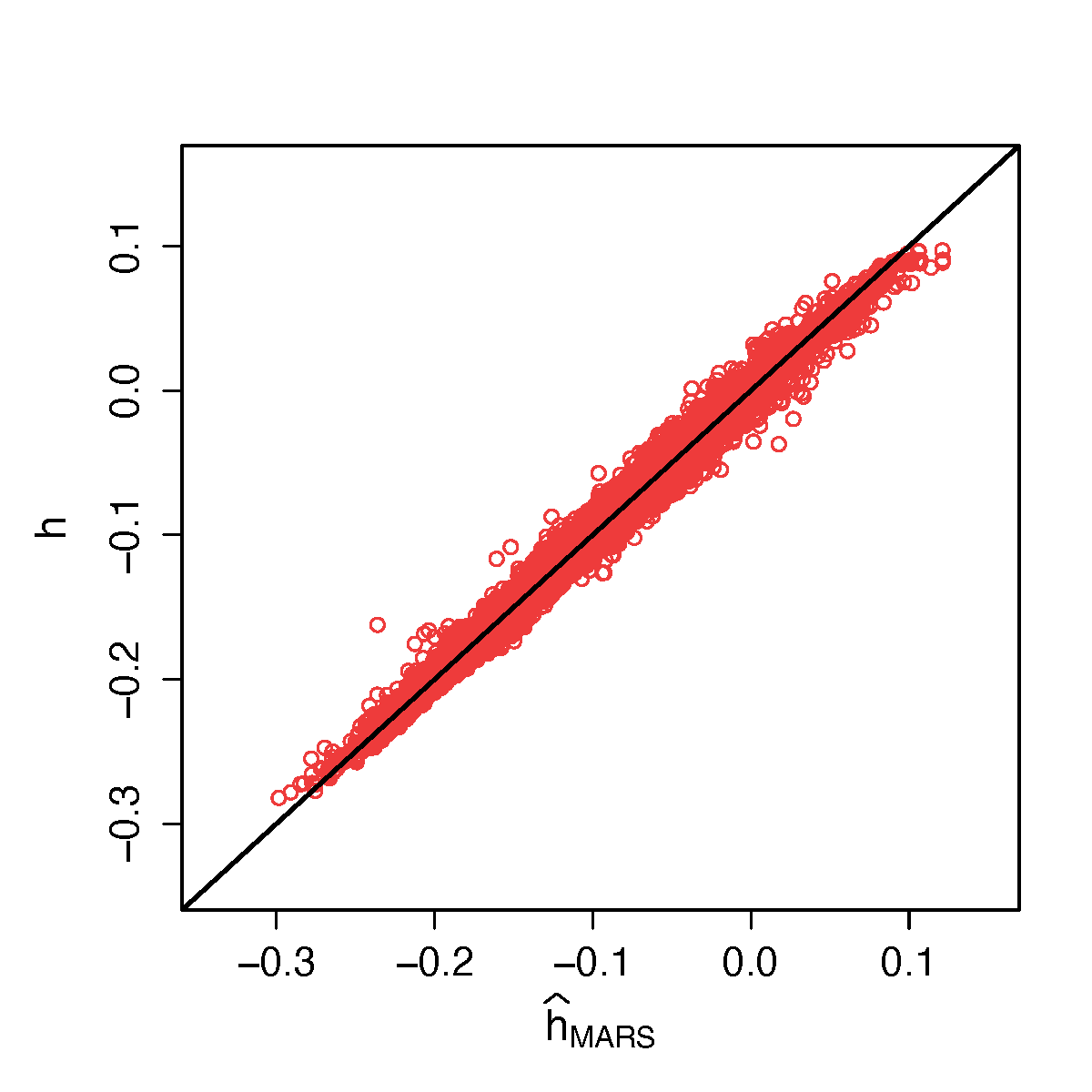}
		\caption{True forward model values for $h_{G}$ plotted against $\widehat{\bh}_{\text{MARS}}$.}
		\label{fig:G_sys_err}
	\end{subfigure}	\label{fig:sys_err_bf}
\hfill
	\begin{subfigure}[b]{0.17\textwidth}
		\centering
		\includegraphics[width=\textwidth]{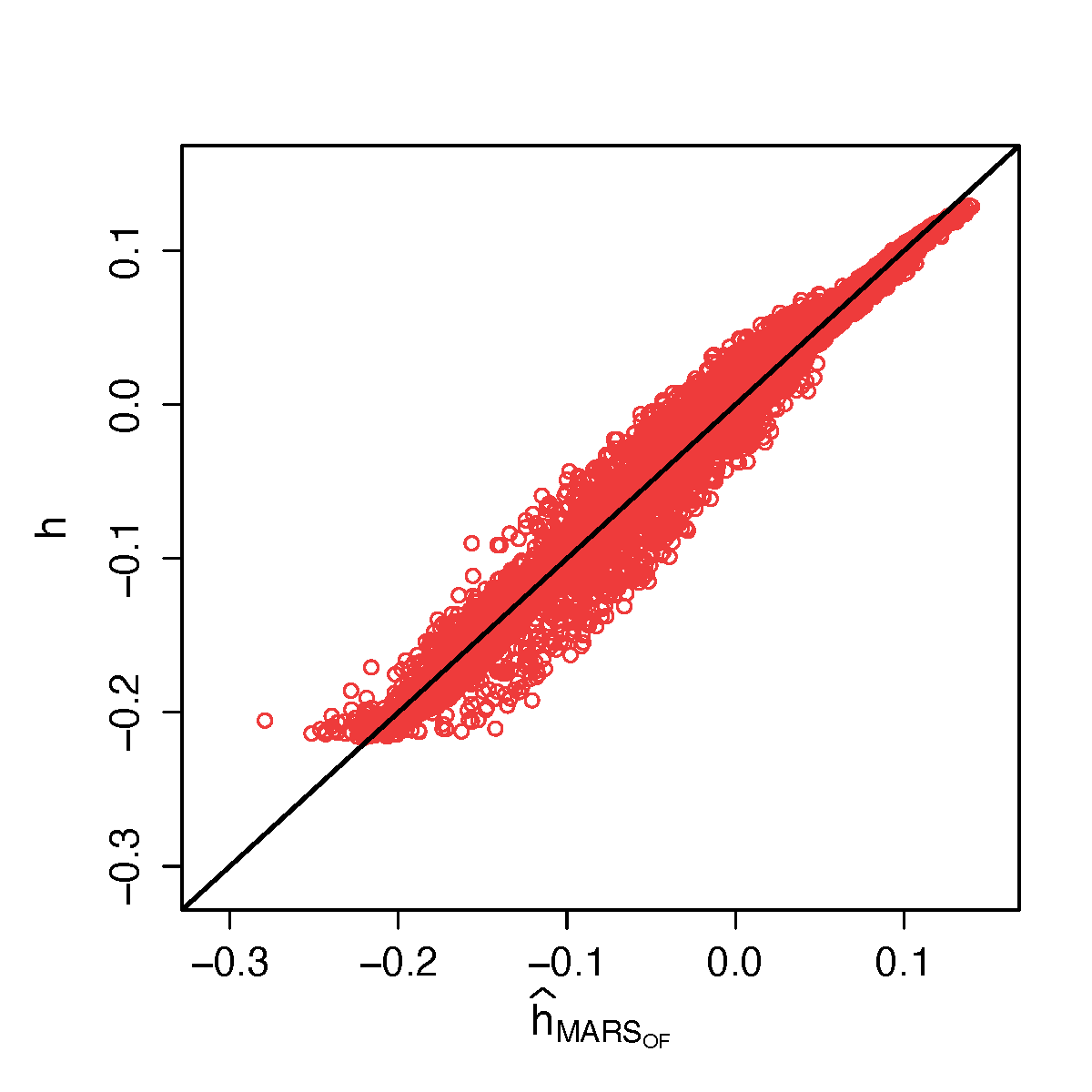}
		\caption{True forward model values for $h_{R_0}$ plotted against $\widehat{\bh}_{\text{MARS}_{\text{OF}}}$.}
		\label{fig:R0_sys_err_of}
	\end{subfigure}
\hfill
	\begin{subfigure}[b]{0.17\textwidth}
		\centering
		\includegraphics[width=\textwidth]{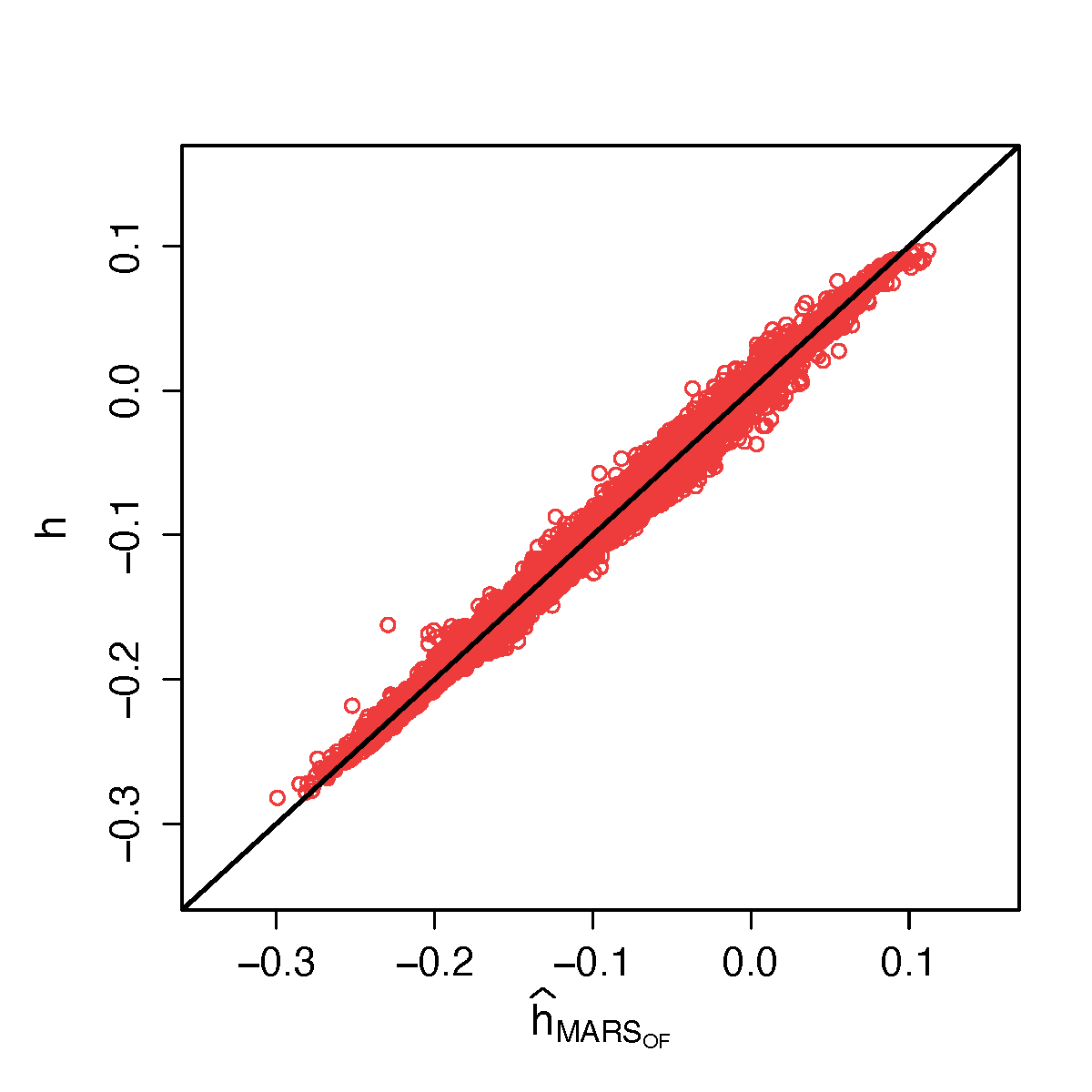}
		\caption{True forward model values for $h_{G}$ plotted against $\widehat{\bh}_{\text{MARS}_{\text{OF}}}$.}
		\label{fig:G_sys_err_of}
	\end{subfigure}
	\caption{True forward model plotted as a function of predicted forward models on test data.}
	\label{fig:sys_err_of}
\end{figure}

\begin{figure}[htb]
	\centering
	\includegraphics[width=0.17\textwidth]{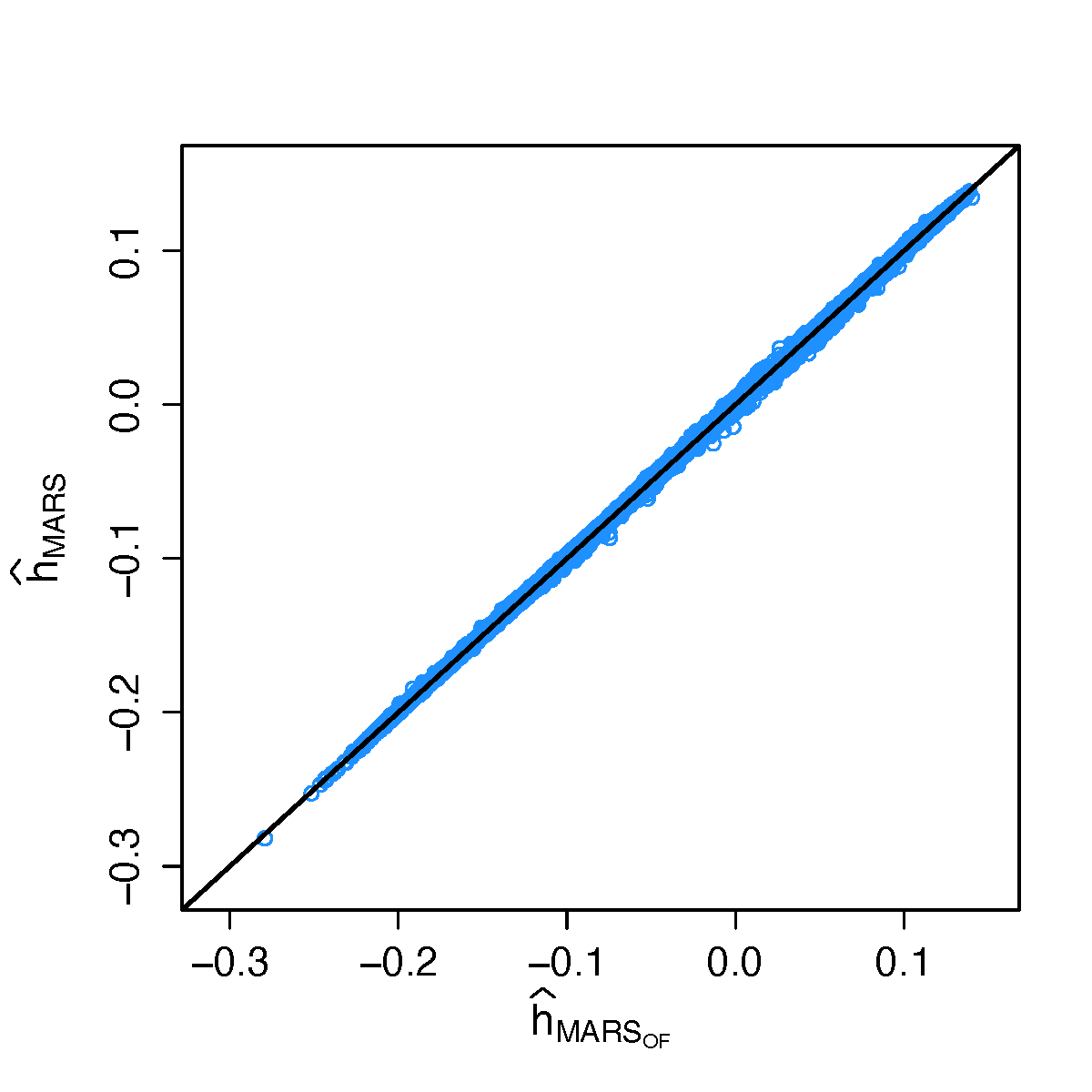}
	\caption{Predictions of $h_{R_0}$ from $\widehat{\bh}_{\text{MARS}}$ plotted as a function of predictions of $h_{R_0}$ from $\widehat{\bh}_{\text{MARS}_{\text{OF}}}$. The black line is $\widehat{\bh}_{\text{MARS}}$ = $\widehat{\bh}_{\text{MARS}_{\text{OF}}}$.}
	\label{fig:mars_R0_of_vs_bf}
\end{figure}

Figure \ref{fig:sys_err_of} shows a scatter plot of the $\bh$ versus $\widehat{\bh}_{\text{MARS}}$, $\widehat{\bh}_{\text{MARS}_{\text{OF}}}$ on the test data. Overall the figures indicate that the errors are slightly smaller for the predictions of the seismic gradient than for the zero offset. Neither $\widehat{\bh}_{\text{MARS}}$ nor $\widehat{\bh}_{\text{MARS}_{\text{OF}}}$ seems to make grave systematic errors, except perhaps for the lowest values of $h_{R_0}$. The predictions in Figures \ref{fig:R0_sys_err} and \ref{fig:R0_sys_err_of} look almost identical. This is confirmed by Figure  \ref{fig:mars_R0_of_vs_bf}, where the predictions of $\widehat{\bh}_{\text{MARS}}$ are plotted against the predictions of $\widehat{\bh}_{\text{MARS}_{\text{OF}}}$. The $14$ more terms in model $\widehat{\bh}_{\text{MARS}_{\text{OF}}}$ seems excessive when predictions are close to identical.

Section \ref{sc_result_realdata} gives results from MCMC simulations at the Alvheim field using $\widehat{\bh}_{\text{MARS}}$ as a replacement of $\bh$. This model was chosen because it was the fastest, and had marginally higher MSE and lower correlation than $\widehat{\bh}_{\text{MARS}_{\text{OF}}}$. Considering correlation and MSE, both MARS models would serve as good substitutes for the forward model, while reducing the computation time considerably. 

In pursuit of computational efficiency, the NPKR models investigated in this study do not serve as optimal replacements for the forward model. An inherent drawback of the NPKR model is its diminished performance with an increased volume of training data, thereby decelerating model execution. The reduction in training data quantity directly correlates with decreased computation time for predictions. For instance, conducting predictions on 50 test data points  utilizing an NPKR model trained on 100 data points resulted in an average duration of 0.674 seconds, which is swifter compared to the forward model ($\bh$). Nonetheless, this pace remains significantly slower than that achieved by MARS models, and it comes with the trade-off of elevated MSE and diminished correlation. Implementing a more strategic data sampling methodology may enhance correlation and mitigate MSE; however, constructing a model based on four thousand data points exhibited a correlation proximate to MARS models while incurring substantially higher MSE.

\subsection{Comparison of the surrogate and the exact forward models} \label{sec:compare_small_area}

In Section \ref{sc_evaluate_MARS_against_true}, the analysis demonstrates that utilizing the MARS surrogate resulted in a speed enhancement of approximately $32$ times when employing the true forward function. Furthermore, the MARS surrogate exhibited a low MSE and a strong correlation. However, given its nature as an approximation, the MCMC samples are drawn from a posterior distribution which serves as an approximation to the true posterior distribution, denoted as $\pi$. For clarity, the approximated posterior, where $\bh$ is substituted by $\widehat{\bh}_{\text{MARS}}$, is labeled as $\widehat{\pi}$. To assess the resemblance between $\widehat{\pi}$ and $\pi$, MCMC samples are generated using both $\bh$ and $\widehat{\bh}_{\text{MARS}}$ as forward models. Additionally, MCMC samples employing $\widehat{\bh}_{\text{MARS}}$ as the forward model and $\widetilde{\bOmega}$, as defined in equation \eqref{Omega_tilde}, as the covariance matrix of the likelihood model, are examined to determine whether samples from this posterior are more akin to those from $\pi$ than those from $\widehat{\pi}$. The posterior with the forward function $\widehat{\bh}_{\text{MARS}}$ and the covariance matrix $\widetilde{\bOmega}$ in the likelihood model is denoted as $\widetilde{\pi}$. The details regarding the approximation $\widehat{\bh}_{\text{MARS}}$ and the covariance matrix $\widetilde{\bOmega}$ are provided in \ref{sec_omega_tilde}.

\begin{figure}[htb]
	\centering
	\includegraphics[width=0.3\textwidth]{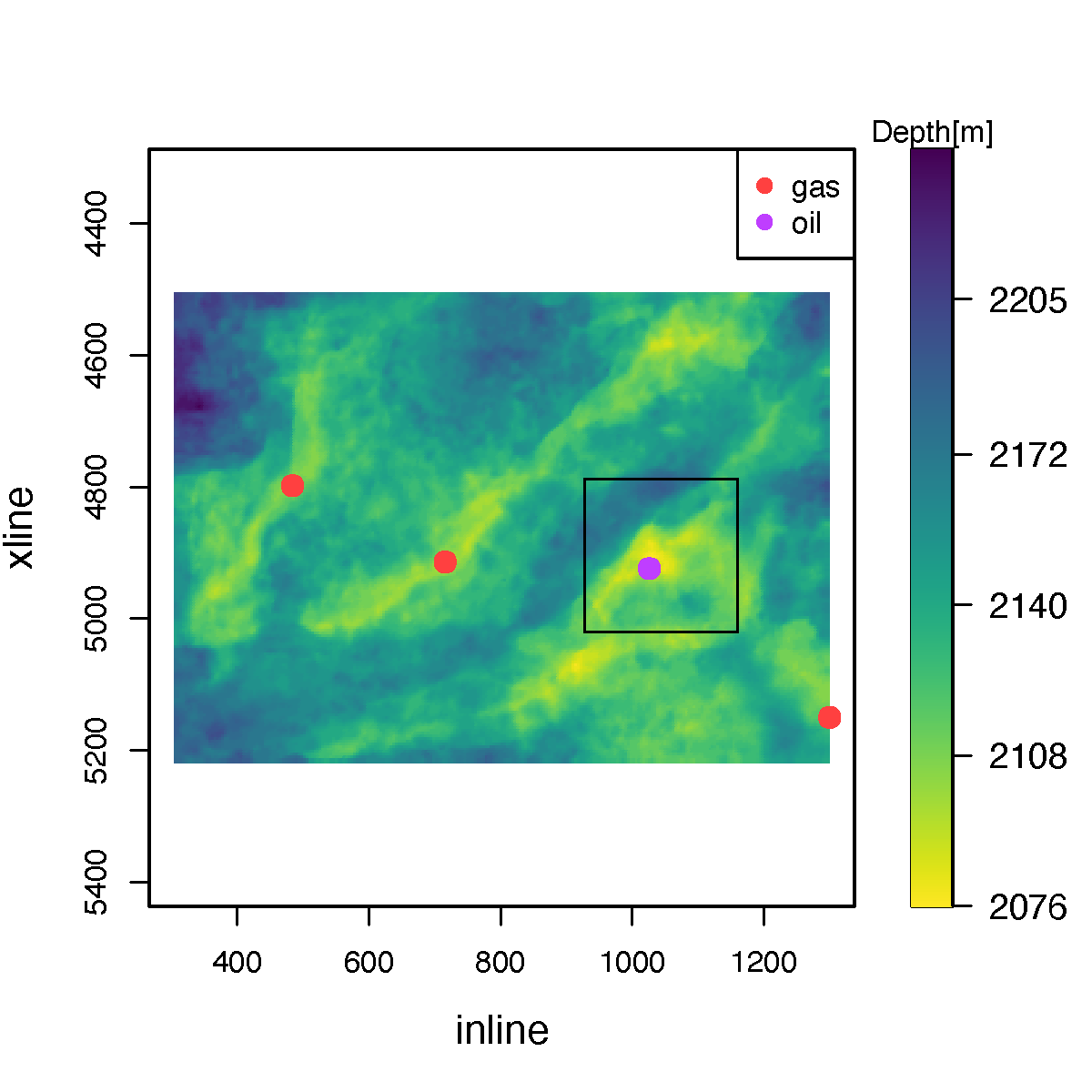}
	\caption{The black square is a smaller area where MCMC with the true forward function is performed.}
	\label{fig:depth_w_area}
\end{figure}

The MH algorithm with proposal distribution $q_2$ is used to sample the posteriors $\pi$, $\widehat{\pi}$ and  $\widetilde{\pi}$. The tuning parameter $s$ is tuned such that the acceptance rate is approximately $23.4\%$, \cite{gelman1997weak}. The value $s=0.027$ was used, which gave the acceptance rate $23.5\%$ when using the true forward model, $23.4\%$ when using $\widehat{\bh}_{\text{MARS}}$ and $23.5\%$ when using $\widehat{\bh}_{\text{MARS}}$ and $\widetilde{\bOmega}$ as covariance matrix of the likelihood. The initial sample was the mean of the prior distribution. The number of iterations was $10^6$ where every $10$th sample is saved. The initial portion of the MCMC samples is discarded as burn-in, ensuring that the subsequent $50,000$ samples are presumed to be drawn from the stationary distributions of the Markov chains.

Due to computation time, this was only done for a smaller area located at inlines$\times$crosslines $ [928, 1160]\times[4788, 5020]$, where every other inline and crossline is omitted, such that the location of the measurements are approximately $100$ meters apart. This means that the smaller area is discretized by $30 \times 30 = 900$ grid points. Hence $\bx_g, \bx_o, \bx_{c} \in \mathbb{R}^{900}$. Figure \ref{fig:depth_w_area} shows the smaller area at the Alvheim field where the MCMC was performed. In the vicinity of the well, the samples are subject to the influence of the likelihood model for the well-log data. Therefore, it is of particular interest to scrutinize the similarities and dissimilarities, with a focus on locations farther away from the well.

\begin{figure}[H]
	\centering
	\begin{subfigure}[b]{0.3\textwidth}
		\centering\captionsetup{width=.7\linewidth}
		\includegraphics[width=.7\textwidth]{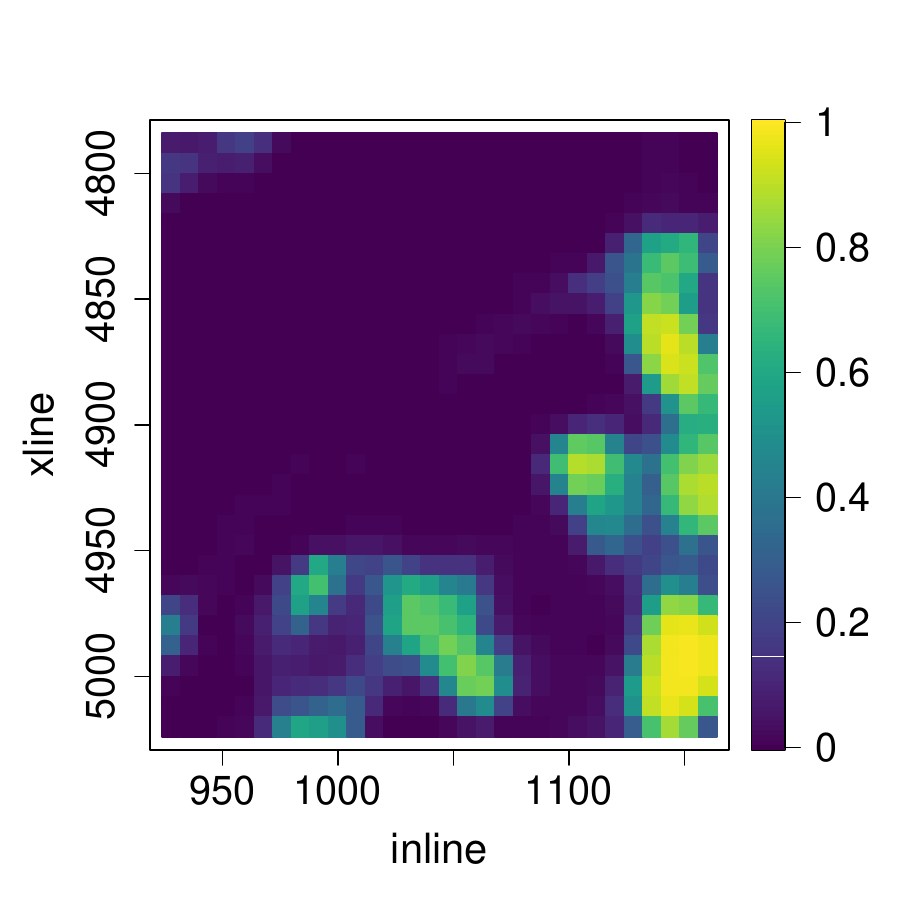}
		\caption{Mean gas saturation from MCMC samples from $\pi$.}
		\label{fig:small_area_gas_sfm}
	\end{subfigure}
	\begin{subfigure}[b]{0.3\textwidth}
		\centering\captionsetup{width=.7\linewidth}
		\includegraphics[width=.7\textwidth]{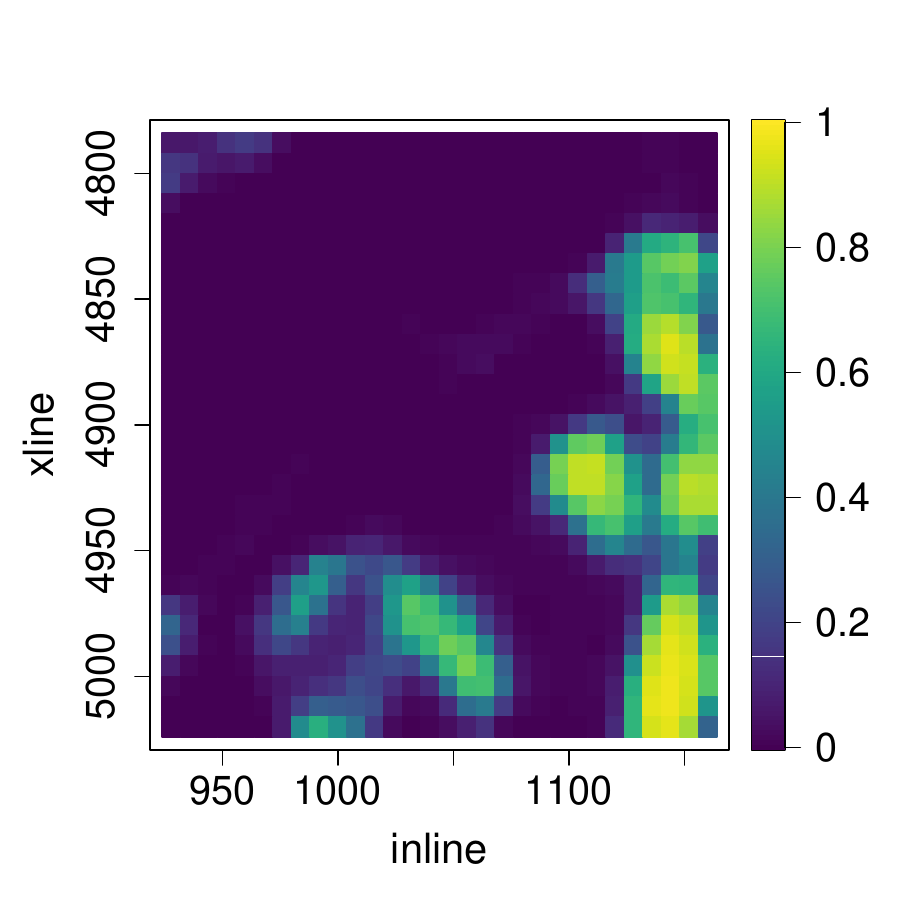}
		\caption{Mean gas saturation from MCMC samples from $\widehat{\pi}$.}
		\label{fig:small_area_gas_mars}
	\end{subfigure}
	\begin{subfigure}[b]{0.3\textwidth}
		\centering\captionsetup{width=.7\linewidth}
		\includegraphics[width=.7\textwidth]{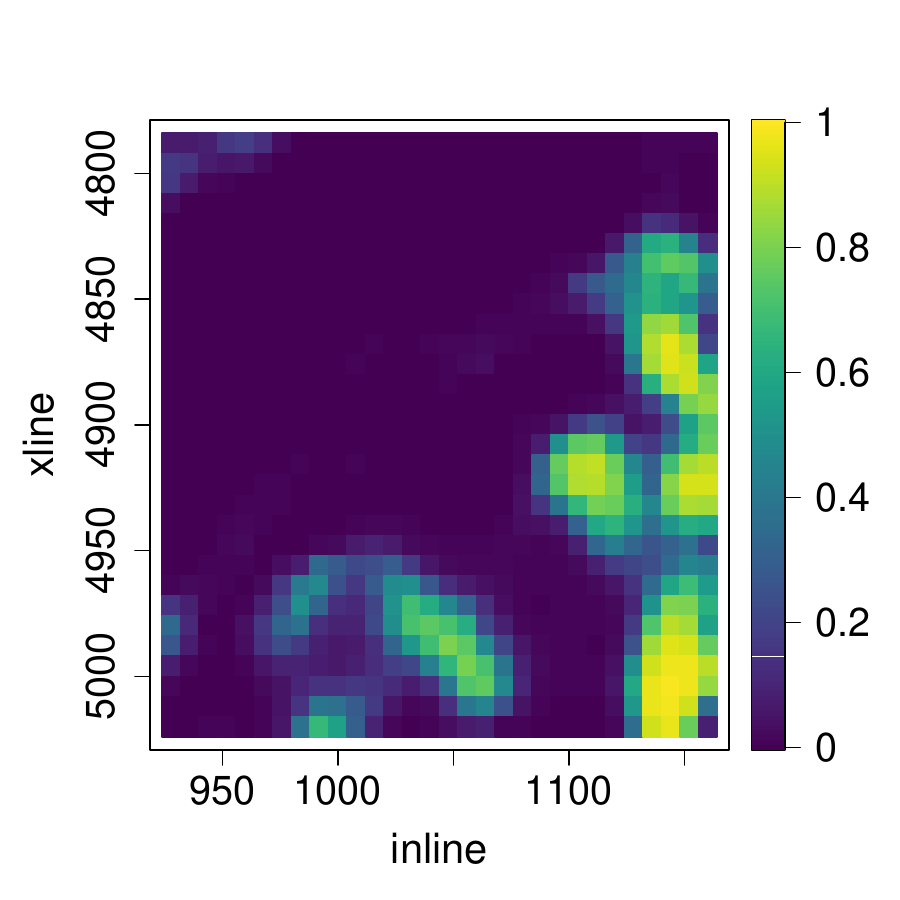}
		\caption{Mean gas saturation from MCMC samples from $\widetilde{\pi}$.}
		\label{fig:small_area_gas_mars_tilde}
	\end{subfigure}
	\begin{subfigure}[b]{0.3\textwidth}
		\centering\captionsetup{width=.7\linewidth}
		\includegraphics[width=.7\textwidth]{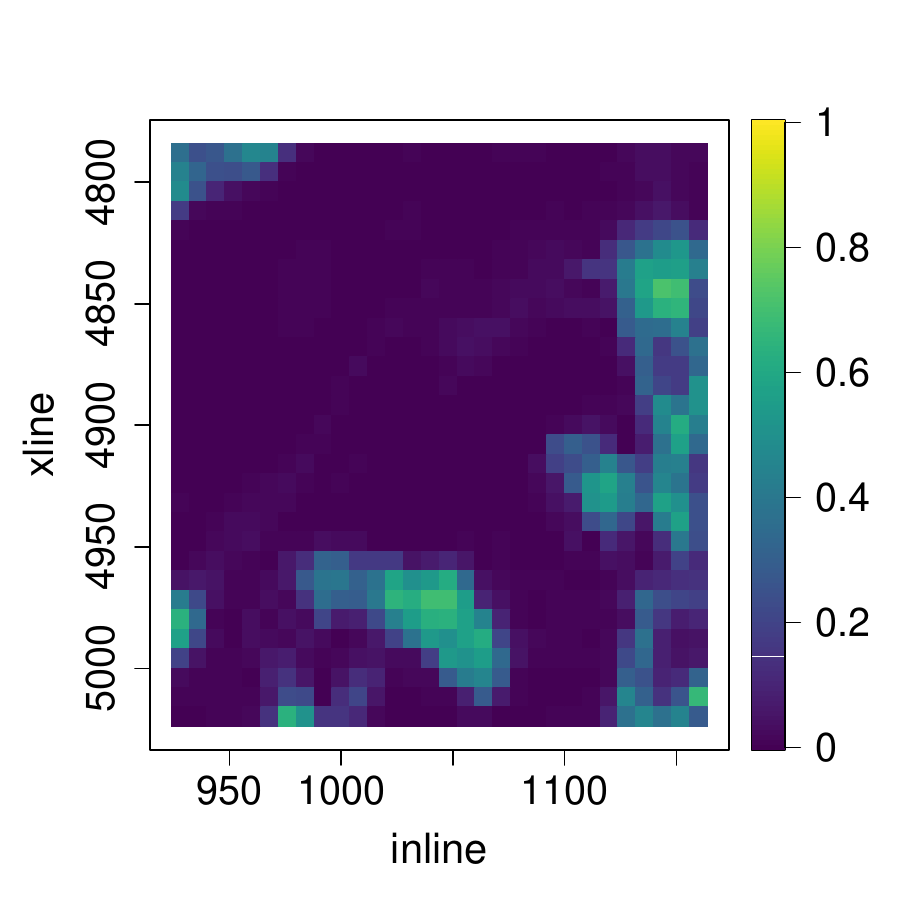}
		\caption{Uncertainty of gas saturation from MCMC samples from $\pi$.}
		\label{fig:small_area_uncertainty_gas_sfm}
	\end{subfigure}
	\begin{subfigure}[b]{0.3\textwidth}
		\centering\captionsetup{width=.7\linewidth}
		\includegraphics[width=.7\textwidth]{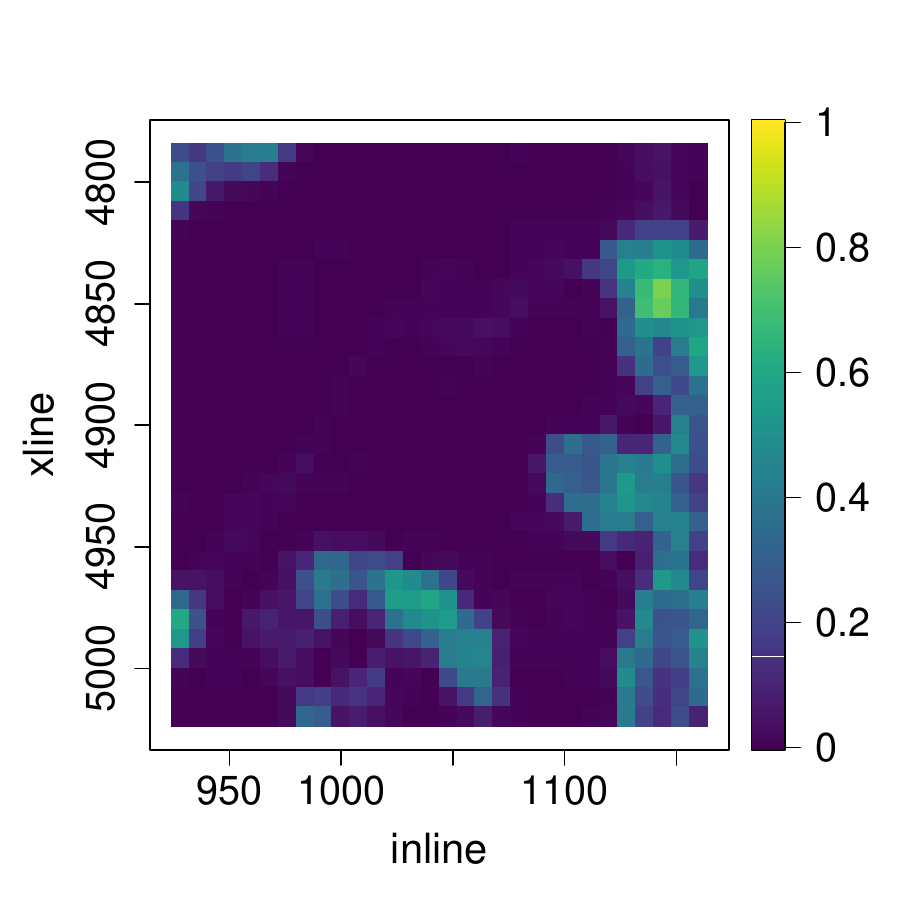}
		\caption{Uncertainty of gas saturation from MCMC samples from $\widehat{\pi}$.}
		\label{fig:small_area_uncertainty_gas_mars}
	\end{subfigure}
	\begin{subfigure}[b]{0.3\textwidth}
		\centering\captionsetup{width=.7\linewidth}
		\includegraphics[width=.7\textwidth]{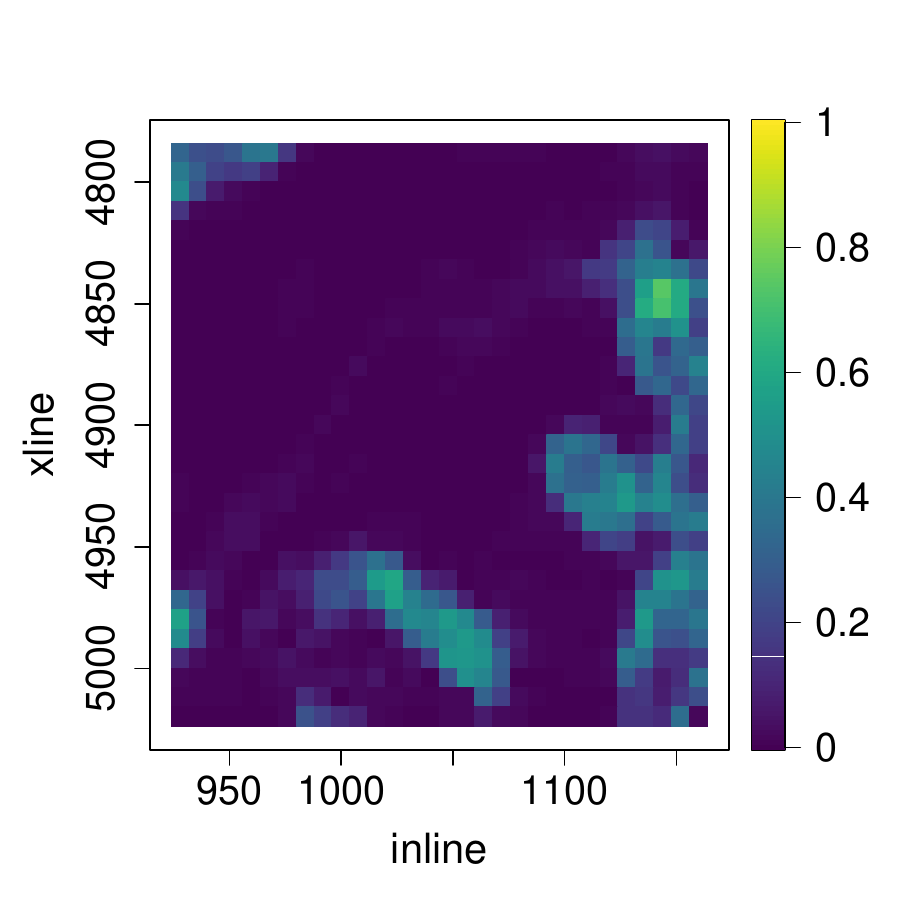}
		\caption{Uncertainty of gas saturation from MCMC samples from $\widetilde{\pi}$.}
		\label{fig:small_area_uncertainty_gas_mars_tilde}
	\end{subfigure}
	\caption{Mean gas saturations and uncertainties from MCMC samples from three posteriors.  The posterior where $\widehat{\bh}_{\text{MARS}}$ is surrogate is $\widehat{\pi}$. In the last posterior, $\widetilde{\pi}$, both the surrogate $\widehat{\bh}_{\text{MARS}}$ and the error correcting covariance matrix $\widetilde{\bOmega}$ are used. The MH algorithm with proposal distribution $q_2$ with $s=0.027$ was used in all three cases. Uncertainty is expressed as the difference between the 90th and 10th quantiles.}
	\label{fig:gas_small_area}

	\centering
	\begin{subfigure}[b]{0.3\textwidth}
		\centering\captionsetup{width=.7\linewidth}
		\includegraphics[width=.7\textwidth]{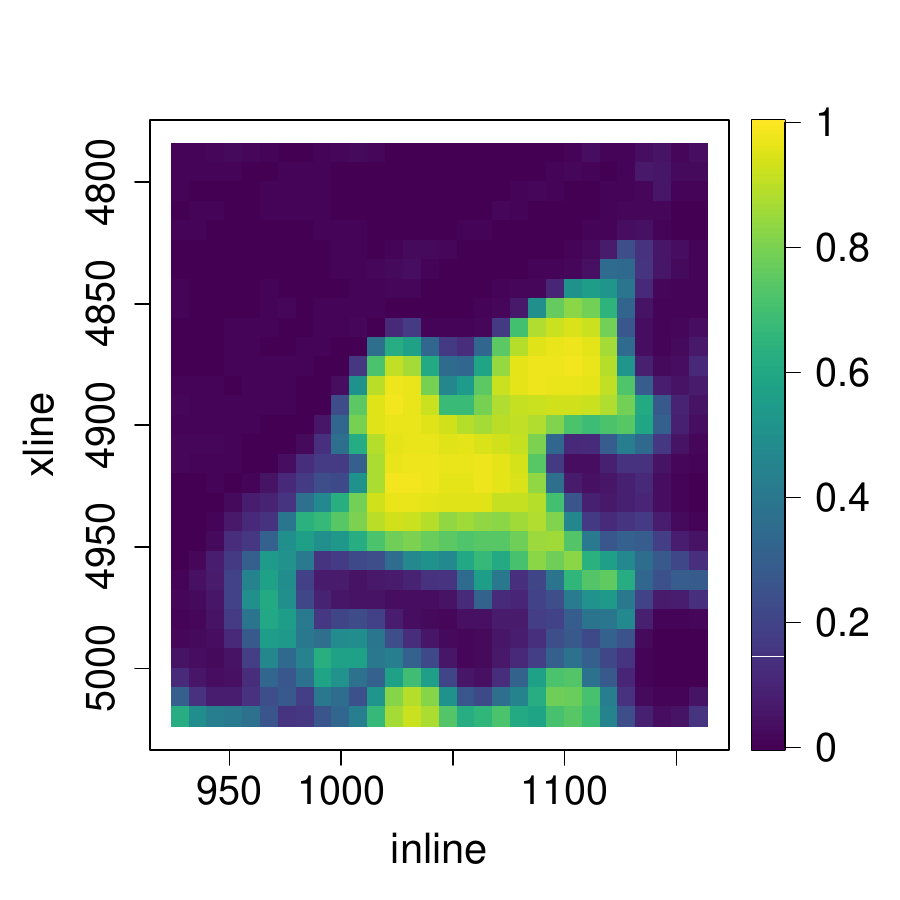}
		\caption{Mean oil saturation from MCMC samples from $\pi$.}
		\label{fig:small_area_oil_sfm}
	\end{subfigure}
	\begin{subfigure}[b]{0.3\textwidth}
		\centering\captionsetup{width=.7\linewidth}
		\includegraphics[width=.7\textwidth]{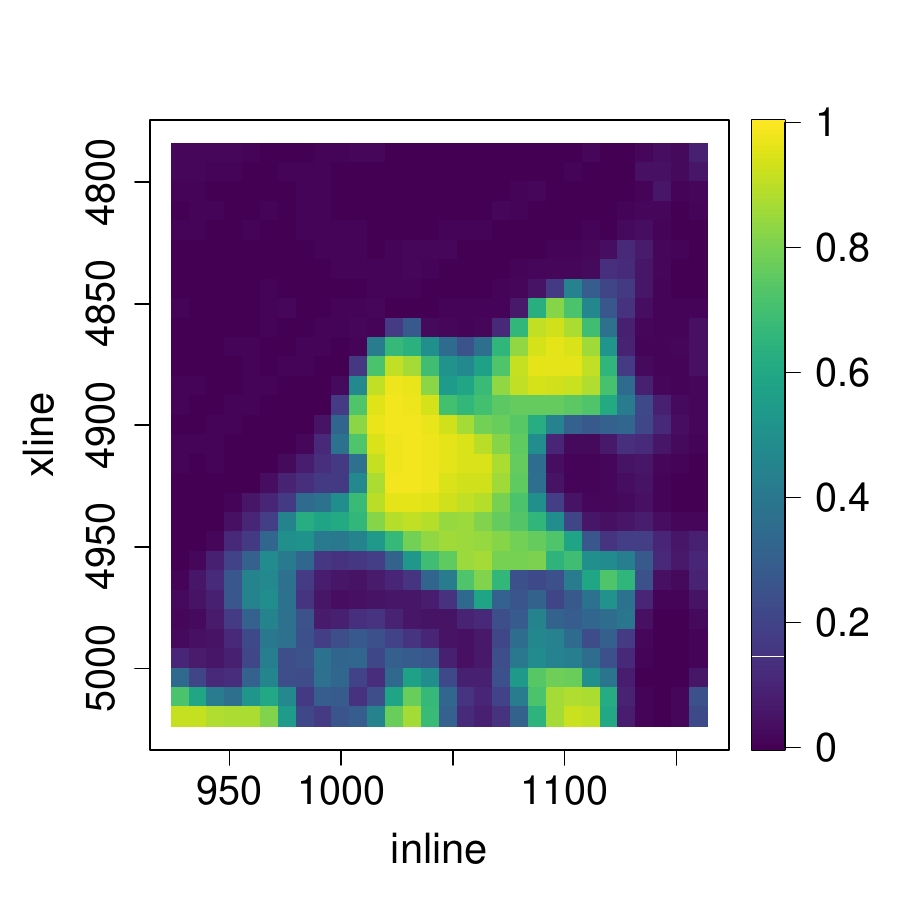}
		\caption{Mean oil saturation from MCMC samples from $\widehat{\pi}$.}
		\label{fig:small_area_oil_mars}
	\end{subfigure}
	\begin{subfigure}[b]{0.3\textwidth}
		\centering\captionsetup{width=.7\linewidth}
		\includegraphics[width=.7\textwidth]{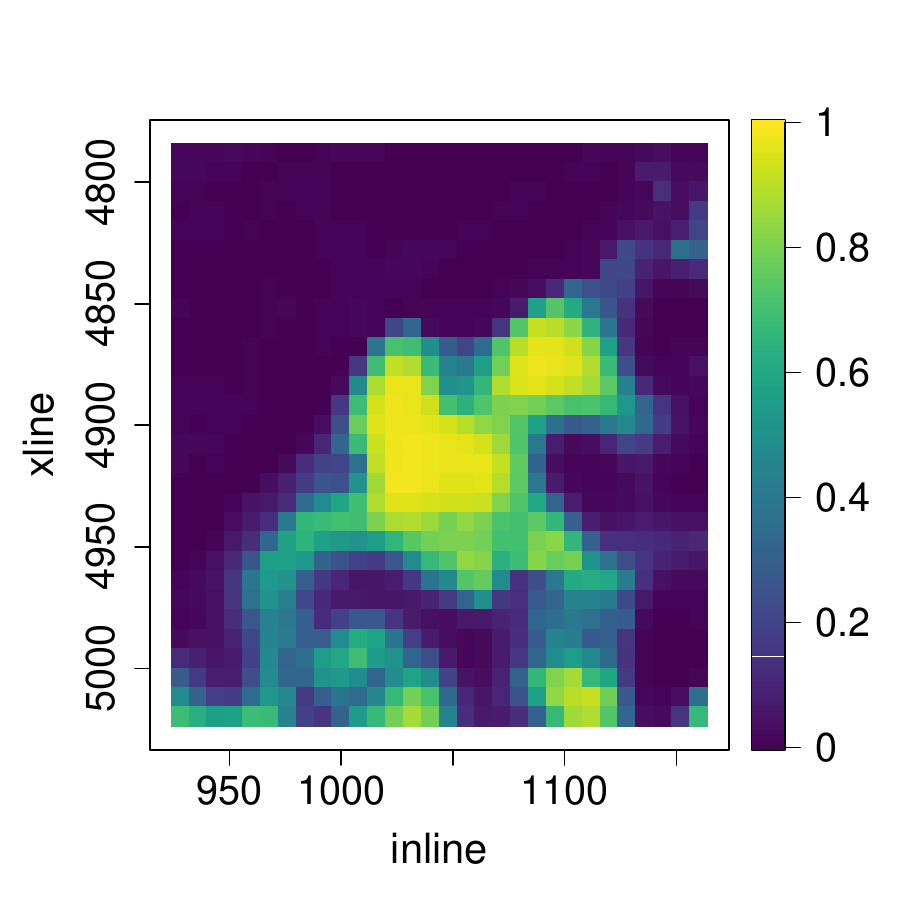}
		\caption{Mean oil saturation from MCMC samples from $\widetilde{\pi}$.}
		\label{fig:small_area_oil_mars_tilde}
	\end{subfigure}
	\begin{subfigure}[b]{0.3\textwidth}
		\centering\captionsetup{width=.7\linewidth}
		\includegraphics[width=.7\textwidth]{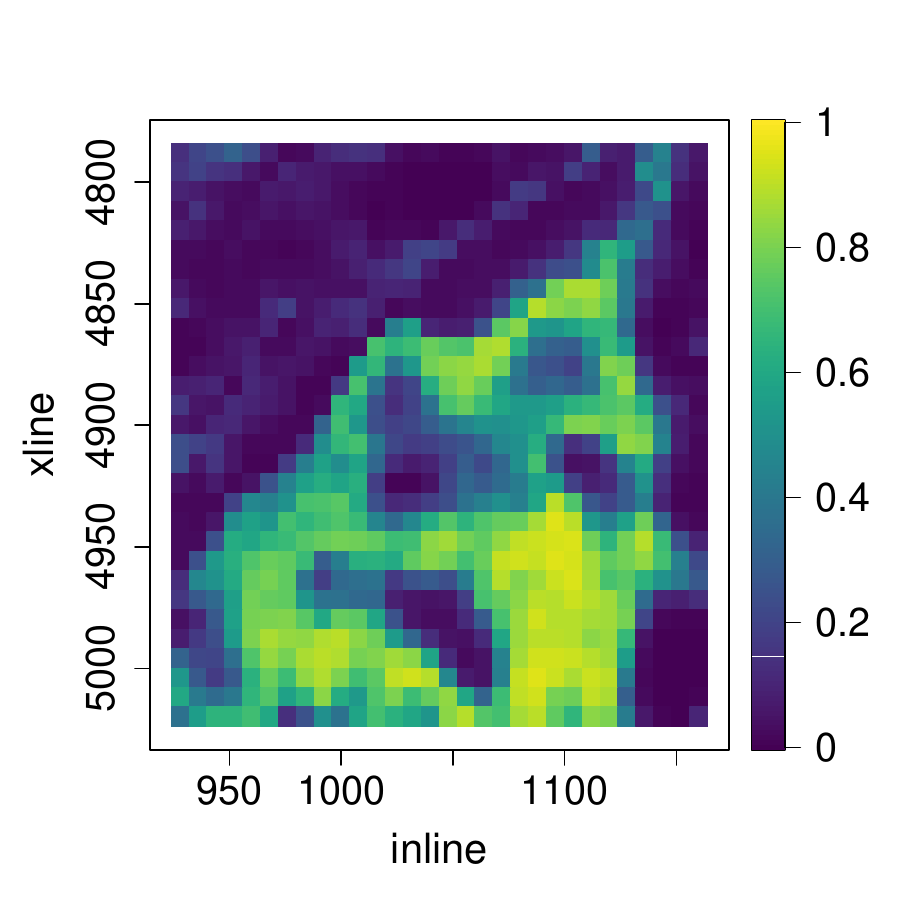}
		\caption{Uncertainty of oil saturation from MCMC samples from $\pi$.}
		\label{fig:small_area_uncertainty_oil_sfm}
	\end{subfigure}
	\begin{subfigure}[b]{0.3\textwidth}
		\centering\captionsetup{width=.7\linewidth}
		\includegraphics[width=.7\textwidth]{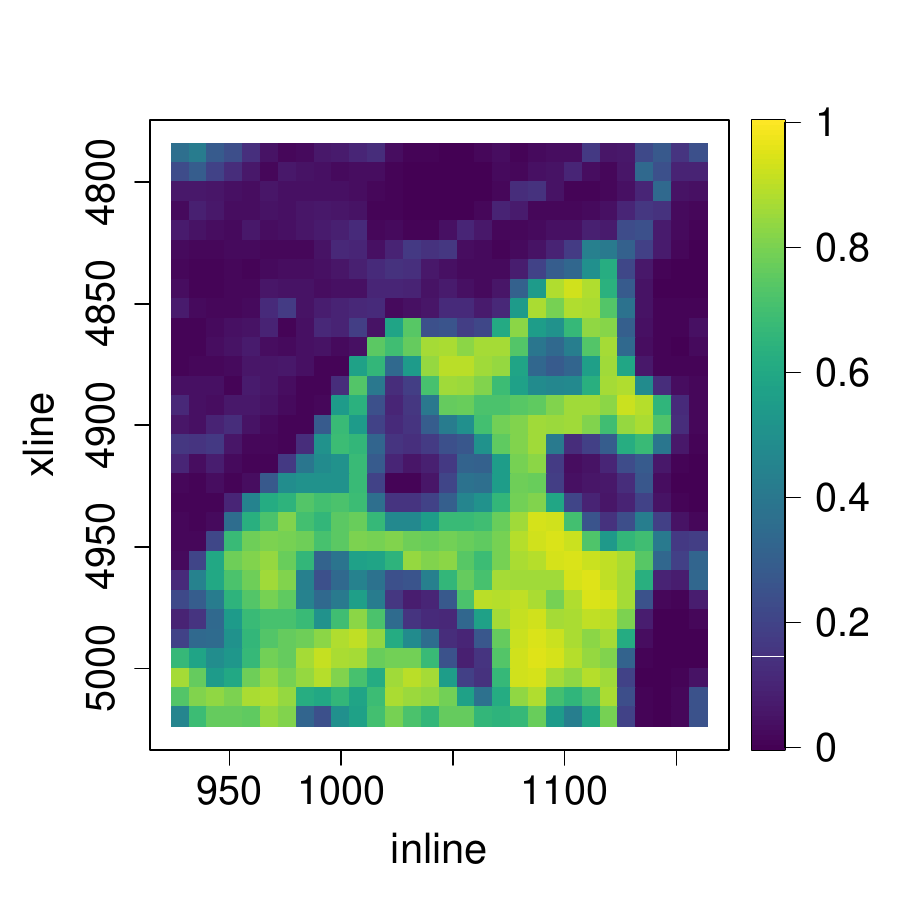}
		\caption{Uncertainty of oil saturation from MCMC samples from $\widehat{\pi}$.}
		\label{fig:small_area_uncertainty_oil_mars}
	\end{subfigure}
	\begin{subfigure}[b]{0.3\textwidth}
		\centering\captionsetup{width=.7\linewidth}
		\includegraphics[width=.7\textwidth]{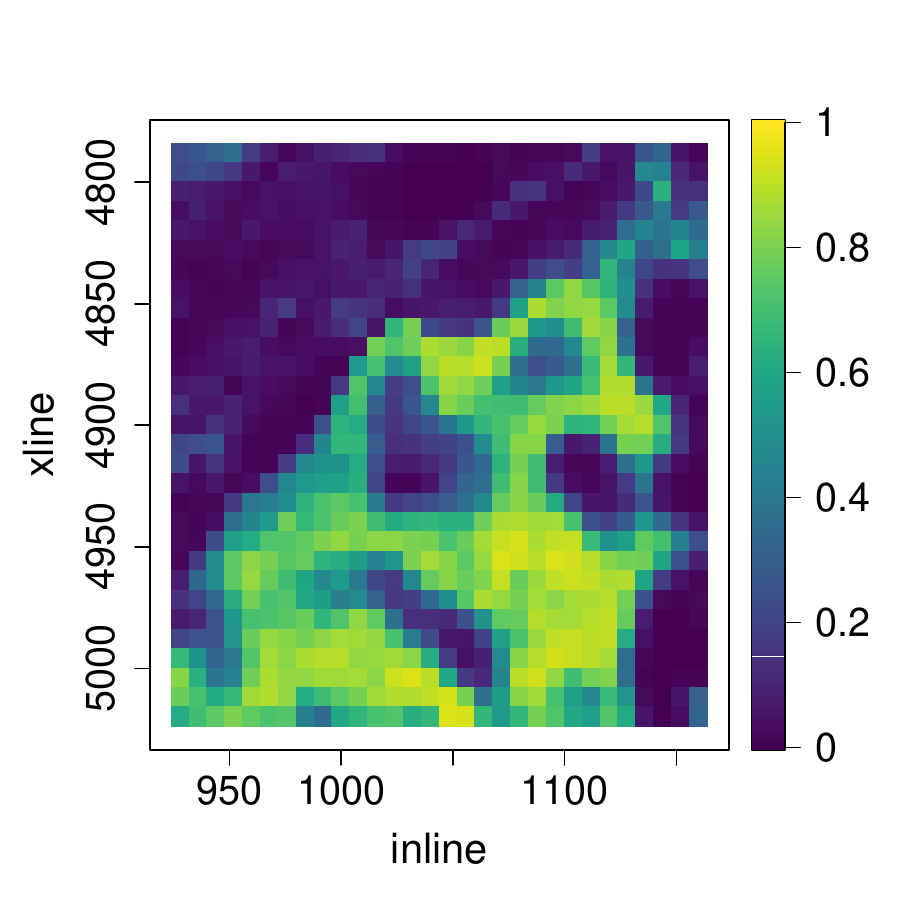}
		\caption{Uncertainty of oil saturation from MCMC samples from $\widetilde{\pi}$.}
		\label{fig:small_area_uncertainty_oil_mars_tilde}
	\end{subfigure}
	\caption{Mean oil saturations and uncertainties from MCMC samples from three posteriors. The posterior where $\widehat{\bh}_{\text{MARS}}$ is surrogate is $\widehat{\pi}$. In the last posterior, $\widetilde{\pi}$, both the surrogate $\widehat{\bh}_{\text{MARS}}$ and the error correcting covariance matrix $\widetilde{\bOmega}$ are used. The MH algorithm with proposal distribution $q_2$ with $s=0.027$ was used in all three cases. Uncertainty is expressed as the difference between the 90th and 10th quantiles.}
	\label{fig:oil_small_area}
\end{figure}

\begin{figure}[H]
	\centering
	\begin{subfigure}[b]{0.3\textwidth}
		\centering\captionsetup{width=.7\linewidth}
		\includegraphics[width=.7\textwidth]{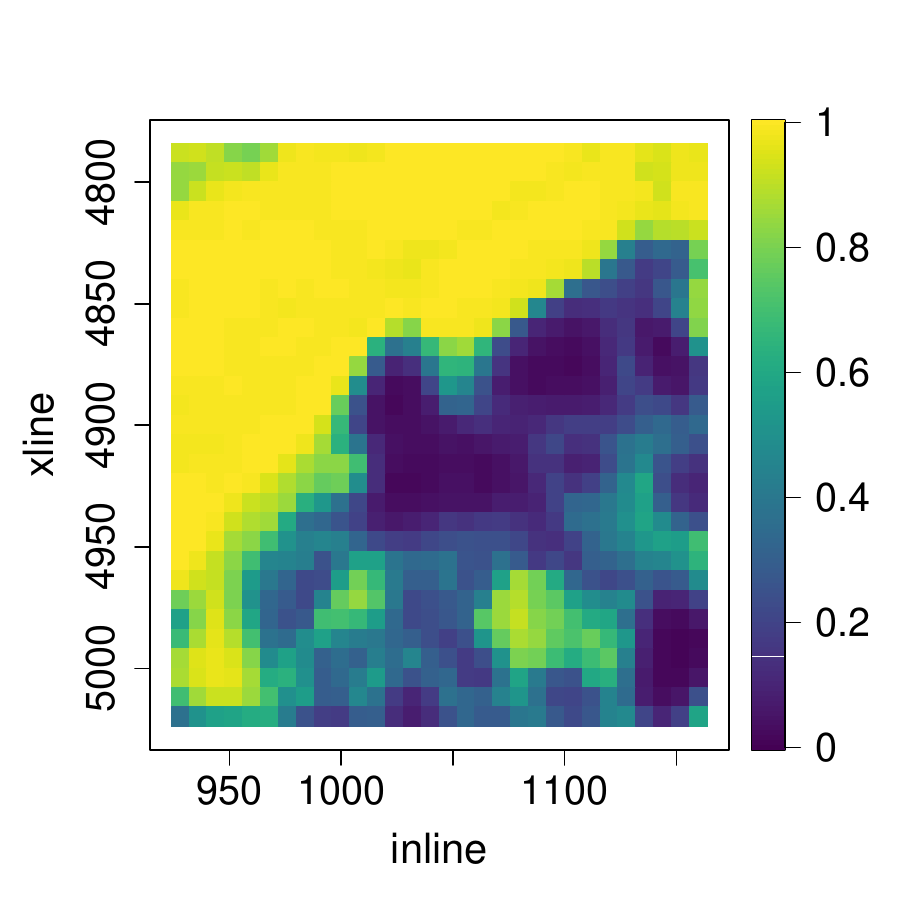}
		\caption{Mean brine saturation from MCMC samples from $\pi$.}
		\label{fig:small_area_brine_sfm}
	\end{subfigure}
	\begin{subfigure}[b]{0.3\textwidth}
		\centering\captionsetup{width=.7\linewidth}
		\includegraphics[width=.7\textwidth]{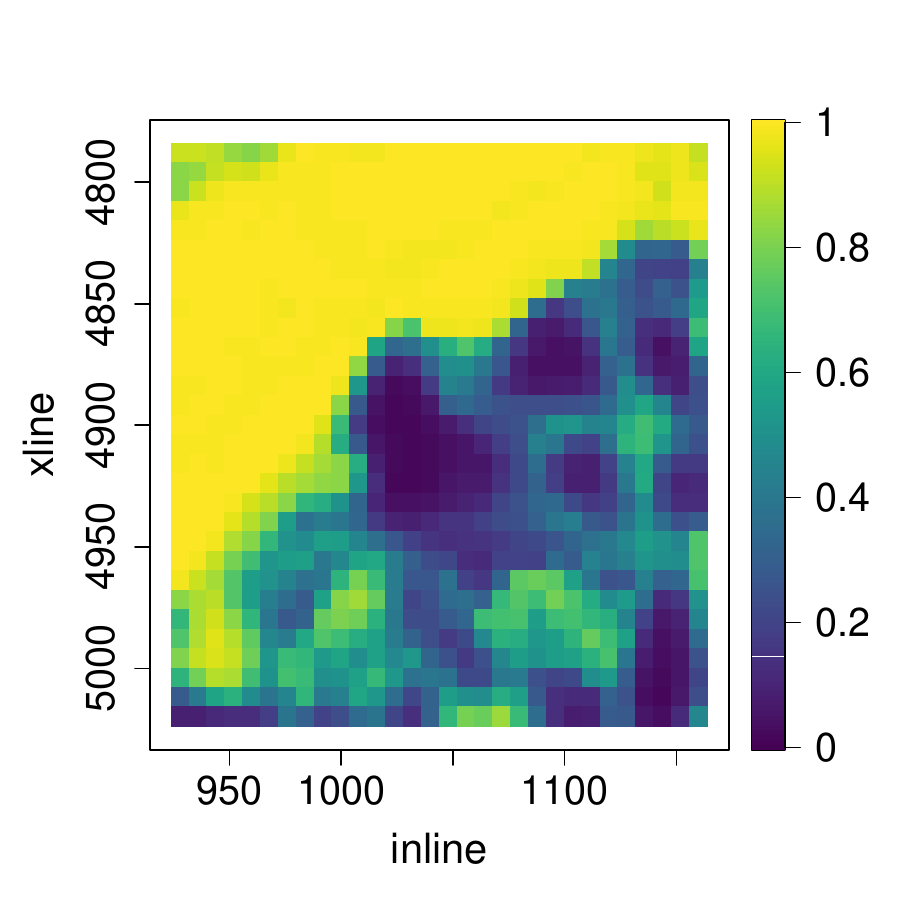}
		\caption{Mean brine saturation from MCMC samples from $\widehat{\pi}$.}
		\label{fig:small_area_brine_mars}
	\end{subfigure}
	\begin{subfigure}[b]{0.3\textwidth}
		\centering\captionsetup{width=.7\linewidth}
		\includegraphics[width=.7\textwidth]{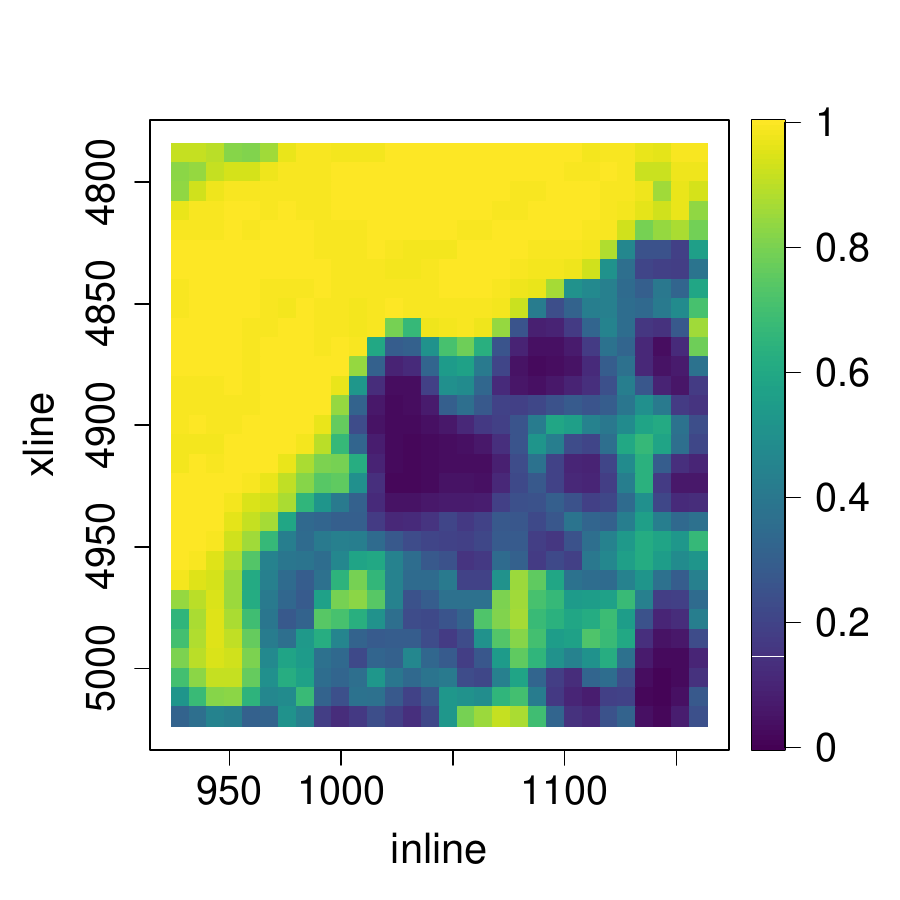}
		\caption{Mean brine saturation from MCMC samples from $\widetilde{\pi}$.}
		\label{fig:small_area_brine_mars_tilde}
	\end{subfigure}
	\begin{subfigure}[b]{0.3\textwidth}
		\centering\captionsetup{width=.7\linewidth}
		\includegraphics[width=.7\textwidth]{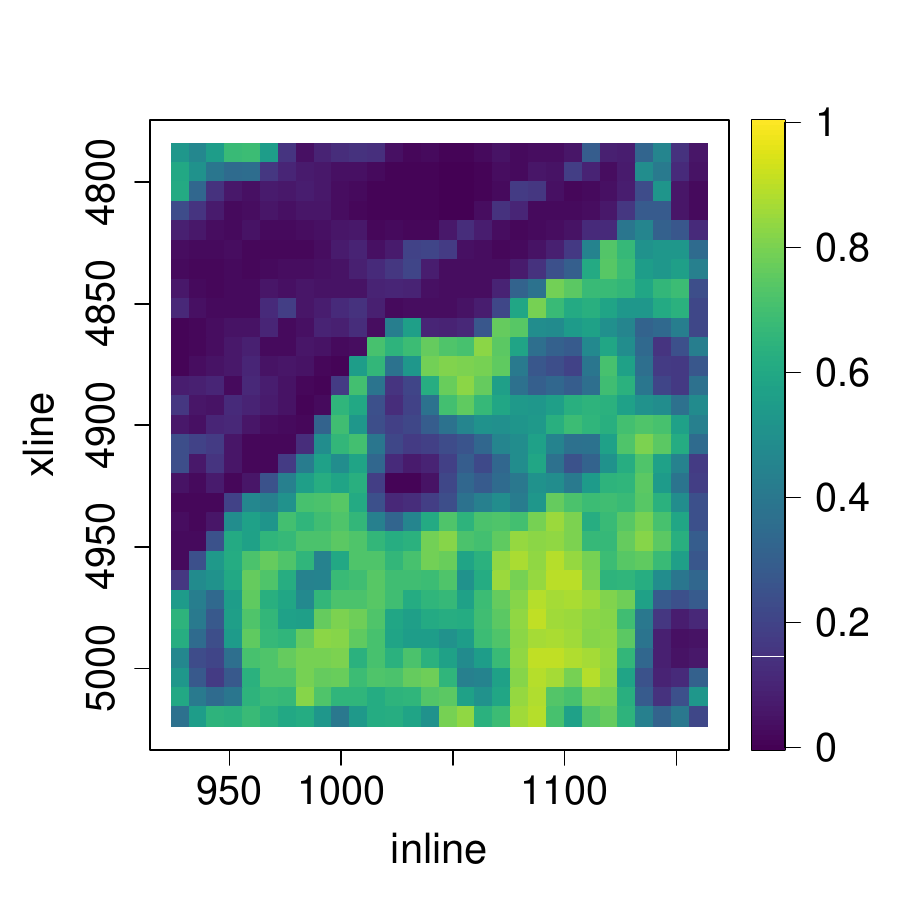}
		\caption{Uncertainty of brine saturation from MCMC samples from $\pi$.}
		\label{fig:small_area_uncertainty_brine_sfm}
	\end{subfigure}
	\begin{subfigure}[b]{0.3\textwidth}
		\centering\captionsetup{width=.7\linewidth}
		\includegraphics[width=.7\textwidth]{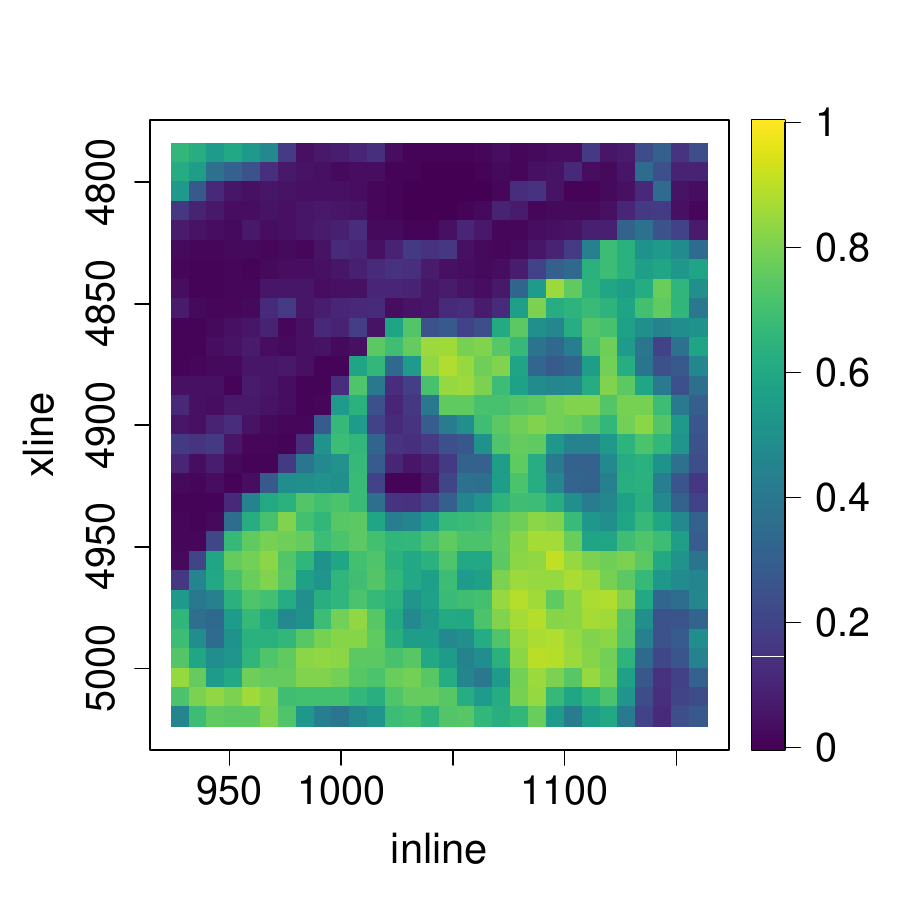}
		\caption{Uncertainty of brine saturation from MCMC samples from $\widehat{\pi}$.}
		\label{fig:small_area_uncertainty_brine_mars}
	\end{subfigure}
	\begin{subfigure}[b]{0.3\textwidth}
		\centering\captionsetup{width=.7\linewidth}
		\includegraphics[width=.7\textwidth]{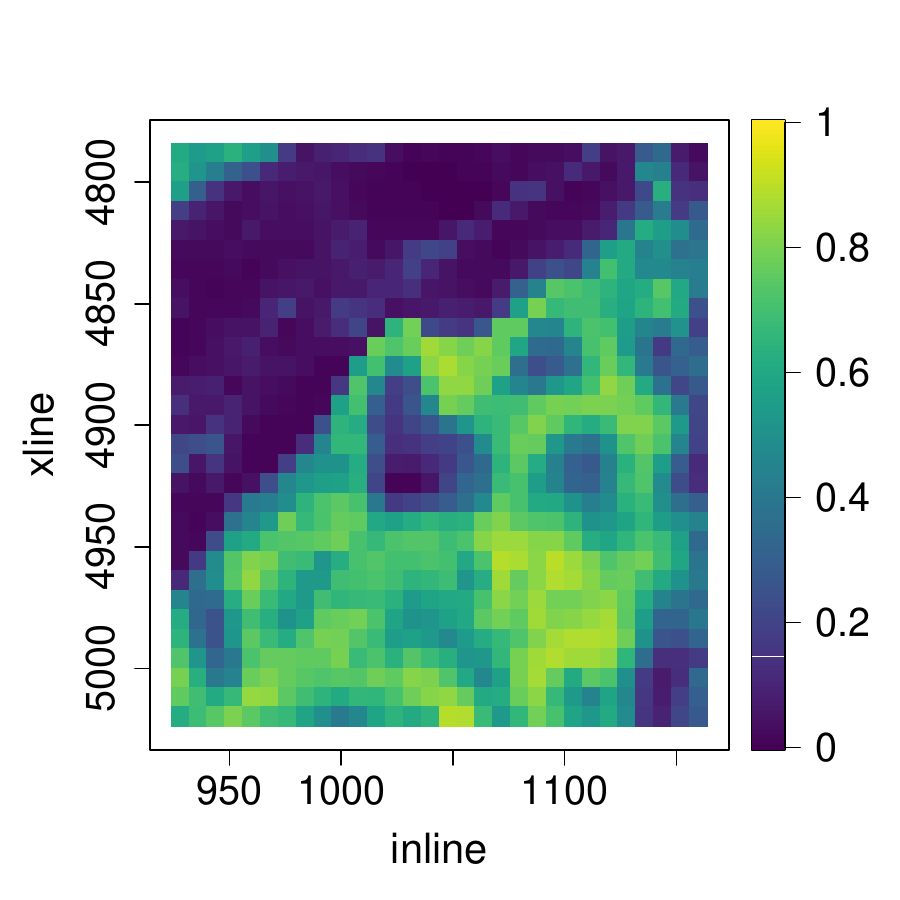}
		\caption{Uncertainty of brine saturation from MCMC samples from $\widetilde{\pi}$.}
		\label{fig:small_area_uncertainty_brine_mars_tilde}
	\end{subfigure}
	\caption{Mean brine saturations and uncertainties from MCMC samples from three posteriors. The posterior where $\widehat{\bh}_{\text{MARS}}$ is surrogate is $\widehat{\pi}$. In the last posterior, $\widetilde{\pi}$, both the surrogate $\widehat{\bh}_{\text{MARS}}$ and the error correcting covariance matrix $\widetilde{\bOmega}$ are used. The MH algorithm with proposal distribution $q_2$ with $s=0.027$ was used in all three cases. Uncertainty is expressed as the difference between the 90th and 10th quantiles.}
	\label{fig:brine_small_area}
\end{figure}

\begin{figure}[htb]
	\centering
	\begin{subfigure}[b]{0.3\textwidth}
		\centering\captionsetup{width=.7\linewidth}
		\includegraphics[width=.7\textwidth]{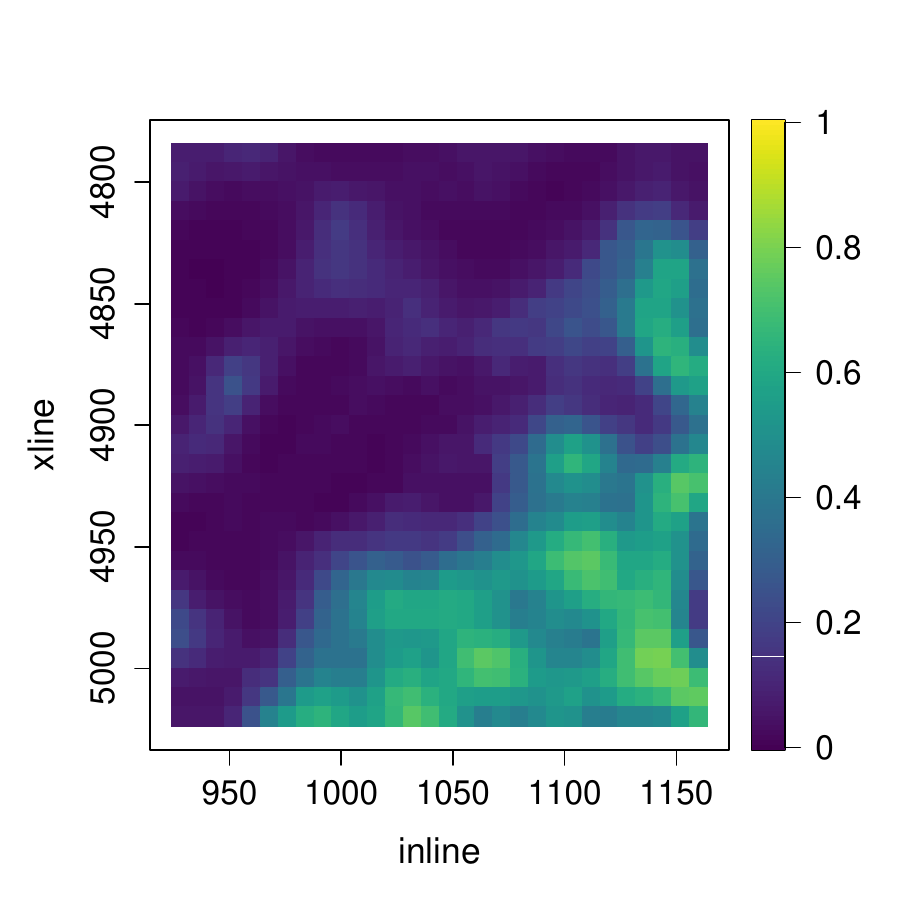}
		\caption{Mean clay content from MCMC samples from $\pi$.}
		\label{fig:small_area_clay_sfm}
	\end{subfigure}
	\begin{subfigure}[b]{0.3\textwidth}
		\centering\captionsetup{width=.7\linewidth}
		\includegraphics[width=.7\textwidth]{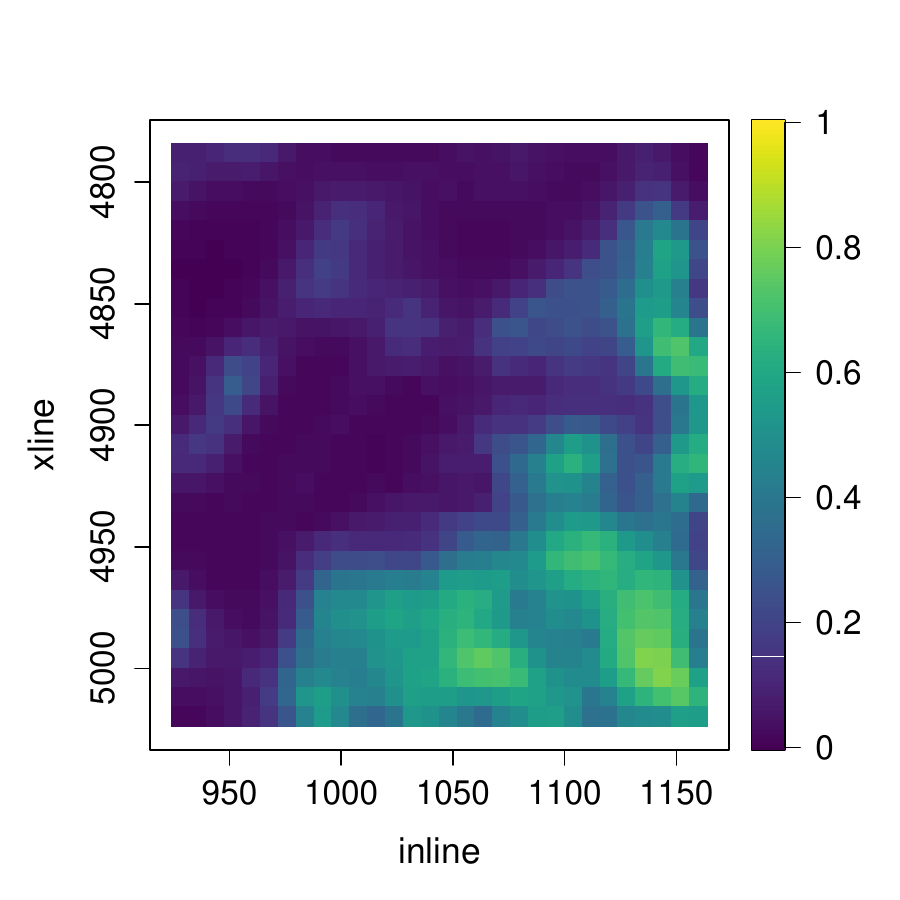}
		\caption{Mean clay content from MCMC samples from $\widehat{\pi}$.}
		\label{fig:small_area_clay_mars}
	\end{subfigure}
	\begin{subfigure}[b]{0.3\textwidth}
		\centering\captionsetup{width=.7\linewidth}
		\includegraphics[width=.7\textwidth]{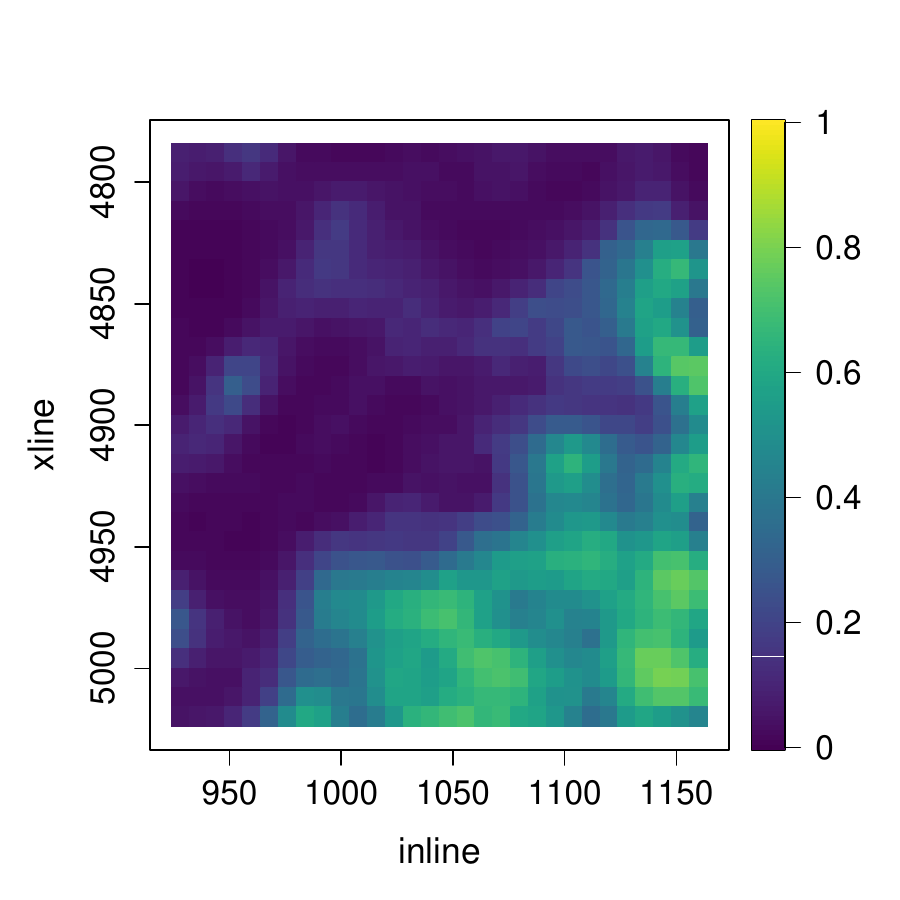}
		\caption{Mean clay content from MCMC samples from $\widetilde{\pi}$.}
		\label{fig:small_area_clay_mars_tilde}
	\end{subfigure}
	\begin{subfigure}[b]{0.3\textwidth}
		\centering\captionsetup{width=.7\linewidth}
		\includegraphics[width=.7\textwidth]{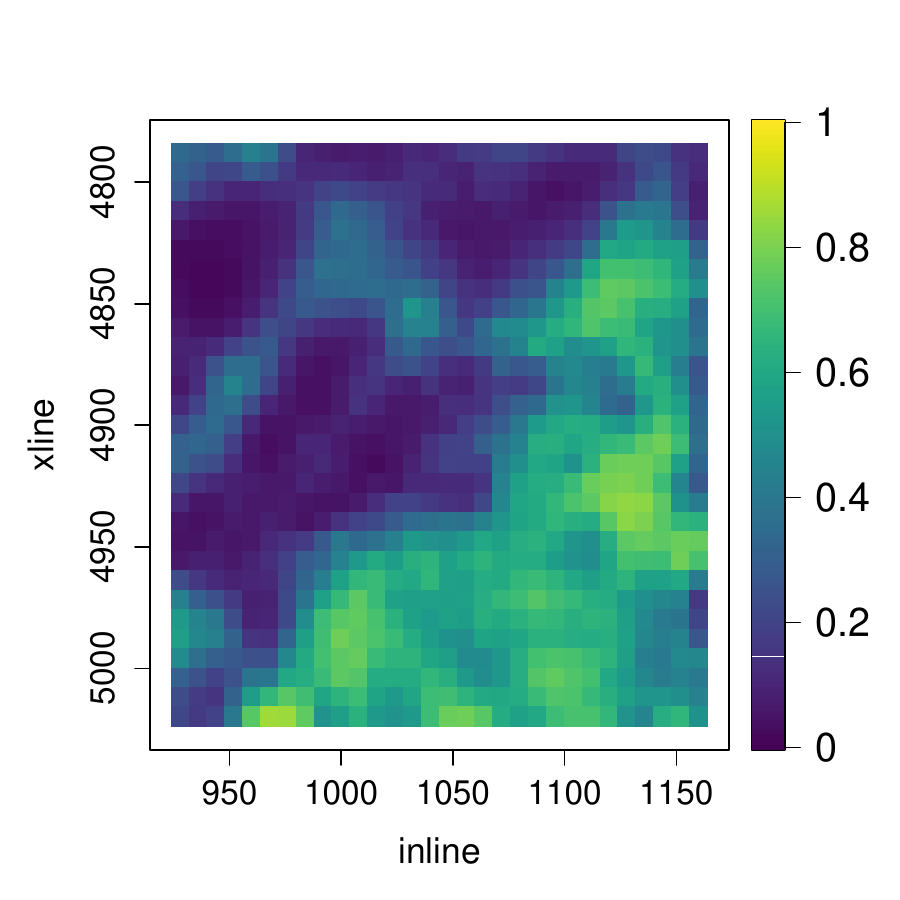}
		\caption{Uncertainty of clay content from MCMC samples from $\pi$.}
		\label{fig:small_area_uncertainty_clay_sfm}
	\end{subfigure}
	\begin{subfigure}[b]{0.3\textwidth}
		\centering\captionsetup{width=.7\linewidth}
		\includegraphics[width=.7\textwidth]{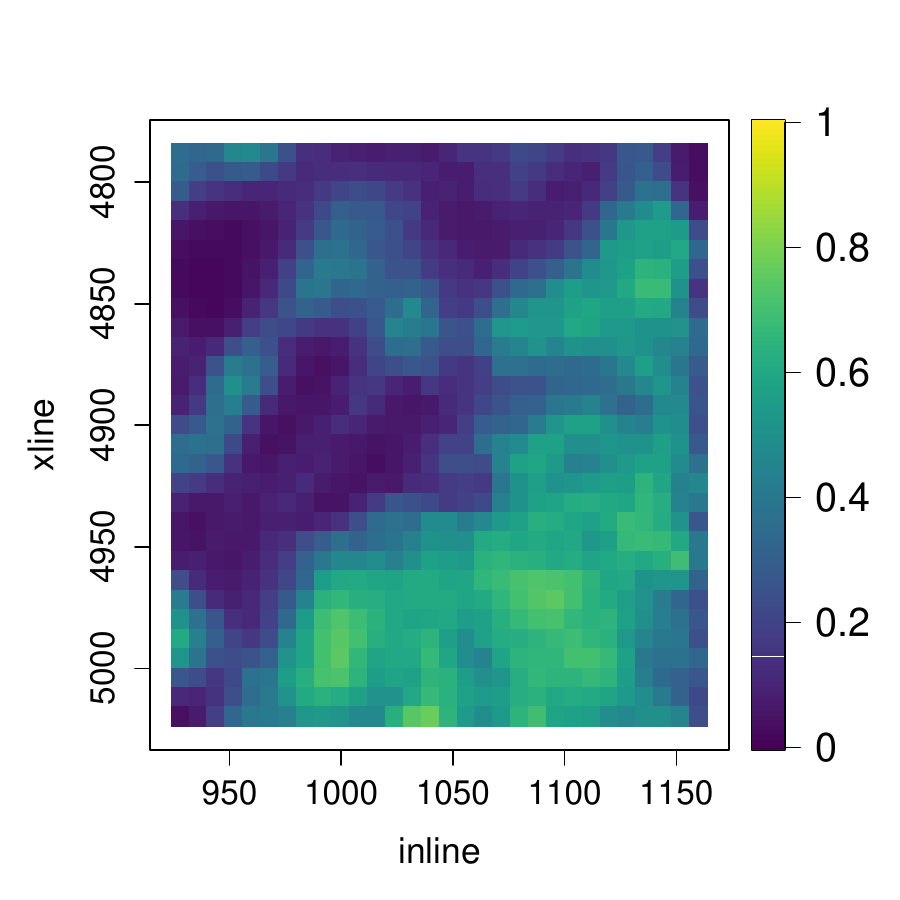}
		\caption{Uncertainty of clay content from MCMC samples from $\widehat{\pi}$.}
		\label{fig:small_area_uncertainty_clay_mars}
	\end{subfigure}
	\begin{subfigure}[b]{0.3\textwidth}
		\centering\captionsetup{width=.7\linewidth}
		\includegraphics[width=.7\textwidth]{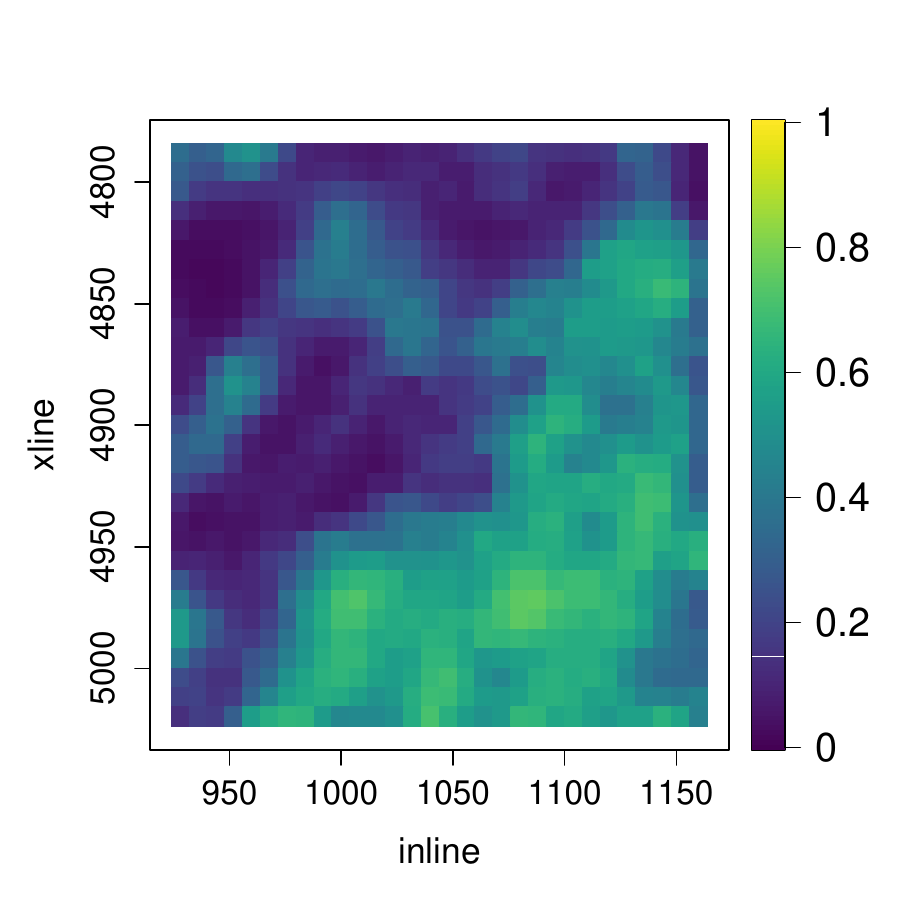}
		\caption{Uncertainty of clay content from MCMC samples from $\widetilde{\pi}$.}
		\label{fig:small_area_uncertainty_clay_mars_tilde}
	\end{subfigure}
	\caption{Mean clay content and uncertainties from MCMC samples from three posteriors. The posterior where $\widehat{\bh}_{\text{MARS}}$ is surrogate is $\widehat{\pi}$. In the last posterior, $\widetilde{\pi}$, both the surrogate $\widehat{\bh}_{\text{MARS}}$ and the error correcting covariance matrix $\widetilde{\bOmega}$ are used. The MH algorithm with proposal distribution $q_2$ with $s=0.027$ was used in all three cases. Uncertainty is expressed as the difference between the 90th and 10th quantiles.}
	\label{fig:clay_small_area}
\end{figure}

Figure \ref{fig:gas_small_area} illustrates mean gas saturations and corresponding uncertainties from three sets of posterior samples, which exhibit remarkable similarity overall. A notable distinction is the wider light green area at approximately inline $1050$ and crossline $4955$ in Figure \ref{fig:small_area_gas_sfm} compared to Figures \ref{fig:small_area_gas_mars} and \ref{fig:small_area_gas_mars_tilde}, with a corresponding increase in uncertainty. Additionally, at inline $1000$ and crossline $4950$, uncertainties in Figures \ref{fig:small_area_uncertainty_gas_sfm} and \ref{fig:small_area_uncertainty_gas_mars} align more closely, while Figure \ref{fig:small_area_uncertainty_gas_mars_tilde} displays lower uncertainty in this region. In the yellow area around inline $1100$ and crossline $4920$, Figures \ref{fig:small_area_gas_mars} and \ref{fig:small_area_gas_mars_tilde} are similar, with Figure \ref{fig:small_area_gas_sfm} showing slightly less intensity. Similarly, differences in uncertainty patterns are observed, especially in the bottom right corner, where samples using the surrogate model are more alike and differ slightly from those using the exact forward model.

In Figure \ref{fig:oil_small_area}, mean oil saturations and uncertainties demonstrate consistent patterns across all cases. However, discrepancies arise, particularly in the lower left corner, where samples from $\pi$ indicate lower oil saturation compared to $\widehat{\pi}$ and $\widetilde{\pi}$. The use of the error-correcting covariance matrix improves the approximation to $\pi$, as evident in Figure \ref{fig:small_area_oil_mars_tilde} compared to Figure \ref{fig:small_area_oil_mars}. Conversely, Figure \ref{fig:small_area_oil_mars_tilde} exhibits a small green area in the lower right corner not present in Figures \ref{fig:small_area_oil_mars} and \ref{fig:small_area_oil_sfm}. Differences in uncertainty are also notable, with higher uncertainty in Figure \ref{fig:small_area_uncertainty_oil_mars_tilde} compared to Figures \ref{fig:small_area_uncertainty_oil_sfm} and \ref{fig:small_area_uncertainty_oil_mars}. Areas where uncertainties in samples from $\widehat{\pi}$ and $\widetilde{\pi}$ resemble each other but differ from $\pi$ are observed around inline $1060$ and crossline $4900$, with lower uncertainty in samples from $\pi$. Similarly, at approximately inline $1060$ and crossline $5020$, oil saturation samples from $\widehat{\pi}$ and $\widetilde{\pi}$ exhibit greater similarity compared to $\pi$.

The mean brine saturation and corresponding uncertainties illustrated in Figure \ref{fig:brine_small_area} exhibit consistency across the three cases. However, a disparity emerges around inline $1060$ and crossline $5020$, where Figures \ref{fig:small_area_brine_mars} and \ref{fig:small_area_brine_mars_tilde} indicate higher brine saturations compared to Figure \ref{fig:small_area_brine_sfm}. This corresponds to the area where differences in oil saturation were observed. The most significant differences in brine saturation uncertainties occur in the bottom left and right corners, where Figures \ref{fig:small_area_brine_mars_tilde} and \ref{fig:small_area_uncertainty_brine_sfm} exhibit stronger similarities compared to Figures \ref{fig:small_area_uncertainty_brine_mars} and \ref{fig:small_area_uncertainty_brine_sfm}.

Regarding the mean clay content depicted in Figures \ref{fig:clay_small_area}, they are largely similar, with slight differences observed towards crossline $4955$. Notably, the mean clay content in Figure \ref{fig:small_area_clay_sfm} is notably lower in this area compared to Figures \ref{fig:small_area_clay_mars} and \ref{fig:small_area_clay_mars_tilde}. Samples from $\widehat{\pi}$ in this region slightly resemble those from $\pi$ compared to their overall similarity. Uncertainty patterns in Figures \ref{fig:clay_small_area} are generally similar. However, uncertainties may be slightly larger in Figure \ref{fig:small_area_uncertainty_clay_sfm}, particularly around inline $900\sim 1000$ at higher crosslines, and around inline $1130$ and crossline $4930$.

Figures  \ref{fig:gas_small_area},  \ref{fig:oil_small_area},  \ref{fig:brine_small_area} and  \ref{fig:clay_small_area} show overall great similarity in the results from the MCMC samples from $\pi$, $\widehat{\pi}$ and $\widetilde{\pi}$ on the smaller area in the Alvheim field. This indicates that substituting $\bh$ with $\widehat{\bh}_{\text{MARS}}$ gives a good approximation to the posterior $\pi$, both when using the covariance matrix $\bOmega$ and when using $\widetilde{\bOmega}$ as covariance matrix in the likelihood model. MARS models were also considered as good surrogates for the time-consuming forward models in \cite{CHEN_mars} and \cite{CHEN_mars_poly}. A MARS model does not assume a functional form of the response, which could be the reason it applied to all three problems.

The use of the adjusted correlation function $\widetilde{\bOmega}$ showed minor improvements in some areas, while in other areas, using $\bOmega$ gave the means and uncertainties closest to the mean and uncertainties of the MCMC samples of $\pi$. The empirical variance and covariance added to the covariance matrix to adjust for the error between $\widehat{\bh}$ and $\bh$ were very small, such that $\widetilde{\bOmega}$ and $\bOmega$ are quite similar. That the covariance matrices are very similar could be the reason for the similar result. 

Most of the differences between the exact and approximate samples are found at the edges of the area. As mentioned, there is a well in that area, such that samples close to the well are affected by the likelihood model for the well-log data in addition to the likelihood model for the seismic AVO data. However, the mean and uncertainty of the saturations are in general very similar, indicating that $\widehat{\pi}$ and $\widetilde{\pi}$ are similar to $\pi$. In MCMC, there is a randomness when proposing and accepting samples, such that the exact same samples might not be considered for the three different Markov chains. This could lead to some differences in the means and uncertainties, however, this difference vanishes with an increasing number of MCMC iterations if $\pi$, $\widehat{\pi}$ and $\widetilde{\pi}$ were the same distributions.

\subsection{Comparing the proposals distributions}
\label{sc_comparing_proposals}
	
Next, the performance of the MH algorithm with the four different proposal distributions, $q_1$, $q_2$, $q_3$ and $q_4$ given in Section \ref{sc_other_proposals}, are compared, where the tuning parameter is denoted by $s$.
The approximation $\text{MARS}_{\nabla}$ is used instead of the gradient of the log posterior in the MALA to reduce computation time. 

Here, we focus the approximate posterior $\widehat{\pi}$ on the small area. That is, $\widehat{\bh}_{\text{MARS}}$ is used as the forward model and the covariance matrix of the likelihood model is $\bOmega$. The objective is to compare how efficiently the four MH algorithms explore the domain of $\widehat{\pi}$. The ESS per computation time is measured to assess which proposal gives the most information about $\widehat{\pi}$ per time. The reported ESS is the mean ESS of all $3N$ stochastic processes $x_g^1, x_o^1, x_c^1, x_g^2, x_o^2, x_c^2, ..., $ $ x_g^N, x_o^N, x_c^N$. The tuning parameter $s$ and acceptance rates, which are connected to the efficiencies of the algorithms are reported. The four MH algorithms were tuned such that the acceptance rates were approximately $23.4\%$ when using the proposal distributions $q_1$, $q_2$ and $q_3$ and approximately $57.4\%$ when using the proposal distribution $q_4$. As the acceptance rates and ESS should be computed on MCMC samples after burn-in, the Markov chain starts at the mean of the posterior samples whose logistic transformation is shown in Figure \ref{fig:small_area_gas_sfm}, Figure \ref{fig:small_area_oil_sfm} and Figure \ref{fig:small_area_clay_sfm}. The MH algorithms are run for $10^6$ iterations where every $100$th sample is saved. This gives $10,000$ MCMC samples from $\widehat{\pi}$. 

{\renewcommand{\arraystretch}{1.6}%
	\begin{table}[htb]
		\centering
		\begin{tabular}{ |c|c|c|c|c|c| } 
			\hline 
			proposal & $s$ & acceptance rate [\%] & computation time [sec] &  ESS [\#] & ESS/time \\
			\hline
			$q_1$ & $0.006$   & $25.5$  & $13546.24$ & $11.01$ & $0.0008 $  \\ 
			$q_2$ & $0.027$  & $23.3$ & $16547.33$  & $93.88$ & $0.0057 $ \\ 
			$q_3$ & $0.045$  & $22.8$ & $11408.03$ & $212.64$ & $0.0186 $ \\ 
			$q_4$ & $0.038$  & $58.9$ & $44418.04$ & $39.72$ & $0.0009 $ \\ 
			\hline
		\end{tabular}
		\caption{Comparison of the efficiency of four MH algorithms. The acceptance rate, computation time and ESS are calculated on $10,000$ MCMC samples.}
		\label{table:ESS_time_q}
	\end{table}

These results, reported in Table \ref{table:ESS_time_q}, show that the MH algorithm with the proposal distribution $q_3$ was the most efficient. The algorithm produced approximately $0.0186$ independent samples per second. This is much higher than the ESS per computation time for the MH algorithm with the other proposals distributions. The second most efficient algorithm was the MH algorithm with the proposal distribution $q_2$.  The ESS of the MH algorithm with proposal distribution $q_3$ is twice better than the ESS for  $q_2$ as proposal distribution. The MH algorithm with proposal distribution $q_3$ was also about 1 hour and 25 minutes faster than the second most efficient algorithm. The slowest algorithm was the MALA, which used more than 12 hours. For comparison, the fastest algorithm used 3 hours and 10 minutes. The MALA used more than three times as much time as $q_1$, but because the samples were less correlated, $q_4$ was slightly more efficient than $q_1$ in this case.

Using the MH algorithm with $q_3$ results in faster MCMC sampling compared to the other three distributions. It is sensible that employing the $q_4$ distribution leads to the slowest sampling, as it lacks symmetry like $q_1$ and $q_2$, and it is not reversible concerning the prior, unlike $q_3$, necessitating gradient evaluations. Similarly, $q_2$ is reasonably the second slowest, given its proposal correlation akin to $q_3$ and the requirement to assess both the prior and likelihood similar to $q_1$. While using $q_1$ does not require proposing correlated samples as $q_3$ does, it does entail evaluating the prior, unlike $q_3$, where only the likelihood needs evaluation. The computation time in Table \ref{table:ESS_time_q} illustrates that proposing correlated samples is faster than evaluating the proposal distribution. In this context, the MALA method notably outperforms the MH algorithm with $q_1$ as the proposal distribution, suggesting the significance of utilizing posterior information as outlined in previous studies. Although MALA's computation time is over three times that of the MH algorithm with $q_1$, its reduced sample correlation enables it to convey slightly more information about $\widehat{\pi}$ per second.

	In seismic AVO inversion within the Alvheim field, testing four MH algorithms revealed that MH with proposal distribution $q_3$ was the most efficient in terms of ESS per computation time. It exhibited both the highest ESS and the smallest computation time among the four algorithms, making it the clear choice for efficiency. However, this does not guarantee its efficiency in other contexts, as finding suitable MCMC methods for various inverse problems can be challenging and problem-specific. Despite this,  $q_3$ has proven effective beyond the Alvheim case, outperforming other methods, like using $q_1$ \cite{rudolf2018generalization}, particularly with increased dimensionality. This efficiency, irrespective of dimensionality, makes  $q_3$ preferable in high-dimensional scenarios like the Alvheim case.

\begin{figure}[htb]
	\centering
	\begin{subfigure}[b]{0.24\textwidth}
		\includegraphics[width=\textwidth]{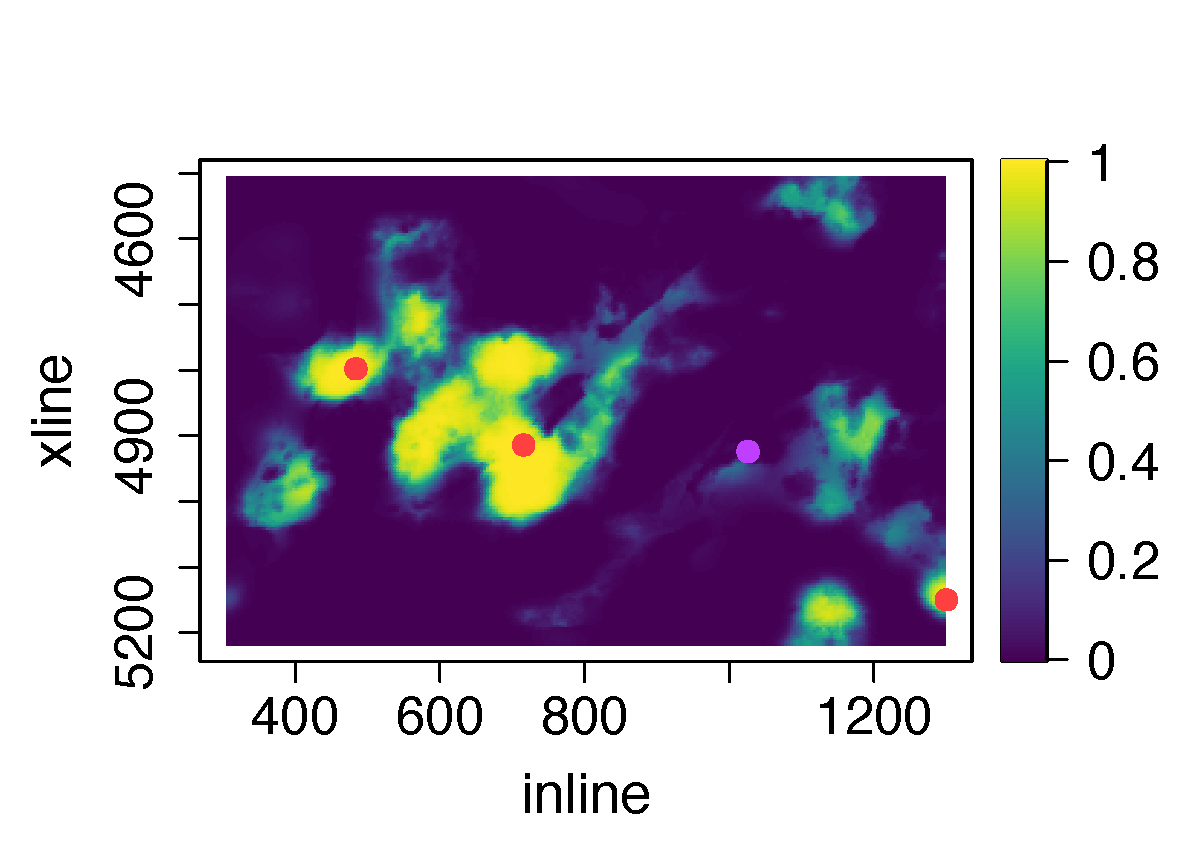}
		\caption{Mean saturation of gas.}
		\label{fig:q3_saturation_gas}
	\end{subfigure}
	\begin{subfigure}[b]{0.24\textwidth}
		\includegraphics[width=\textwidth]{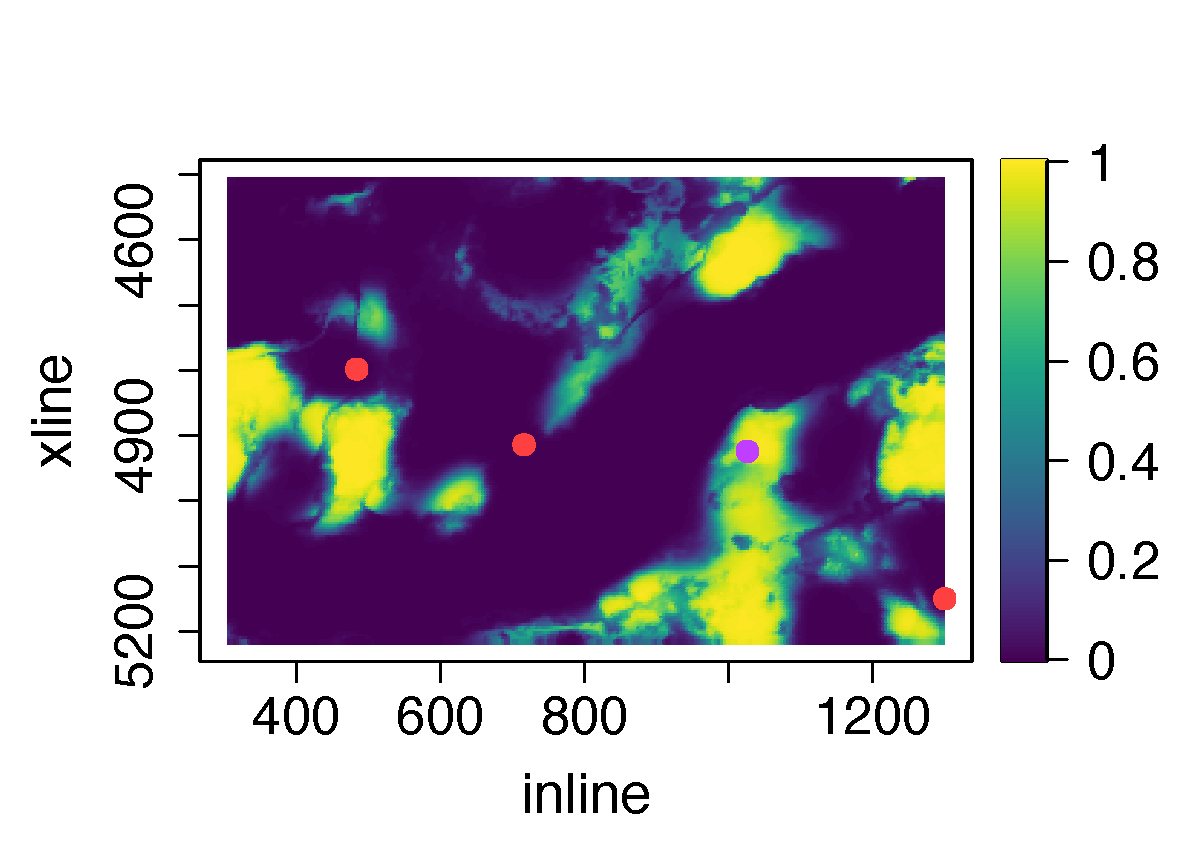}
		\caption{Mean saturation of oil.}
		\label{fig:q3_saturation_oil}
	\end{subfigure}
	\begin{subfigure}[b]{0.24\textwidth}
		\includegraphics[width=\textwidth]{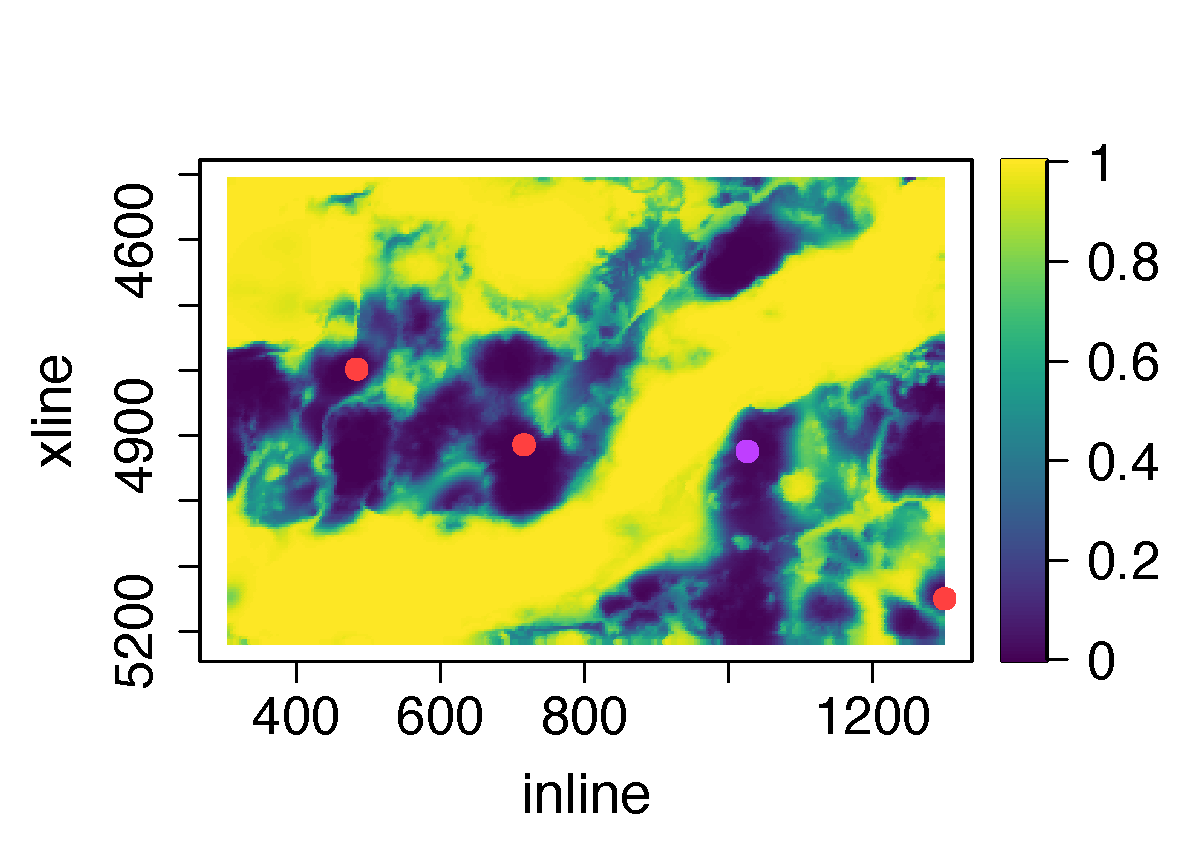}
		\caption{Mean saturation of brine.}
		\label{fig:q3_saturation_brine}
	\end{subfigure}
	\begin{subfigure}[b]{0.24\textwidth}
	\includegraphics[width=\textwidth]{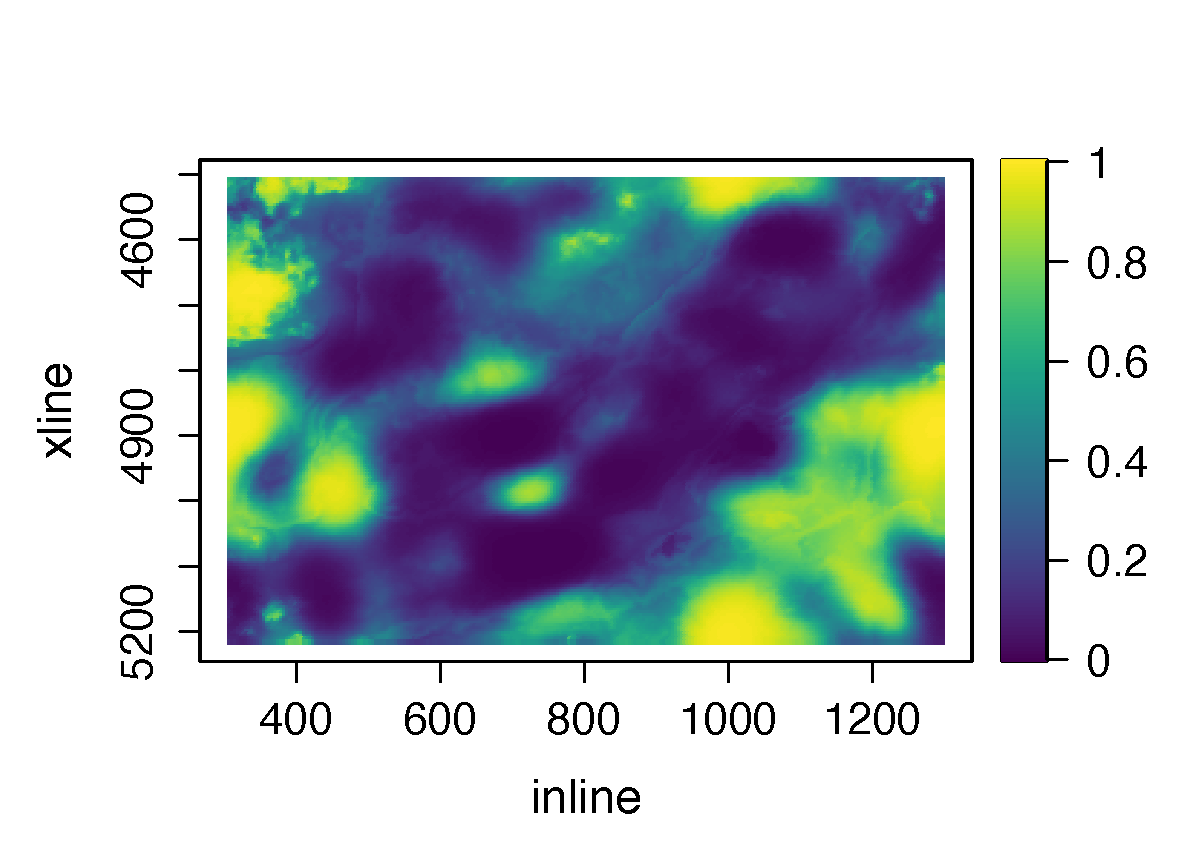}
	\caption{Mean clay content.}
	\label{fig:q3_content_clay}
\end{subfigure}
	\begin{subfigure}[b]{0.24\textwidth}
		\includegraphics[width=\textwidth]{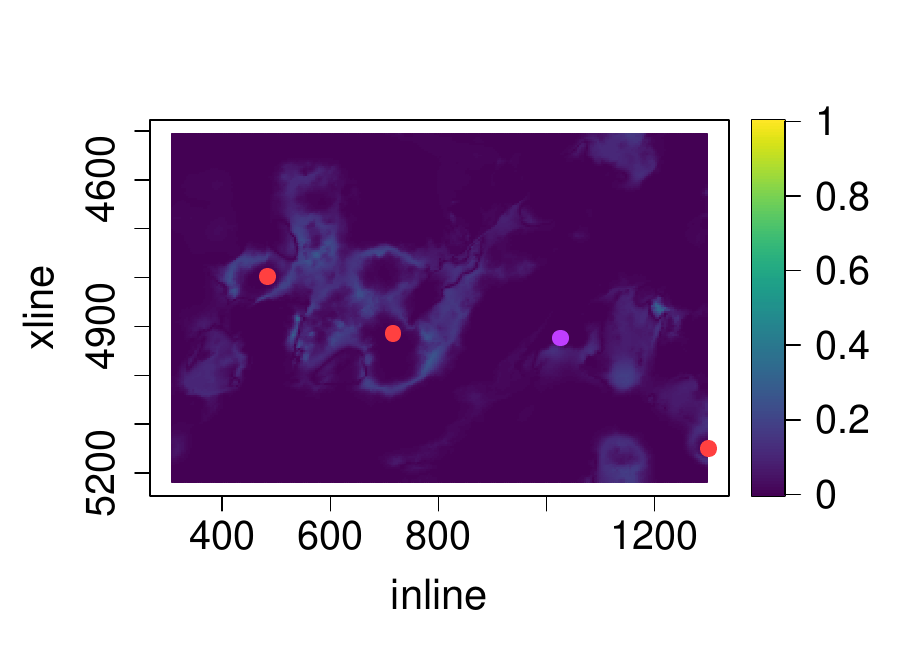}
		\caption{Uncertainty of gas saturation.}
		\label{fig:q3_uncertainty_gas}
	\end{subfigure}
	\begin{subfigure}[b]{0.24\textwidth}
		\includegraphics[width=\textwidth]{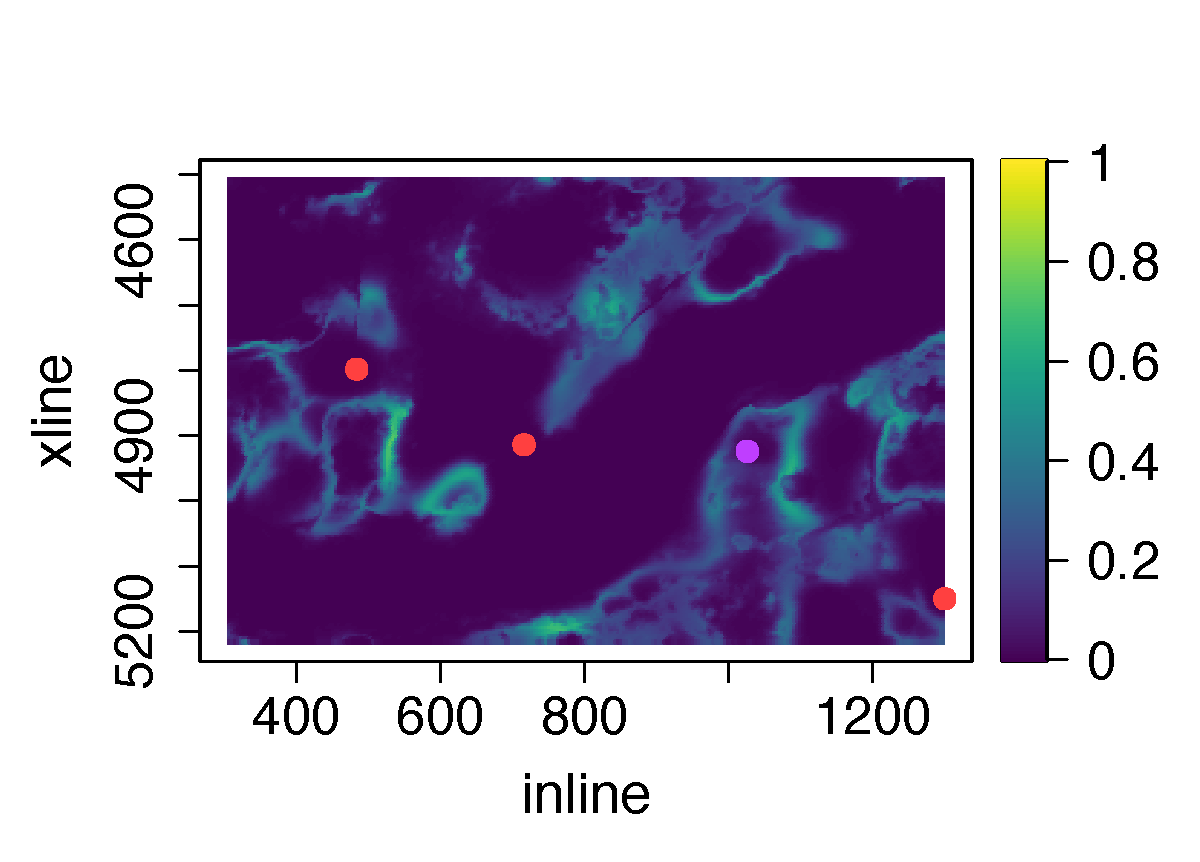}
		\caption{Uncertainty of oil saturation.}
		\label{fig:q3_uncertainty_oil}
	\end{subfigure}
	\begin{subfigure}[b]{0.24\textwidth}
		\includegraphics[width=\textwidth]{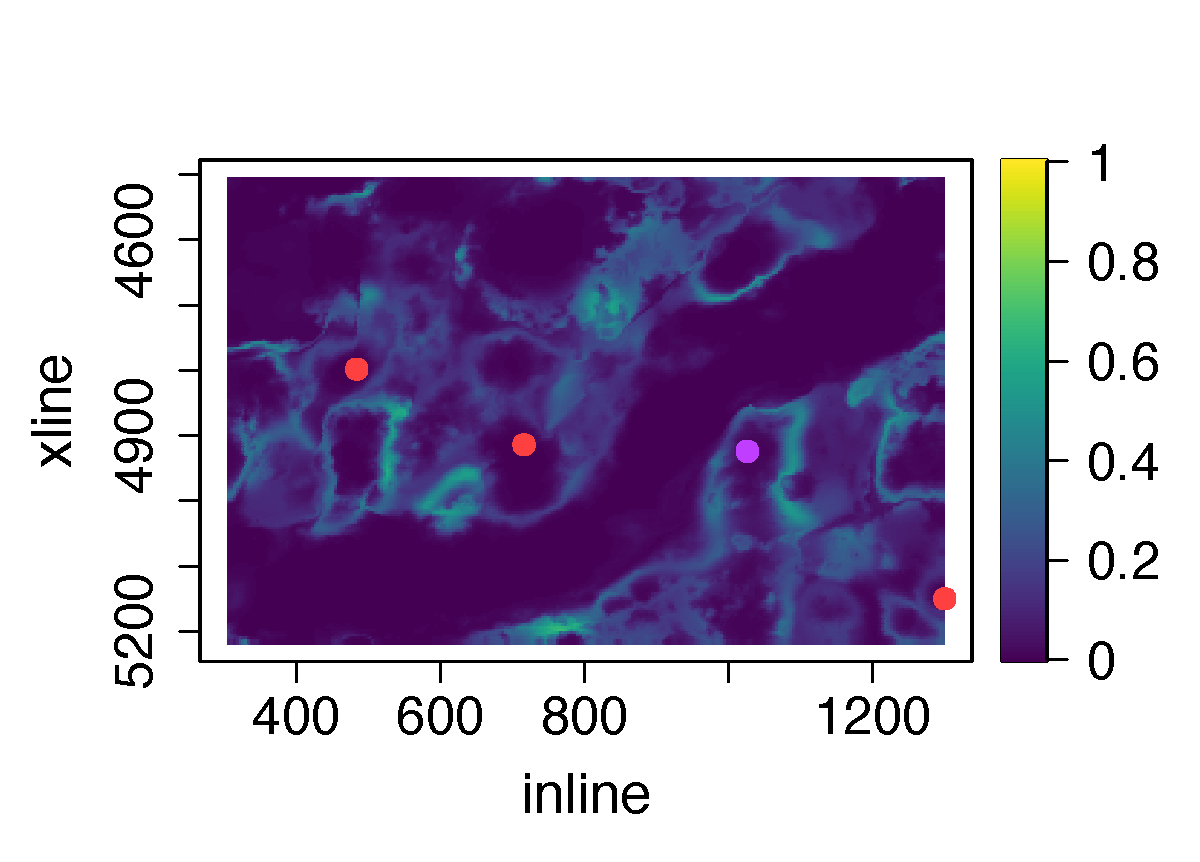}
		\caption{Uncertainty of gas saturation.}
		\label{fig:q3_uncertainty_brine}
	\end{subfigure}
	\begin{subfigure}[b]{0.24\textwidth}
	\includegraphics[width=\textwidth]{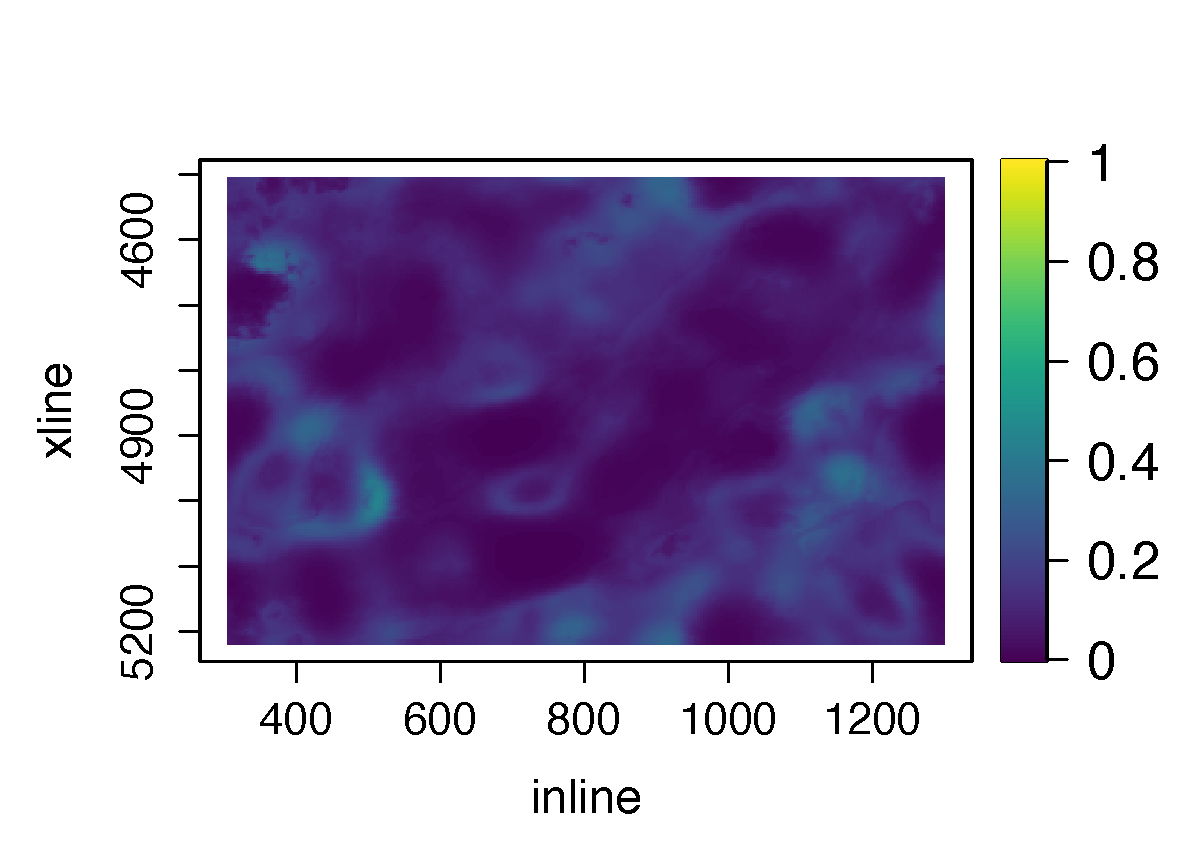}
	\caption{Uncertainty of clay content.}
	\label{fig:q3_uncertainty_clay}
\end{subfigure}
	\caption{Mean and uncertainty of gas, oil, brine and clay saturations from MCMC samples from $\widehat{\pi}$. The uncertainty is the difference in the $90$th and $10$th quantile.}
	\label{fig:q3_sauturations}
\end{figure}

\section{Results with the Alvheim field data}
\label{sc_result_realdata}

As highlighted from simulation studies in Section \ref{sc_result_simulation}, preconditioned Crank-Nicolson algorithm show the best performance in our comparison thus we focus on this algorithm for our case study on the Alvheim field data. Additional results for other proposal are given in \ref{sc_extra_simulation}. 

In this section, MCMC samples from $\widehat{\pi}$ over the entire Alvheim field are compared to the results of \cite{spremic2023bayesian}. The two results are compared by mean saturation of gas, oil and brine, mean clay content, uncertainties and ternary plots from one deep and one shallow location in the Alvheim field. The MH algorithm with $q_3$ as proposal distribution is used to draw $500,000$ MCMC samples after burn-in where every $50$th sample is saved. That gives $10,000$ MCMC samples from the posterior $\widehat{\pi}$. The tuning parameter was $s=0.0088$ and the acceptance rate was $25.4\%$. The Markov chain is started in two locations to check that the Markov chain converges to the same values. 

The MCMC results are shown in Figure \ref{fig:q3_sauturations}. The wells are marked by coloured circles: red circles indicate wells where mainly gas was found and the violet circle indicates the well where the primarily oil was found. Figure \ref{fig:q3_sauturations} shows that near the wells where mainly gas was found, the mean gas saturation is high and the mean oil saturation is low. Likewise, the mean gas saturation is low and the mean oil saturation is high near the well where most oil was retrieved. The uncertainty is generally low near the four wells. The mean gas and oil saturations are generally low at the deeper areas, such as in the bottom left corner, the top left corner, in the top middle and between the gas and oil wells in the area that stretches from approximately (inline,crossline) = (600,5200) to (1300,4500). The uncertainty of the saturations is generally low, however higher for the oil saturations than the gas saturations. The uncertainties are low in the middle of areas which show high oil or gas saturations, and higher where the saturations transition from high to low values, such that the uncertainties look like circles enclosing the areas with high gas or oil saturations.

\subsection*{Some similarities to \citet{spremic2023bayesian}}

There are several similarities between our results and those from  \cite{spremic2023bayesian}. The results in  \cite[Figure 12]{spremic2023bayesian} in  show high mean saturation of gas and oil near the wells where mostly gas and oil were found respectively. There is also low uncertainty close to the four wells. In general, the areas of high mean gas saturation look similar to each other, but the mean saturation of gas from the MCMC samples has more defined shapes compared to the mean gas saturation in \cite[Figure 12]{spremic2023bayesian}, which is more square-shaped and blurred around the edges of high mean saturation areas. Another similarity is the low mean saturation of gas and oil between the gas and oil wells in the bottom left area, the top left area and the area that stretches from approximately (inline,crossline) = (600,5200) to (1300,4500). The uncertainties of the clay content in \cite[Figure 13]{spremic2023bayesian} and Figure \ref{fig:q3_sauturations} also look very similar.

The mean oil saturations in the two results show some dissimilarities. For example, in \cite[Figure 12]{spremic2023bayesian} the well marked by the green triangle, there are modest tendencies toward high oil saturations, whereas in the same area in Figure \ref{fig:q3_saturation_oil}, mean oil saturation values are close to one. There is also an area to the right of the well marked by the violet circle in Figure \ref{fig:q3_saturation_oil} which shows high mean oil saturation. This area shows very low mean oil saturation in \cite[Figure 12]{spremic2023bayesian}. Another example of areas where the MCMC results show higher oil saturation compared to the results of \cite{spremic2023bayesian} are two areas below the well located the furthest to the left at approximately inline $500$ and crossline $4800$. Conversely above this well at approximately crossline $4600\sim 4700$, there is an area where the mean oil saturation from the MCMC results is very low and the mean oil saturation in \cite[Figure 12]{spremic2023bayesian} is high.

\subsection*{Some differencies to \citet{spremic2023bayesian}}

Another significant difference is the difference in the uncertainties for the mean gas and oil saturations. In \cite[Figure 12]{spremic2023bayesian} the uncertainty is high for the mean oil saturations in the areas described in the previous paragraph. That is the area below the well marked by the green triangle, the area below the leftmost well and the area over the leftmost well. In addition, the uncertainty of the mean oil saturation is high above the well between the leftmost well and the well marked by the green triangle. The uncertainty of the gas saturations in \cite[Figure 12]{spremic2023bayesian} are high in the same areas as the mean gas saturation is high, except near the wells. The uncertainty of the mean gas saturation from the MCMC results in Figure \ref{fig:q3_uncertainty_clay} are generally considerably lower. 

There is also a difference in the mean clay content, which shows higher values for the MCMC results compared to the results of \cite{spremic2023bayesian}. The mean clay content is high in the same areas, however, much higher in Figure \ref{fig:q3_uncertainty_clay},\ref{fig:q3_content_clay} compared to the mean clay content in \cite[Figure 13]{spremic2023bayesian}. 

Ternary plots at two locations in the Alvheim field for the two sets of samples from the two approximation methods are shown in Figure \ref{fig:ternary} with $10,000$ values of MCMC samples from $\widehat{\pi}$ at locations $(800, 4800)$ and $(436, 5432)$ respectively. Ternary plots for the $100$ approximate posterior samples from the ensemble based approach at location $(800, 4800)$ and $(436, 5432)$ are shown in Figure \ref{fig:ternary_shallow_Mina} and \ref{fig:ternary_deep_Mina}, respectively. The depth at the location $(800, 4800)$ is $2112$ meters and the depth at the location $(436, 5432)$ is $2190$ meters. 

The ternary plots in Figure \ref{fig:ternary} show that the obtained samples in the two methods give the same composition at the deep location. In the shallow location, however, there are some differences. The MCMC results show less uncertainty in the oil and gas saturations. The oil saturations were also higher in Figure \ref{fig:ternary_shallow_MCMC} compared to Figure \ref{fig:ternary_shallow_Mina}. This agrees with the higher mean oil saturations in the MCMC results shown in Figure \ref{fig:q3_saturation_oil} compared to the mean oil saturations in \cite[Figure 12]{spremic2023bayesian}. The uncertainty of the brine saturations in Figure \ref{fig:ternary_shallow_MCMC} and Figure \ref{fig:ternary_shallow_Mina} is about the same. The ternary plots are only from two locations, such that general conclusions cannot be drawn. However, they indicate that the MCMC samples had more oil and lower uncertainty in the shallow area compared to the results of \cite{spremic2023bayesian}, which agrees with results in Figure \ref{fig:q3_sauturations} and \cite[Figure 12]{spremic2023bayesian}.

Even though there are dissimilarities between the results from the two different methods for approximating the posterior $\pi$, the results show the same tendencies. For most of the areas where the MCMC result shows a high mean oil saturation and the ensemble-based approach shows a higher mean oil saturation, the ensemble-based approach also has a higher uncertainty of the mean oil saturation. It means that the two results are compatible with each other. There is an exception around inline $1250$ and crossline $4900$ where the MCMC results show a high mean oil saturation and the ensemble-based approximation to the posterior shows a low mean oil saturation and low uncertainty. A possible reason could be that this area is located at the edge of the area such that there is no AVO data to the right of this area in the conditioning in the ensemble-based approximation. 

\begin{figure}[htb]
	\centering
	\begin{subfigure}[b]{0.3\textwidth}
		\centering\captionsetup{width=0.85\linewidth}
		\includegraphics[width=0.9\textwidth]{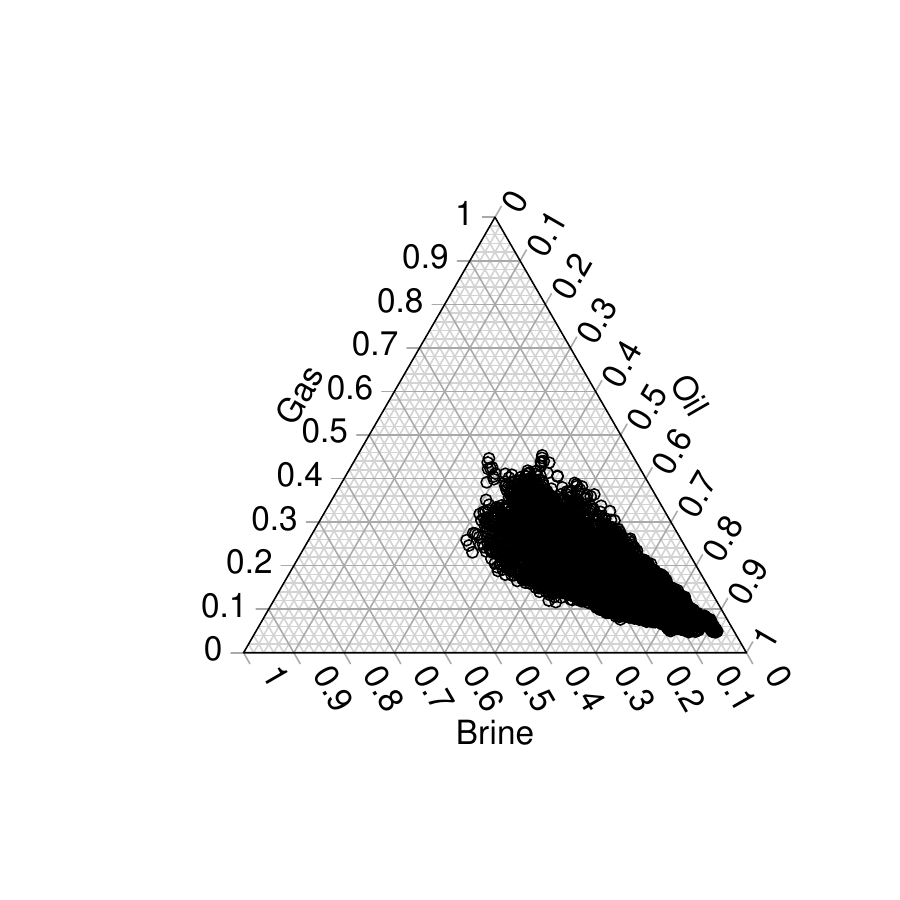}
		\caption{Ternary plots for the MCMC samples from $\widehat{\pi}$.}
		\label{fig:ternary_shallow_MCMC}
	\end{subfigure}
	\begin{subfigure}[b]{0.3\textwidth}
		\centering\captionsetup{width=0.85\linewidth}
		\includegraphics[width=0.9\textwidth]{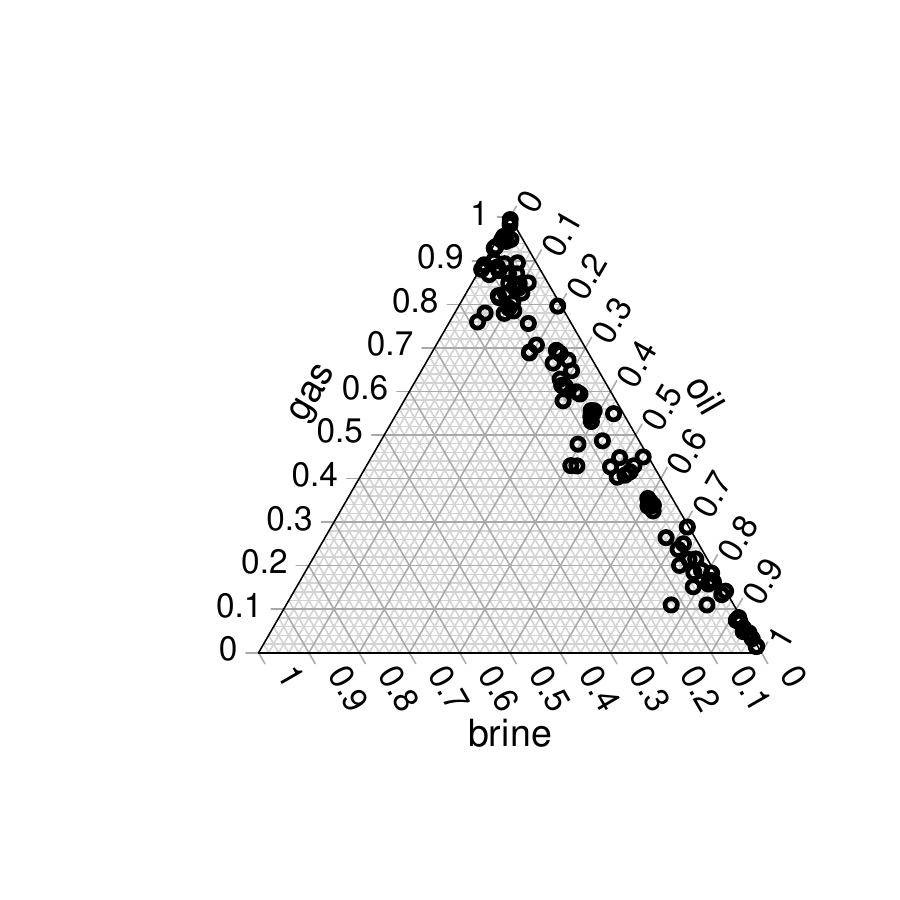}
		\caption{Ternary plots for the approximate posterior samples from \cite{spremic2023bayesian}. }
		\label{fig:ternary_shallow_Mina}
	\end{subfigure}
	\caption{Ternary plots for samples from the two approximations to the posterior $\pi$ at location $(800, 4800)$ in the Alvheim field, where the depth is $2112$ meters.}
	\label{fig:ternary}
\end{figure}

In the approach described in \cite{spremic2023bayesian}, the prior samples from the total area are divided into smaller patches when conditioning on the AVO data, to prevent false correlations between the AVO data which can occur when using such methods \citep{spremic2023bayesian}. Two patch sizes were compared to each other, and when using larger patches, there was more oil in the results, which resembles the MCMC results more. 

\cite{spremic2023bayesian} added an iterative loop of the approximate posterior samples to improve the posterior approximation. The results indicated a possible increase in the integrated oil saturation, which resembles the tendencies in the results from MCMC.

Another explanation could be that $\widehat{\pi}$ has less uncertainty than $\pi$. Figure \ref{fig:clay_small_area} showed that the approximate posterior samples had slightly lower uncertainty of the clay content. However, the uncertainty of the clay content is one of the similarities between the results from the MCMC and the method described in \cite{spremic2023bayesian}. The uncertainties of the gas and oil saturations in Figure \ref{fig:gas_small_area} and in Figure \ref{fig:oil_small_area} did not indicate that using the surrogate leads to a lower uncertainty. If it is the case that $\widehat{\pi}$ has less uncertainty than $\pi$ it is not clear from the results in Section \ref{sec:compare_small_area}.

\section{Discussion and conclusion}
\label{sc_discuss_conclusion}
In this paper, we tackle some challenges in Bayesian reservoir inversion methods including computational complexities due to forward model evaluations and high-dimensional Gaussian random fields. Innovative statistical approaches based on the generalized Bayesian method and efficient techniques such as the fast Fourier transform (FFT) for simulating Gaussian random fields are employed, making large-scale computations manageable even on standard hardware. Novel sampling methodologies using the preconditioned Crank-Nicolson method offer efficient exploration of high-dimensional parameter spaces, promising substantial computational speed enhancements in MCMC sampling and streamlining Bayesian reservoir inversion methods for practical applications.

The forward function of the likelihood model in the Alvheim case was approximated by a MARS model to speed up the MCMC algorithm. The approximation was about $32$ times faster. It was also confirmed by a comparison between MCMC samples obtained using the exact forward model. Using the MARS model, the MH algorithms with different proposal distributions (random walk, preconditional Crank-Nicolson, MALA) were compared on a smaller part of the Alvheim field. A comparison of ESS per computation time showed that the preconditional Crank-Nicolson algorithm was the most efficient algorithm in the Alvheim case. This algorithm was therefore used with the surrogate to perform MCMC on the entire Alvheim field. The MCMC results had similarities to the results in \cite{spremic2023bayesian}, however, the MCMC results showed more oil and lower uncertainty for the gas and oil saturations. 

Another method of employing the MARS approximation involves utilizing samples derived from $\widehat{\pi}$ to construct an informed independent proposal. Research by \cite{khoshkholgh2021informed} discovered that devising an independent proposal distribution based on posterior information enhanced efficiency compared to employing a dependent standard random walk proposal distribution. Concurrently, an independent proposal could be executed in parallel to expedite computations. We have performed a study to concurrently sample from $\pi$ and $\widehat{\pi}$. That is the Metropolis-Hastings (MH) algorithm with proposal distribution $q_3$ was utilized to generate 500,000 MCMC samples from $\widehat{\pi}$. Moreover, every 1000th iteration involved subjecting the proposed sample to an acceptance/rejection step utilizing the acceptance probability computed using the exact forward model rather than the forward model approximation.  Unfortunately, the acceptance rate was zero for exact MCMC sampling. A challenge arising from combining dependent and independent proposals lies in their differing optimal tuning. An independent proposal demonstrates greater efficiency when the acceptance rate nears unity. However, for the MH algorithm with proposal  $q_3$ tuned to attain an acceptance rate of around 23.4\%, the proposed samples may be excessively extreme or distant from the posterior, potentially leading to rejection by the independent proposal. Consequently, this approach might be more suitable for the MALA, which boasts a higher optimal acceptance rate.

Since the MARS model is non-parametric, it can in principle model functions of arbitrary shape. This suggests that MARS can approximate other forward models in other problems as well. The forward model in the Alvheim example takes four continuous input parameters. If there are more covariates, one probably need more terms in the model, which would increase the computation time a bit. However, the input space of the MARS model which served as a surrogate in \cite{CHEN_mars} had dimension $15$, and the MARS model still reduced the computation in that problem as well. The conclusion that MARS is a good surrogate model for any forward model in any problem can not be made. However, as the model is non-parametric, the approximation could work for similar problems with complicated and computationally inefficient forward models in a Bayesian setting with similar input spaces.

		\subsection*{Acknowledgments}
		We acknowledge support from the Centre for Geophysical Forecasting - CGF (Norwegian Research Council grant no. 309960) and the GAMES consortium (grant no. 294404) at the Norwegian University of Science and Technology (NTNU). 
		
		\subsection*{Conflicts of interest/Competing interests}
		The authors declare no potential conflict of interests.

\clearpage
		\appendix
		\section{Efficient sampling and evaluation of Gaussian distribution using FFT}
		
		\label{sc_detail_FFT}
		This section outlines the theory and application of circulant matrices and fast Fourier transform (FFT) in generating Gaussian random field samples and efficiently evaluating probability density functions. 
		
		\subsubsection*{Circulant matrices and FFT}
		
		The general idea is to use structure of circulant matrices and perform matrix-vector multiplication in a lower-dimensional space in the Fourier domain. 	A circulant matrix has the following structure
		
		$$ \boldsymbol{C} = \begin{pmatrix}
		c_{0} & c_{n-1} & \dots & c_{1} \\
		c_{1} & c_{0}  &\dots & c_{2} \\
		c_{2} & c_{1} & \dots & c_{3} \\
		\vdots & \vdots &\ddots & \vdots\\
		c_{n-1} & c_{n-2} & \dots & c_{0}\\
		\end{pmatrix}
		$$
		
		Eigenvalues of a circulant matrix can be obtained by the discrete Fourier transform (DFT) of the first column of $\bC$, and matrix $\bC$ is fully specified by its one row or column, meaning all the information that we need is stored in its basis. 
		Matrix of eigenvalues of a matrix such as $\bC$, is denoted as $\bLambda$.
		
		We now assume our covariance matrix $\bSigma$ is a circulant matrix. 
		[nolist, inline]{Check the Block circulant assumption}
		
		We know there exists a decomposition 
		$
		\bSigma = \bC^{1/2}\bC^{1/2},
		$ 
		since $\bSigma$ is assumed to be symmetric, positive definite matrix. Same is true for the inverse, 
		$ 
		\bSigma^{-1} = \bC^{-1/2}\bC^{-1/2}. 
		$
		Let $z \sim N(0,1)$.
		Using the unitary or Fourier matrix, $F$, one can then write,
		\begin{equation*}
		\bC^{1/2} = \bF \bLambda^{1/2}\bF^H.
		\end{equation*}
		To compute the product $\bC^{-1/2}\bz$,
		\begin{equation*}
		\bC^{1/2}\bz = \bF \bLambda^{-1/2} \bF^H 
		= \frac{1}{\sqrt{n}}(\text{idft2}(\bz) \odot \sqrt{\text{dft2}(\bc)})
		lin    
		\end{equation*}
		where $\bc$ is the base of the circulant matrix. Here dft2$(\cdot)$ and idft2$(\cdot)$ are respectively two-dimensional discrete Fourier transform and inverse discrete Fourier transform. For more details, we encourage the readers to have a look at \cite{rue2005gaussian}.

		\subsubsection{Constructing the distance matrix}\label{sc:dist_mx}
		
		Here we explain how the distance matrix can be constructed such that it represents a base of a circulant matrix. 
		We make use of topological knowledge, namely a torus. For a more detailed explanation we encourage a curious reader to go deeper into reading the references, for example \cite{davies2013circulant} provides some illustrations and detailed explanations. 
		
		Assuming our covariance matrix has a circulant structure it is fully defined by its first row or column. This is called the basis of the matrix and all of the information that we need is stored there. Covariance matrix is made up of variances and correlation structure. For the correlation structure, it is necessary to define how we will measure distances between the points. And this is a crucial step. We imagine we are on a torus and construct the distances accordingly. 
		
		We measure distance from our corner point, corresponding to the first element of the matrix, $C_{1,1}$ to all the other points (this will correspond to a row of a distance matrix, later used in our covariance matrix). And since we are on a torus, we are looking to measure the minimum distances. 
		We are on a circulant shaped geometric body, hence, once we are halfway over across the ellipse/circle (cross-section of an empty torus), we are closer to the points of choice from the ``other side" of the circular body.
		
		One can either choose to implement this as a double for-loop, where one goes over the grid points on a torus and calculates the distance (and also correlation function values), or one can represent this in a vector form and perform calculations through vector and matrix calculations. Details of the latter can be found in \cite{davies2013circulant}.
		Additionally, depending on the programming language one chooses to use, documentation should be thoroughly read. This is because implementation of the discrete FFT, both one and multi-dimensional, differs between programming languages and the way normalization is performed or included can be different. 
		Packages and libraries for performing two-dimensional FFT are available in both \texttt{R} \cite{Rsoftware} and \texttt{Python}. 
		
		\subsubsection*{Sampling}
		
		Depending on the programming language, we can make use of the different packages available in for example R or python, which enable us to calculate the one or two-dimensional dft, necessary for computing the vector-matrix product and obtaining our samples.

		Assuming the size of the grid, $\mathcal{D}$, to be $n_x \times n_y$, covariance $\bSigma = \bsigma^2\rho(\bD)$, where $\bD$ is the distance matrix computed as described in \ref{sc:dist_mx}. The algorithm for simulating a sample of a Gaussian random field is shown in Algorithm \ref{alg:samp_fft}.

		\begin{algorithm}[]
			\caption{Sampling from $N(\bmu, \bSigma)$ using circulant matrices and FFT.}
			\label{alg:samp_fft}
			\begin{algorithmic}[1]
				\State Construct distance matrix $\bD$ as in \ref{sc:dist_mx}
				\State $\tilde{c} = \sigma^2\rho(\bD)$ \Comment{Constructing the base of a circulant matrix}
				\State $\bz \sim N(0, 1)$, \Comment{Generating an independent sample}
				\State $\bLambda^{sr} = \sqrt{\text{Re}(\text{dft2}(\tilde{c}))}$  \Comment{Finding the eigenvalues through DFT}
				\State $\bx^*= \bmu + \text{Re}(\text{dft2}(\bLambda^{sr} \odot \text{idft2}(\bz)))$  \Comment{Constructing a sample $\bx^* \sim N(\bmu,\bSigma)$}
			\end{algorithmic}
		\end{algorithm}
		
		\subsubsection*{Evaluating}
		
		We wish to evaluate Gaussian pdf, which is given as,
		
		$$ p(\mathbf{x} ; \mathbf{\mu}, \mathbf{\Sigma}) = \frac{1}{\sqrt{(2\pi)^k |\mathbf{\Sigma}|}} \exp\left(-\frac{1}{2} (\mathbf{x}-\mathbf{\mu})^T \mathbf{\Sigma}^{-1} (\mathbf{x}-\mathbf{\mu})\right).
		$$
		
		This gives the log-pdf being: 
		
		$$ \log(p(\mathbf{x})) \propto -\frac{1}{2} (\mathbf{x}-\mathbf{\mu})^T \mathbf{\Sigma}^{-1} (\mathbf{x}-\mathbf{\mu}) $$
		
		If we for notational simplicity assume $\mu = 0$, we have that
		$ \log(p(\mathbf{x})) \propto -\frac{1}{2} \mathbf{x}^T \mathbf{\Sigma}^{-1} \mathbf{x}$
		which can then be written as 
		\begin{equation}\label{eq:loglik}
		\log(p(\mathbf{x})) \propto -\frac{1}{2} \mathbf{x}^T \mathbf{\bC}^{-1/2}\mathbf{\bC}^{-1/2} \mathbf{x}.
		\end{equation}
		Thus, We can compute equation \eqref{eq:loglik} in an efficient way, making use of the theory presented above  for sampling. Details can be found in Algorithm \ref{alg:eval_fft}. Here, we are evaluating a $N(\bzero, \bSigma)$, $\bmu=\bzero$, otherwise when we have $N(\bmu, \bSigma)$, we let $\bv=(\bx^c-\bmu)$. 
		
		\begin{algorithm}[]
			\caption{Evaluating a Gaussian pdf using circulant matrices and FFT.}
			\label{alg:eval_fft}
			\begin{algorithmic}[1]
				\State Construct distance matrix $\bD$ as in \ref{sc:dist_mx}
				\State $\tilde{c} = \sigma^2\rho(\bD)$ \Comment{Previously constructed base of a circulant matrix}
				\State  $\bv=(\bx^c-\bmu)$
				\State $\bLambda^{sr} = \sqrt{\text{Re}(\text{dft2}(\tilde{c}))}$  \Comment{Finding the eigenvalues through DFT}
				\State $\bu= \text{Re}(\text{dft2}(\bLambda^{-sr} \odot \text{idft2}(\bv)))$  \Comment{Computing $\bC^{-1/2}\bv$}
				\State $ \log(p(\mathbf{x}))_{FFT} = - \frac{1}{2} \bu^T\bu$ \Comment{Equivalent to equation \eqref{eq:loglik}}
			\end{algorithmic}
		\end{algorithm}

\section{Non-parametric kernel regression} 
\label{sec:npkr}

NPKR (Non-parametric kernel regression) approximates the response $f$ at a new data point by a weighted average of the data points in the training set. There are numerous ways to determine these weights. In this work, the Nadaraya-Watson estimator
\begin{equation}
\widehat{f}(\bx) = \sum_{i=1}^{n} w_i(\bx) f_i
\label{Nadaraya_Watson}
\end{equation}
will be used with weights
\begin{equation}
w_i(\bx) = \frac{K_{\bb}(\bx-\bx_i)}{ \sum_{l=1}^{n}K_{\bb}(\bx-\bx_l)} \quad i=1, 2, ..., n. 
\label{Nadaraya_Watson_Wi}
\end{equation}
The function $K_{\bb}(\bx-\bx_i)$ is a Gaussian kernel with fixed bandwidths for each predictor $\bb = [b_1, b_2, ..., b_p ]$. When using the Gaussian kernel, the bandwidths are the standard deviation of the density. The weight in equation \eqref{Nadaraya_Watson_Wi} decrease with the distance between the new data point and the data points in the training set. For $p$ covariates, $K_{\bb}$ is chosen to be the product of $p$ one dimensional Gaussian kernels:
\begin{align*}
K_{\bb}(\bx-\bx_i) 
= 
\prod_{j= 1}^{p} k_{b_j}(x_j-x_{ji}) 
= 
\prod_{j = 1}^{p} \frac{1}{b_j} k \left( \frac{x_j-x_{ji}}{b_j} \right) 
= 
\prod_{j= 1}^{p}  \frac{1}{\sqrt{2 \pi} b_j} e^{-\frac{(x_j-x_{ji})^2}{2 b_j^2}}.
\end{align*}
The Gaussian kernel is a popular non-compact kernel \citep{ESL}. The function \texttt{npreg} from the \texttt{R} package ``\texttt{np}" \cite{np_package} is used to fit the NPKR model. The reader can refer to \cite{NPR_Li} for an in-depth exploration and description of the model

The bias-variance trade-off asserts itself when selecting the bandwidths. Selecting small bandwidths tends to lower the bias and increase the variance as the nearest points are valued a lot. On the other hand, selecting too large bandwidths leads to a high bias and low variance, since this tends to include values far away from the new data points in the estimate. For the NPKR model used in this work, the bandwidths are found using least squares cross-validation. This is the only predetermined property of an NPKR model because the weights need to be determined every time a prediction of a new data point is made. Much of the work is therefore done at evaluation time \citep{ESL}. 

Using an NPKR model has both advantages and disadvantages. An advantage lies in its flexibility as it does not assume a specific functional form for the response, which is particularly beneficial in high-dimensional scenarios. Additionally, the intuitive nature of averaging nearby data points contributes to its appeal. The model, represented by equations \eqref{Nadaraya_Watson} and \eqref{Nadaraya_Watson_Wi}, constitutes a linear combination of smooth functions, resulting in an overall smooth model.

However, a notable disadvantage of NPKR is its susceptibility to bias at boundaries, as noted in prior literature \cite{CASI,ESL}. When making predictions near boundaries, the weighted average tends to favor data points away from the boundary, potentially leading to inaccurate predictions, especially if the actual response is decreasing. This bias becomes more pronounced in higher dimensions due to the increased proportion of data points near the boundary. Although one approach to mitigate this bias involves using locally weighted linear regression instead of a weighted average, this method is not explored in this work. The boundary bias issue also extends to unevenly distributed training data, where predictions may be skewed towards the direction with more data points. The reliance of NPKR on the specific data points in the training set further underscores the importance of evenly spaced training data to potentially enhance the model's performance.

\subsection{Approximating the gradient of the forward model} \label{sec:h_hatt_gradient}
To use the MALA introduced in Section \ref{sc_other_proposals}, the gradient $\nabla \bh$ is needed. However, the forward model is treated as a black box. Consequently, the partial derivatives are unavailable. A numerical approximation to the gradient 
can be found by adding small perturbations, $\varepsilon$, to only one of the covariates while the rest are fixed. The difference in the forward model with and without the perturbation is divided by $\varepsilon$. For the numerical partial derivative of $h_{R_0}$ with respect to the covariate $x_g$ at data point $\bx_i = [x_{ig}, x_{io}, x_{ic}, x_{id}]$ this is 
\begin{equation*}
\frac{\partial h_{R_0}}{\partial x_g}(\bx_i) = \frac{h_{R_0}(\bx_i + [\varepsilon, 0, 0, 0]) - h_{R_0}(\bx_i) }{\varepsilon},
\end{equation*}
with $\varepsilon = 0.0001$. However, the forward function needs to be evaluated to calculate the numerical partial derivatives, which require a lot of computation time. In this thesis, the gradients of the four models discussed so far are tested for approximating the gradient of the forward model, $\nabla \bh$, in order to reduce the computation time. The gradient of the MARS models and NPKR models are described in   \ref{sec:npkr}. Because the analytical gradient is unavailable, the approximations are compared to the numerical approximations of the gradient instead. For simplicity, the numerical approximation to the gradient is denoted by $\nabla \bh$ and referred to as the gradient of the forward model  throughout this section.

In addition, MARS models are created with partial derivatives as responses. That is, six MARS models are fitted with each of the partial derivatives $\frac{\partial h_{R_0}}{\partial x_g}$, $\frac{\partial h_{R_0}}{\partial x_o}$, $\frac{\partial h_{R_0}}{\partial x_c}$, $\frac{\partial h_{G}}{\partial x_g}$, $\frac{\partial h_{G}}{\partial x_o}$, and $\frac{\partial h_{G}}{\partial x_c}$ as responses. As MCMC is only performed with $\bx = [\bx_g, \bx_o, \bx_c]$ the partial derivative of $\bh$ with respect to the depth is not needed. The six MARS models fitted to the partial derivatives are collectively denoted as $\text{MARS}_{\nabla}$.

The K-means approximation to the gradient of an NPKR model is again tested. The procedure is described in \ref{sec:npkr}. A simple grid search of the number of clusters $\in [3, 5, 10]$ and the number of closest neighbours to average $\in [3, 5, 10]$ was performed. The computation time decreased as the number of clusters increased. Therefore, the number of clusters was set to $10$. When the number of clusters was $10$, the lowest MSE and the highest correlation was achieved when using $5$ neighbours for both $\widehat{\bh}_{\text{NPKR}_{1000}}$ and $\widehat{\bh}_{\text{NPKR}_{4000}}$.

The accuracy of the approximations to the gradient is the mean correlation 
\begin{equation*}
\text{Corr}\left(\nabla \bh, \widehat{ \nabla \bh}\right) := \frac{1}{6} \sum_{j \in \{g, o, c\}} \left( \text{Corr}\left(\frac{\partial h_{R_0}}{\partial x_j}, \widehat{\frac{\partial h_{R_0}}{\partial x_j}}\right) + \text{Corr}\left(\frac{\partial h_G}{\partial x_j}, \widehat{\frac{\partial h_G}{\partial x_j}}\right) \right)
\end{equation*}
and mean MSE 
\begin{equation*}
\text{MSE}\left(\nabla \bh, \widehat{ \nabla \bh}\right) := \frac{1}{6} \sum_{j \in \{g, o, c\}} \left( \text{MSE}\left(\frac{\partial h_{R_0}}{\partial x_j}, \widehat{\frac{\partial h_{R_0}}{\partial x_j}}\right) + \text{MSE}\left(\frac{\partial h_{G}}{\partial x_j}, \widehat{\frac{\partial h_{G}}{\partial x_j}}\right) \right)
\end{equation*}
for all the covariates $j$. The correlation, MSE and computation time are reported in Table \ref{table:gradients}. Again, the computation time is the average of $50$ computation times. One should keep in mind when reading Table \ref{table:gradients}, that the correlation and MSE are calculated using the numerical approximations as $\nabla \bh$.

As per Table \ref{table:gradients}, $\text{MARS}_{\nabla}$ emerges as the most effective in approximating the forward model gradient, displaying superior correlation, MSE, and computation time. In Chapter 5, MALA employs this approximation. The gradients of the MARS models, $\nabla \widehat{\bh}_{\text{MARS}}$ and $\nabla \widehat{\bh}_{\text{MARS}_{\text{OF}}}$, rank second and third highest in correlation, and second and third lowest in MSE, following $\text{MARS}_{\nabla}$. Notably, in contrast to the 2D case, $\nabla \widehat{\bh}_{\text{MARS}}$ outperforms $\nabla \widehat{\bh}_{\text{MARS}_{\text{OF}}}$ marginally in correlation and MSE. However, computation time for these gradients is considerably high. Integration of binary search in the partial derivative extraction algorithm reduces computation time, yet its efficiency varies with interaction terms in the MARS model. For a MARS model with approximately 40 non-interaction terms, computation time is around 10\% that of $\nabla \widehat{\bh}_{\text{MARS}}$. However, reducing the MARS model's degree exacerbates MSE and correlation. Increasing data points in the NPKR model enhances correlation and reduces MSE for its gradient but extends computation time. While using the K-means approach expedites approximating $\nabla \widehat{\bh}_{\text{NPKR}_{4000}}$, it slows down approximating $\nabla \widehat{\bh}_{\text{NPKR}_{1000}}$.

The computation time of $\nabla \widehat{\bh}_{\text{NPKR}_{1000}}$ was close to the computation time of $\nabla \widehat{\bh}_{\text{MARS}}$ and lower than the computation time for the gradient of the larger MARS model. The K-means approach with $10$ clusters and $5$ neighbours is not particularly fast. It was however, a bit better at predicting the gradient of the forward model compared to $\nabla \widehat{\bh}_{\text{NPKR}_{1000}}$ and $\nabla \widehat{\bh}_{\text{NPKR}_{4000}}$. 
{\renewcommand{\arraystretch}{1.6}%
	\begin{table}[H]
		\centering
		\begin{tabular}{ |c|c|c|c| } 
			\hline 
			$ \widehat{\nabla \bh}$ & correlation & MSE $[10^{-5}]$ & computation time [sec] \\
			\hline
			$\text{MARS}_{\nabla}$ & 0.984  & 0.2 & 0.32 \\ 
			$\nabla \widehat{\bh}_{\text{MARS}}$ & 0.879 & 1.7 & 20.59 \\ 
			$\nabla \widehat{\bh}_{\text{MARS}_{\text{OF}}}$ & 0.872 & 1.9 & 33.73 \\
			$\nabla \widehat{\bh}_{\text{NPKR}_{1000}}$ & 0.621 & 6.2 & 22.41 \\ 
			$\widehat{\nabla \bh_{\text{NPKR}_{1000}}}$ (K-means)  & 0.679 & 6.1 & 43.58 \\ 
			$\nabla \widehat{\bh}_{\text{NPKR}_{4000}}$ & 0.702 & 4.7 & 62.19 \\
			$\widehat{\nabla \bh_{\text{NPKR}_{4000}}}$ (K-means) & 0.844 & 3.7 & 54.28 \\ 
			\hline
		\end{tabular}
		\caption{Correlation, MSE and computation time for approximations $\nabla \widehat{\bh}$.}
		\label{table:gradients}
\end{table}}

\begin{figure}[htb]
	\centering
	\begin{subfigure}[b]{0.2\textwidth}
		\centering\captionsetup{width=.8\linewidth}
		\includegraphics[width=\textwidth]{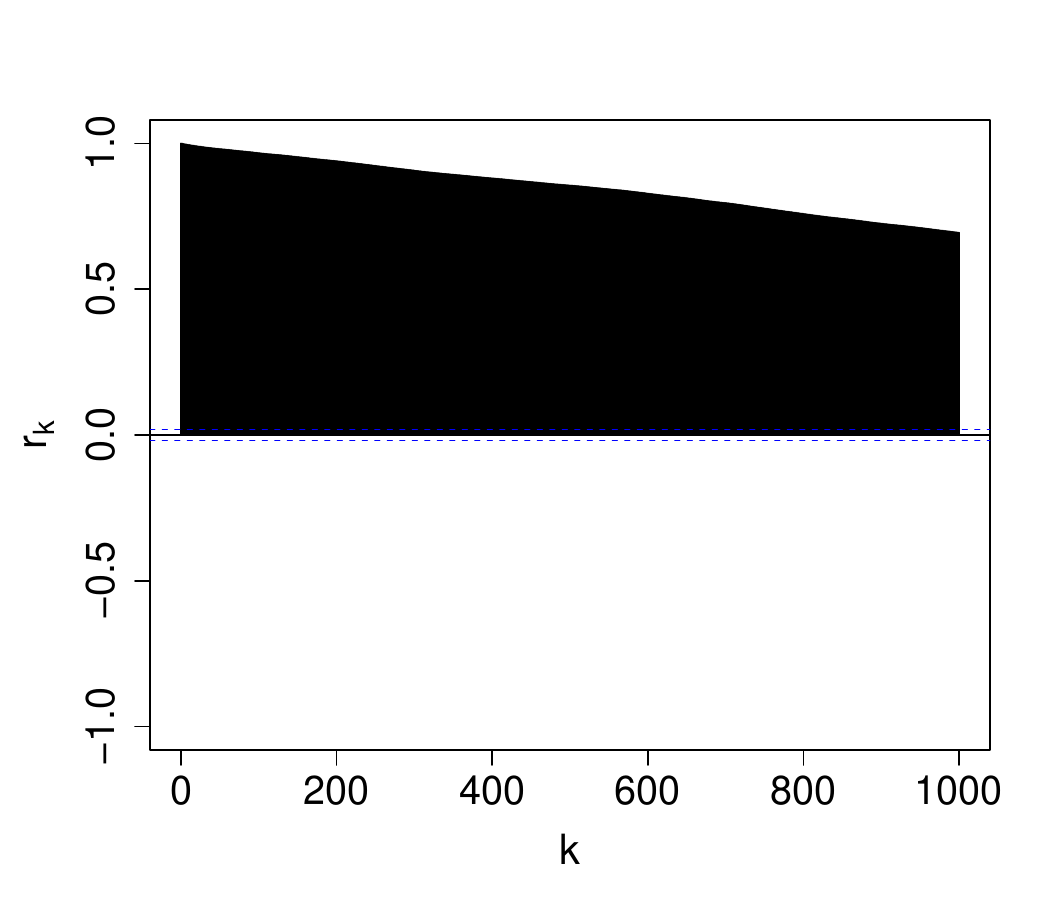}
		\caption{ACF plot from MCMC computations with $q_1$ as proposal distribution.}
		\label{fig:acf_q1_gas}
	\end{subfigure}
	\begin{subfigure}[b]{0.2\textwidth}
		\centering\captionsetup{width=.8\linewidth}
		\includegraphics[width=\textwidth]{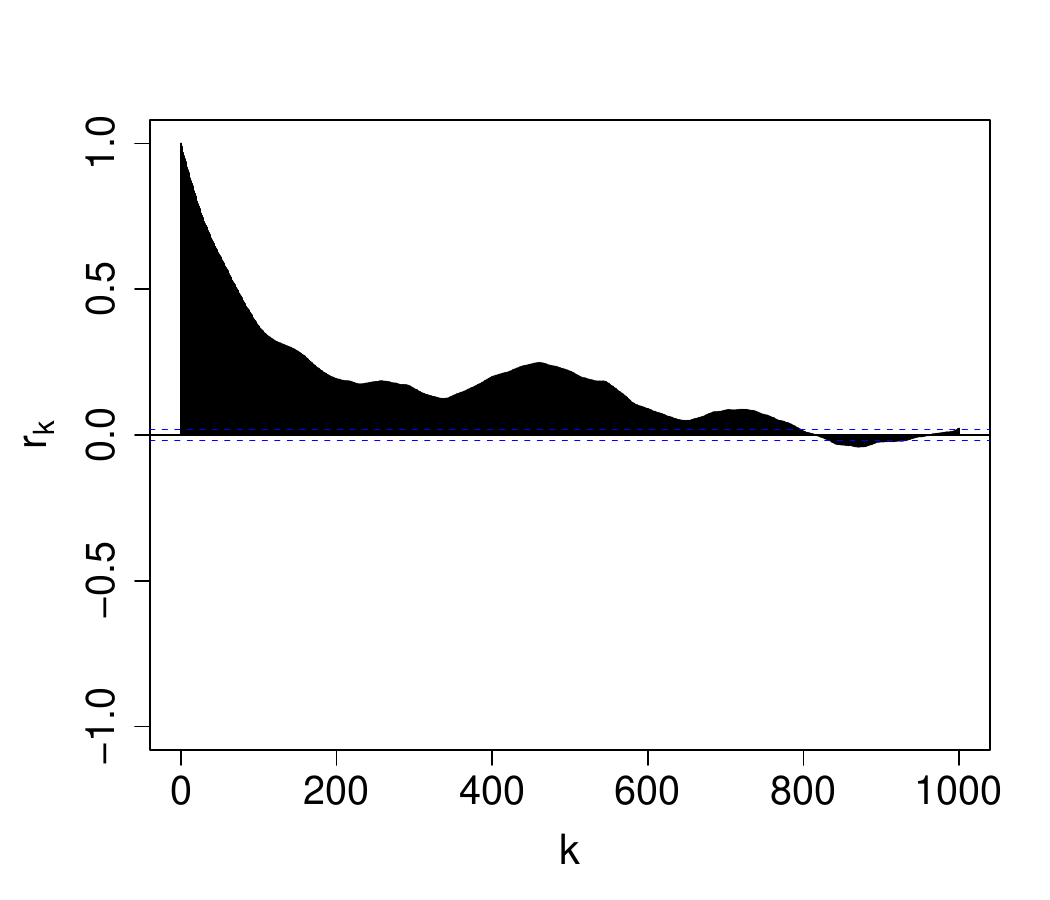}
		\caption{ACF plot from MCMC computations with $q_2$ as proposal distribution.}
		\label{fig:acf_q2_gas}
	\end{subfigure}
	\begin{subfigure}[b]{0.2\textwidth}
		\centering\captionsetup{width=.8\linewidth}
		\includegraphics[width=\textwidth]{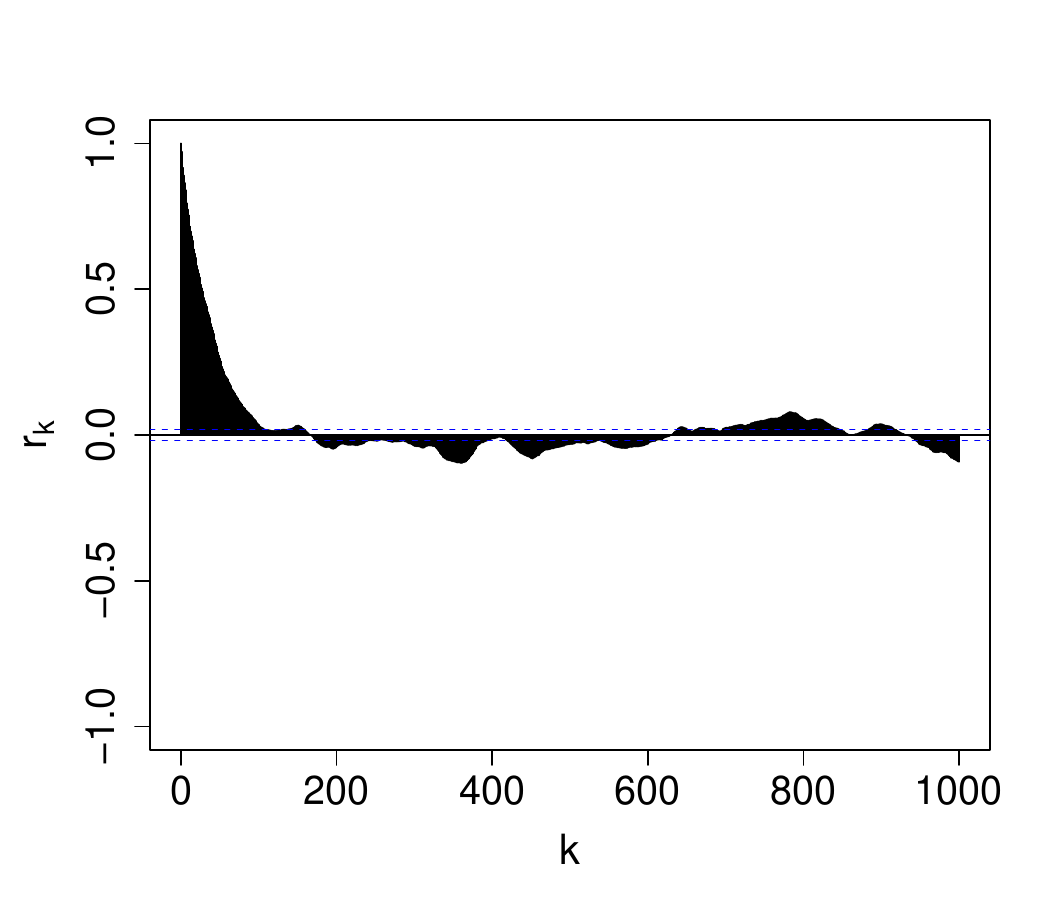}
		\caption{ACF plot from MCMC computations with $q_3$ as proposal distribution.}
		\label{fig:acf_q3_gas}
	\end{subfigure}
	\begin{subfigure}[b]{0.2\textwidth}
		\centering\captionsetup{width=.8\linewidth}
		\includegraphics[width=\textwidth]{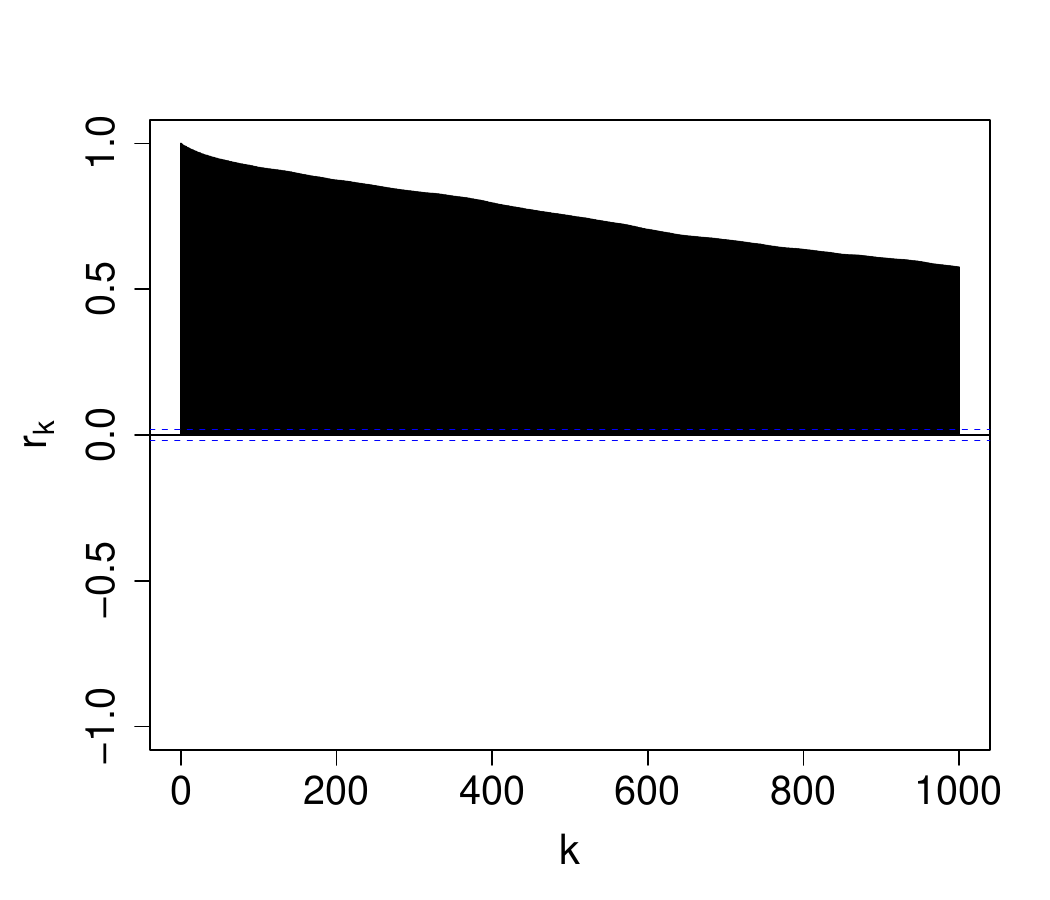}
		\caption{ACF plot from MCMC computations with $q_4$ as proposal distribution.}
		\label{fig:acf_q4_gas}
	\end{subfigure}
	\begin{subfigure}[b]{0.2\textwidth}
		\centering\captionsetup{width=.8\linewidth}
		\includegraphics[width=\textwidth]{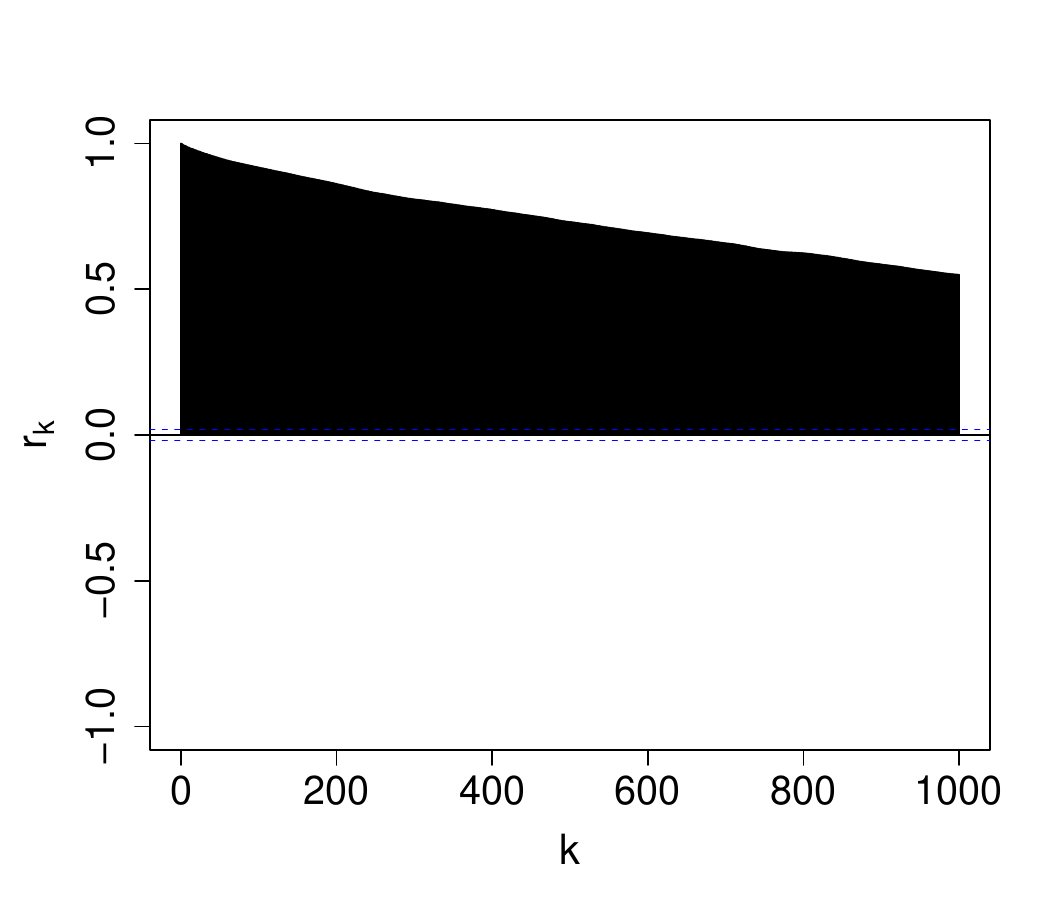}
		\caption{ACF plot from MCMC computations with $q_1$ as proposal distribution.}
		\label{fig:acf_q1_oil}
	\end{subfigure}
	\begin{subfigure}[b]{0.2\textwidth}
		\centering\captionsetup{width=.8\linewidth}
		\includegraphics[width=\textwidth]{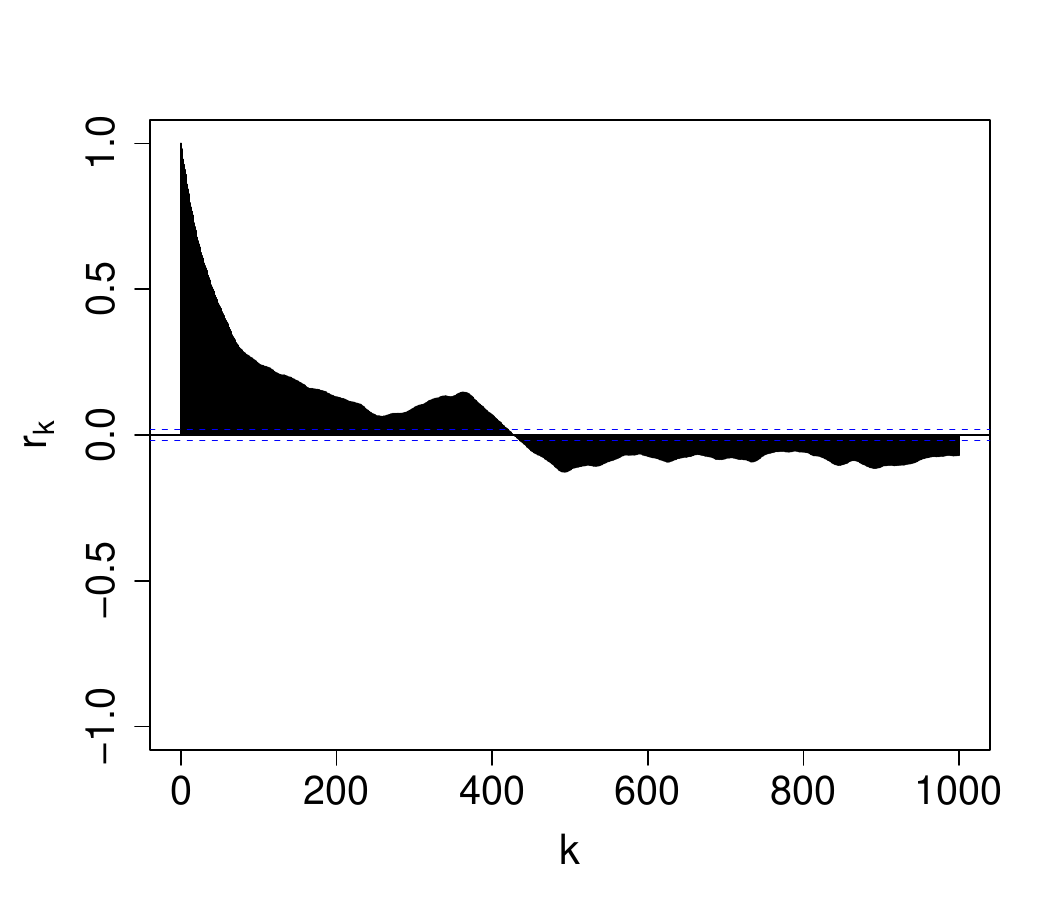}
		\caption{ACF plot from MCMC computations with $q_2$ as proposal distribution.}
		\label{fig:acf_q2_oil}
	\end{subfigure}
	\begin{subfigure}[b]{0.2\textwidth}
		\centering\captionsetup{width=.8\linewidth}
		\includegraphics[width=\textwidth]{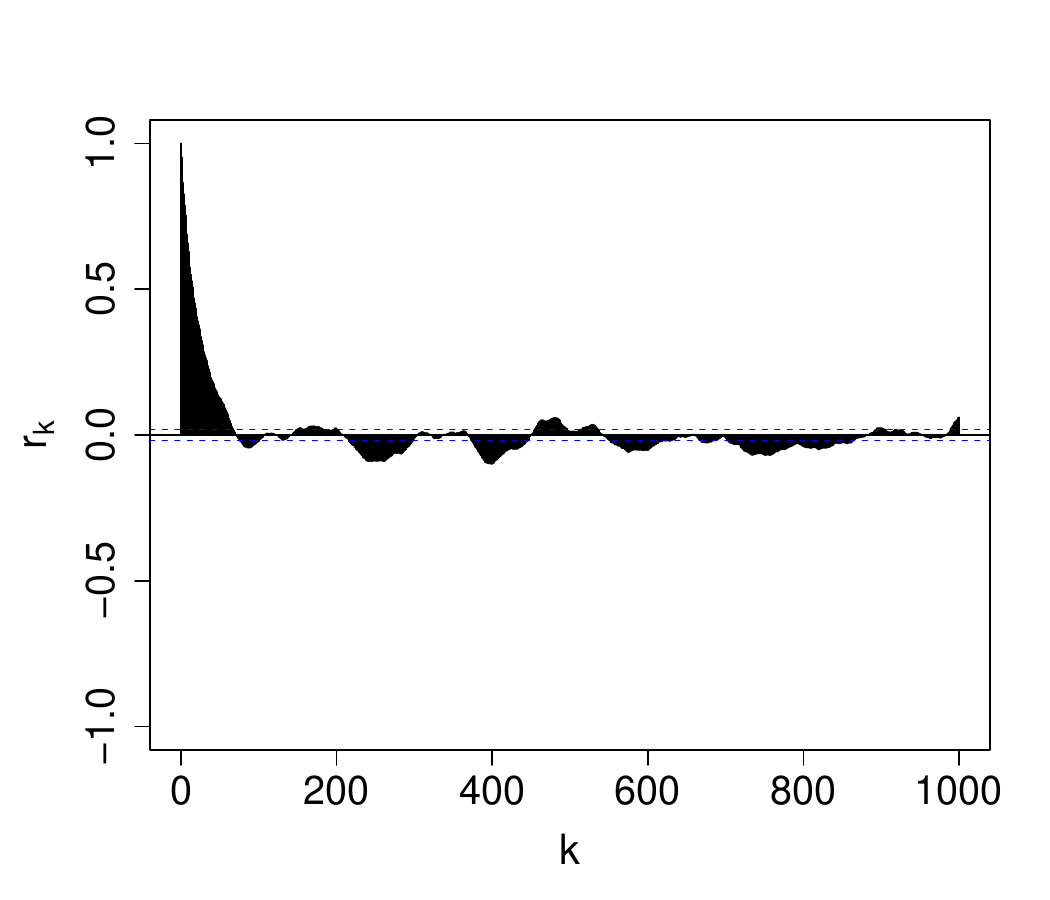}
		\caption{ACF plot from MCMC computations with $q_3$ as proposal distribution.}
		\label{fig:acf_q3_oil}
	\end{subfigure}
	\begin{subfigure}[b]{0.2\textwidth}
		\centering\captionsetup{width=.8\linewidth}
		\includegraphics[width=\textwidth]{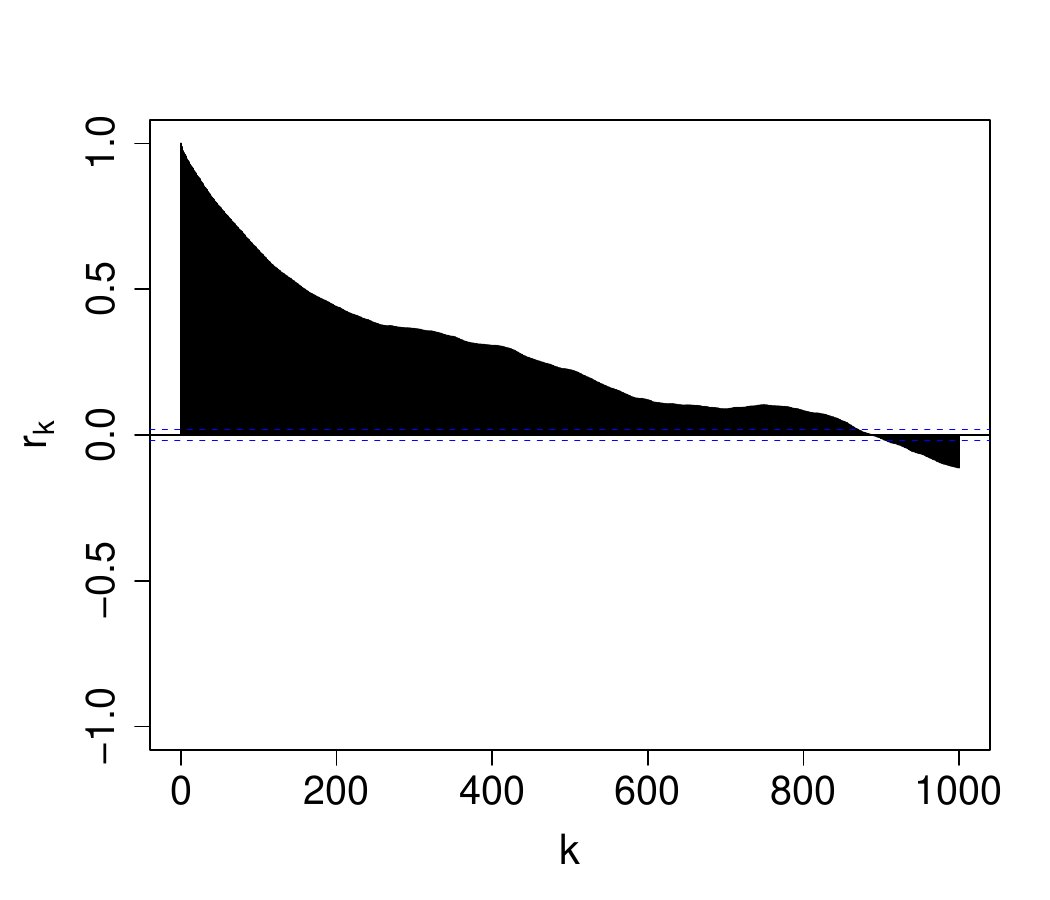}
		\caption{ACF plot from MCMC computations with $q_4$ as proposal distribution.}
		\label{fig:acf_q4_oil}
	\end{subfigure}
	\begin{subfigure}[b]{0.2\textwidth}
		\centering\captionsetup{width=.8\linewidth}
		\includegraphics[width=\textwidth]{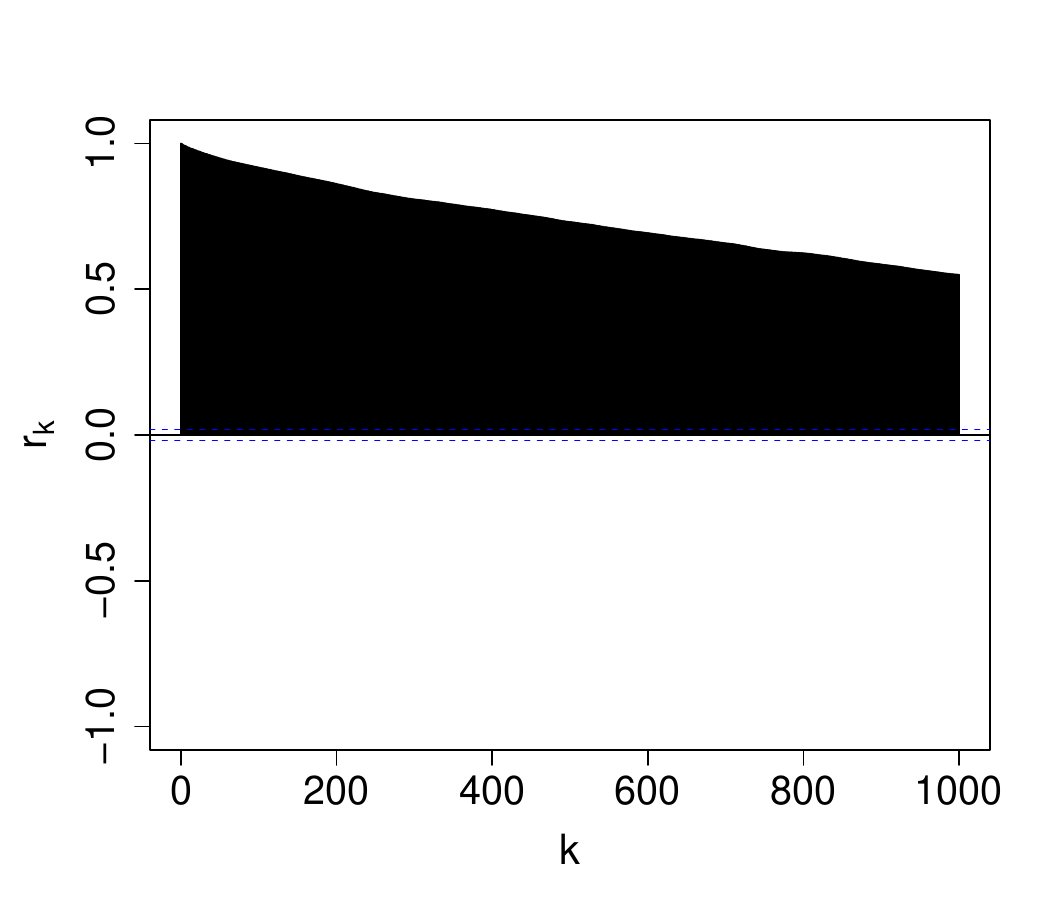}
		\caption{ACF plot from MCMC computations with $q_1$ as proposal distribution.}
		\label{fig:acf_q1_clay}
	\end{subfigure}
	\begin{subfigure}[b]{0.2\textwidth}
		\centering\captionsetup{width=.8\linewidth}
		\includegraphics[width=\textwidth]{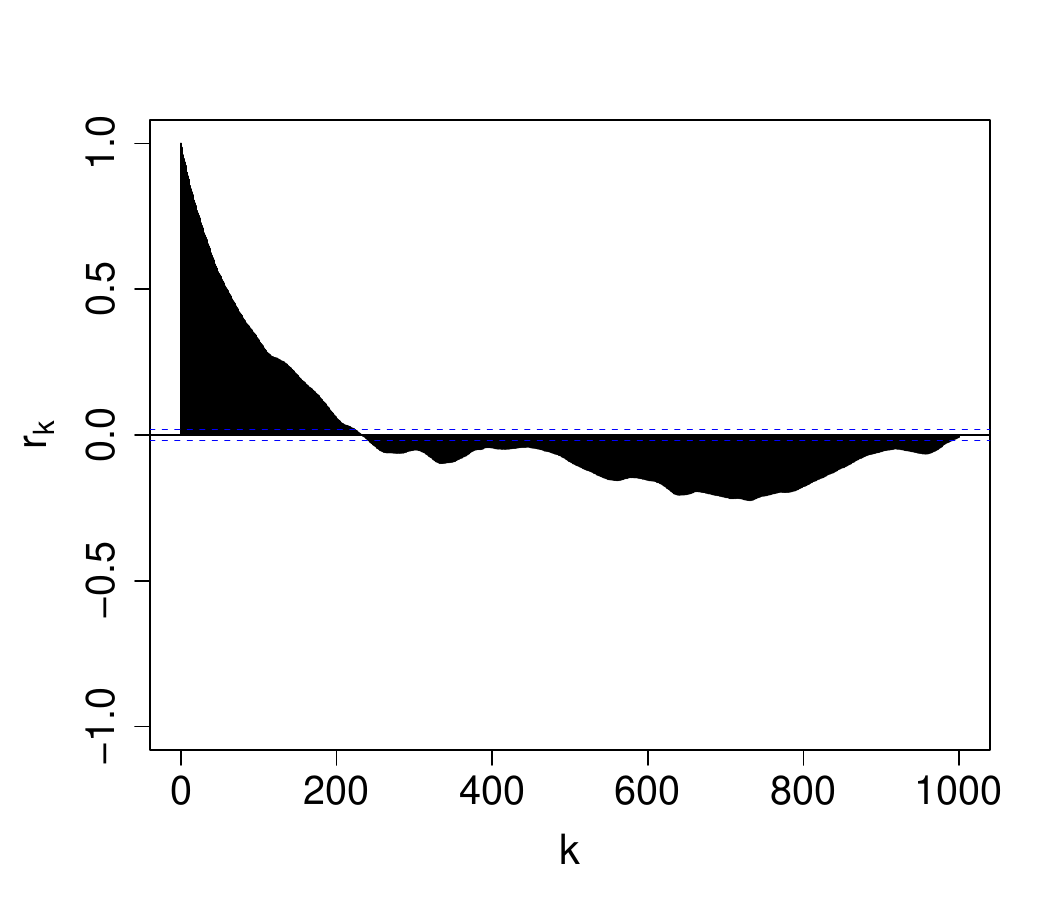}
		\caption{ACF plot from MCMC computations with $q_2$ as proposal distribution.}
		\label{fig:acf_q2_clay}
	\end{subfigure}
	\begin{subfigure}[b]{0.2\textwidth}
		\centering\captionsetup{width=.8\linewidth}
		\includegraphics[width=\textwidth]{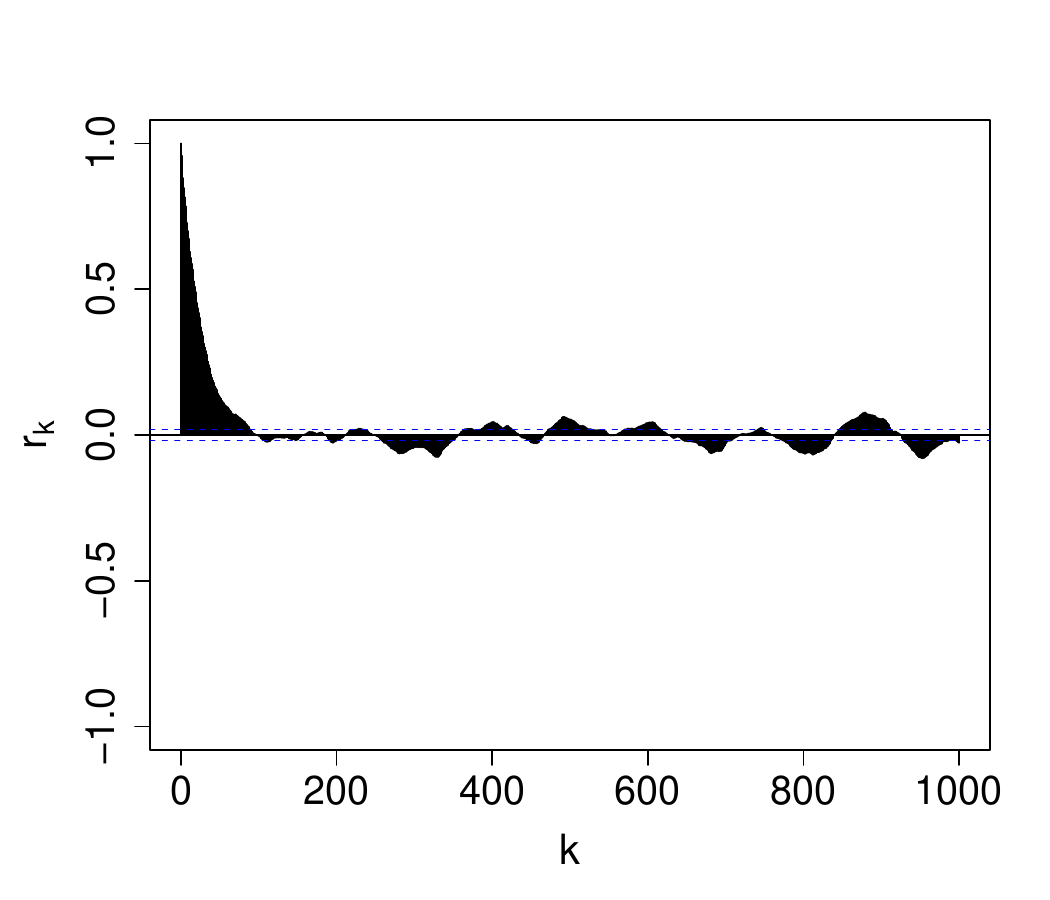}
		\caption{ACF plot from MCMC computations with $q_3$ as proposal distribution.}
		\label{fig:acf_q3_clay}
	\end{subfigure}
	\begin{subfigure}[b]{0.2\textwidth}
		\centering\captionsetup{width=.8\linewidth}
		\includegraphics[width=\textwidth]{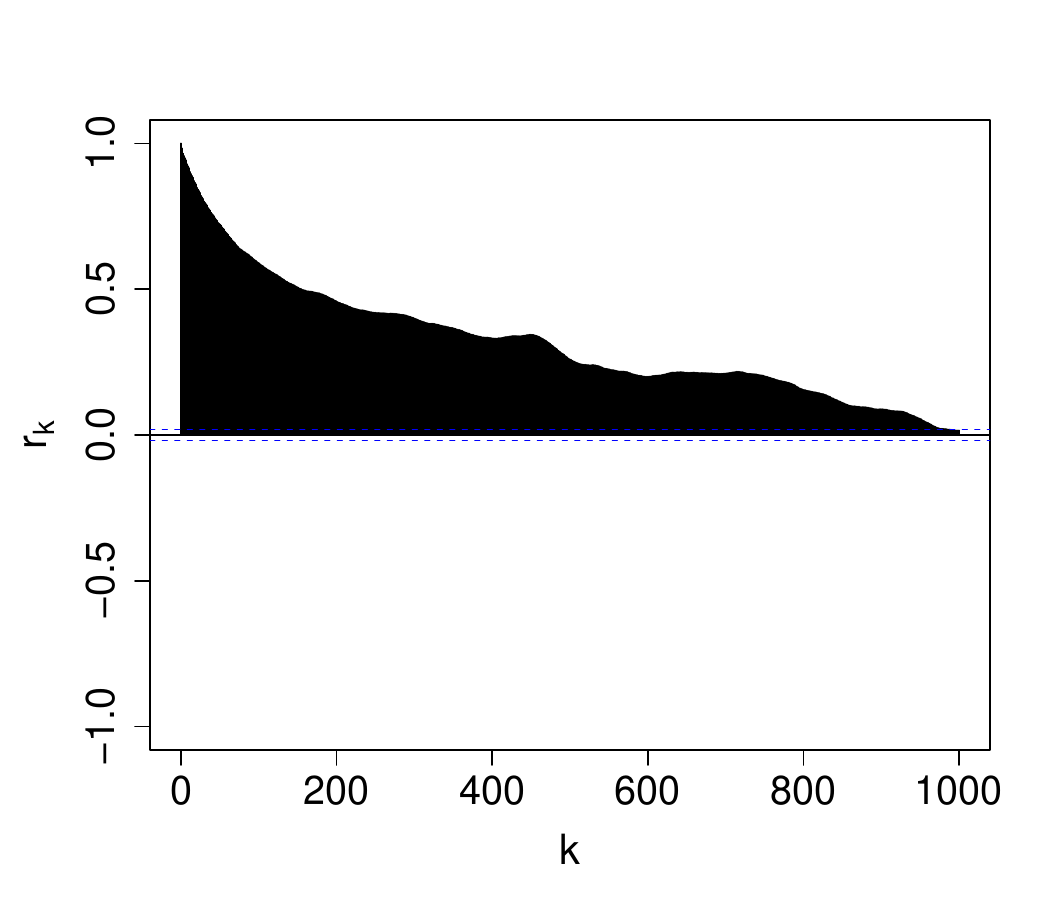}
		\caption{ACF plot from MCMC computations with $q_4$ as proposal distribution.}
		\label{fig:acf_q4_clay}
	\end{subfigure}
	\caption{ACF plot up to lag $k=1000$ of every $100$th value of $\{x_{g,t}^l\}_{t=0}^{1000000}$ (first row), $\{x_{o,t}^l\}_{t=0}^{1000000}$ (second row) and $\{x_{c,t}^l\}_{t=0}^{1000000}$ (third row) at location $l = 435$ using  four MH algorithms. The MH algorithm with proposal distribution $q_1$, $q_2$, $q_3$ and $q_4$ was used to get the results in the first, second, third and fourth columns respectively.}
	\label{fig:acf_post}
\end{figure}

\subsection{Adjusting for the uncertainty added to the forward model} 
\label{sec_omega_tilde}

To adjust for the additional error caused by replacing $\bh(\bx)$ by $\widehat{\bh}_{\text{MARS}}(\bx)$, the relationship between the variable of interest and the data can be modified to
\begin{equation*}
\by_s = \widehat{\bh}_{\text{MARS}}(\bx) + \widetilde{\bomega}, \quad \widetilde{\bomega} \sim \mathcal{N} (0, \widetilde{\bOmega}_0 ),
\end{equation*}
where $\widetilde{\bOmega}_0$ is the covariance matrix $\bOmega_0$ with empirical variance and covariance are added to it. A homoscedastic independence of the covariates is assumed. Plots of $|\bh(\bx) - \widehat{\bh}_{\text{MARS}}(\bx)|$ against the covariates used to investigate dependency between the absolute error and the covariates, indicated that this is not an unreasonable assumption.  The idea is that an average sample variance in the approximation partially compensates for using $\widehat{\bh}_{\text{MARS}}$. The adjusted covariance matrix is, where the subscript MARS is omitted for simplicity, 
\begin{equation*}
\widetilde{\bOmega}_0 = \bOmega_0 + \begin{bmatrix}
\widehat{\text{Var}}\left[\widehat{h_{R_0}}\right] & \widehat{\text{Cov}}\left[\widehat{h_{R_0}}, \widehat{h_{G}}\right] \\
\widehat{\text{Cov}}\left[\widehat{h_{R_0}}, \widehat{h_{G}}\right] & \widehat{\text{Var}}\left[\widehat{h_{G}}\right],
\end{bmatrix}
\end{equation*}
with 
\begin{equation*}
\widehat{\text{Var}}\left[\widehat{h_{R_0}}\right] = \frac{1}{n} \sum_{i=1}^{n} \left(\widehat{h_{{R_0}_i}} - h_{{R_0}_i} \right)^2, 
\quad
\widehat{\text{Var}}\left[\widehat{h_{G}}\right] = \frac{1}{n} \sum_{i=1}^{n} \left(\widehat{h_{G_i}} - h_{G_i} \right)^2,
\end{equation*}
and 
\begin{equation*}
\widehat{\text{Cov}}\left[\widehat{h_{R_0}}, \widehat{h_{G}}\right] = \frac{1}{n} \sum_{i=1}^{n} \left(\widehat{h_{{R_0}_i}} - h_{{R_0}_i} \right)\left(\widehat{h_{G_i}} - h_{G_i} \right). 
\end{equation*}
Here $\widehat{h_{{R_0}_i}}$ is the prediction for test data point $i$ and $h_{h_{{R_0}_i}}$ denotes the response at test data point $i$. The error-adjusted covariance matrix for the seismic AVO log-likelihood is thus 
\begin{equation}
\widetilde{\bOmega} = \begin{bmatrix}
\left[\widetilde{\bOmega}_0 \right] & 0 & \hdots & 0 & 0\\
0 & \left[\widetilde{\bOmega}_0 \right] &  \hdots & 0 & 0\\
\vdots & \vdots &  \ddots & \vdots & \vdots \\
0 &  0 & \hdots & \left[\widetilde{\bOmega}_0 \right] & 0 \\
0 &  0 & \hdots & 0 & \left[\widetilde{\bOmega}_0 \right] \\
\end{bmatrix}. 
\label{Omega_tilde}
\end{equation}

\begin{figure}[htb]
	\centering
	\begin{subfigure}[b]{0.2\textwidth}
		\centering\captionsetup{width=.8\linewidth}
		\includegraphics[width=\textwidth]{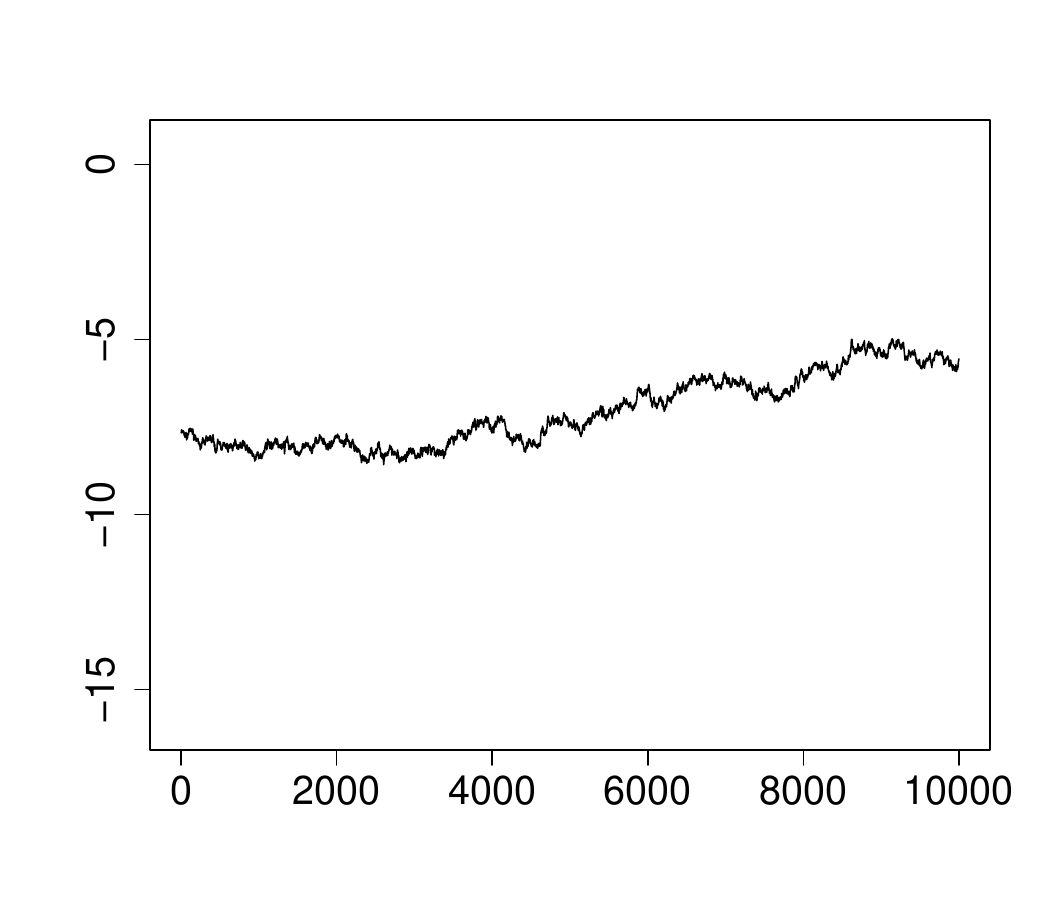}
		\caption{Trace plot from MCMC computations with $q_1$ as proposal distribution.}
		\label{fig:trace_q1_gas}
	\end{subfigure}
	\begin{subfigure}[b]{0.2\textwidth}
		\centering\captionsetup{width=.8\linewidth}
		\includegraphics[width=\textwidth]{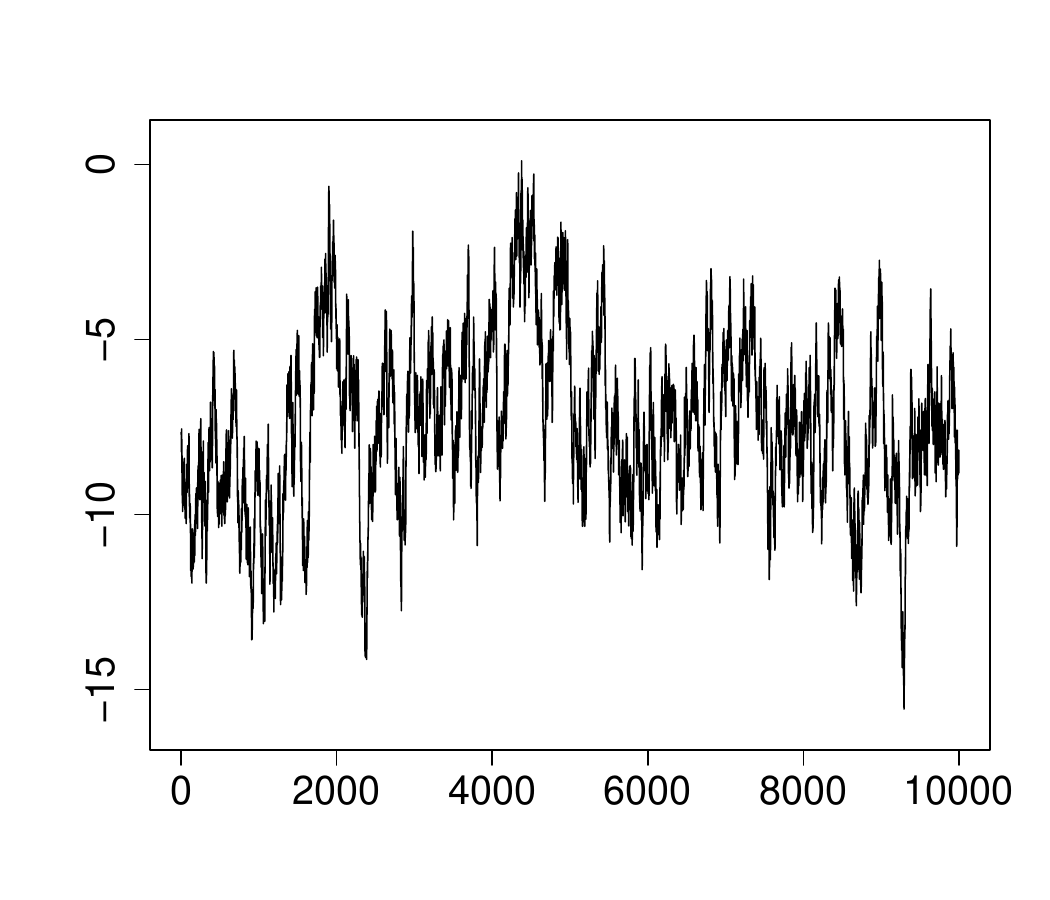}
		\caption{Trace plot from MCMC computations with $q_2$ as proposal distribution.}
		\label{fig:trace_q2_gas}
	\end{subfigure}
	\begin{subfigure}[b]{0.2\textwidth}
		\centering\captionsetup{width=.8\linewidth}
		\includegraphics[width=\textwidth]{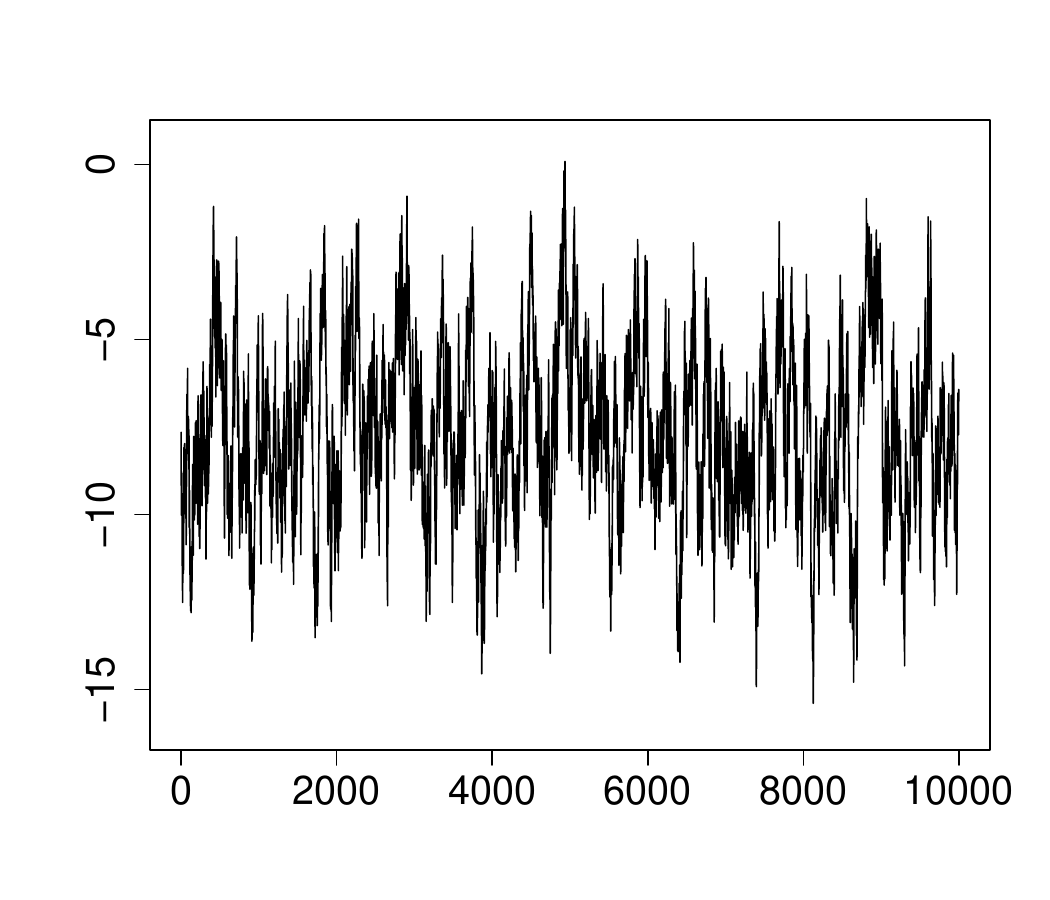}
		\caption{Trace plot from MCMC computations with $q_3$ as proposal distribution.}
		\label{fig:trace_q3_gas}
	\end{subfigure}
	\begin{subfigure}[b]{0.2\textwidth}
		\centering\captionsetup{width=.8\linewidth}
		\includegraphics[width=\textwidth]{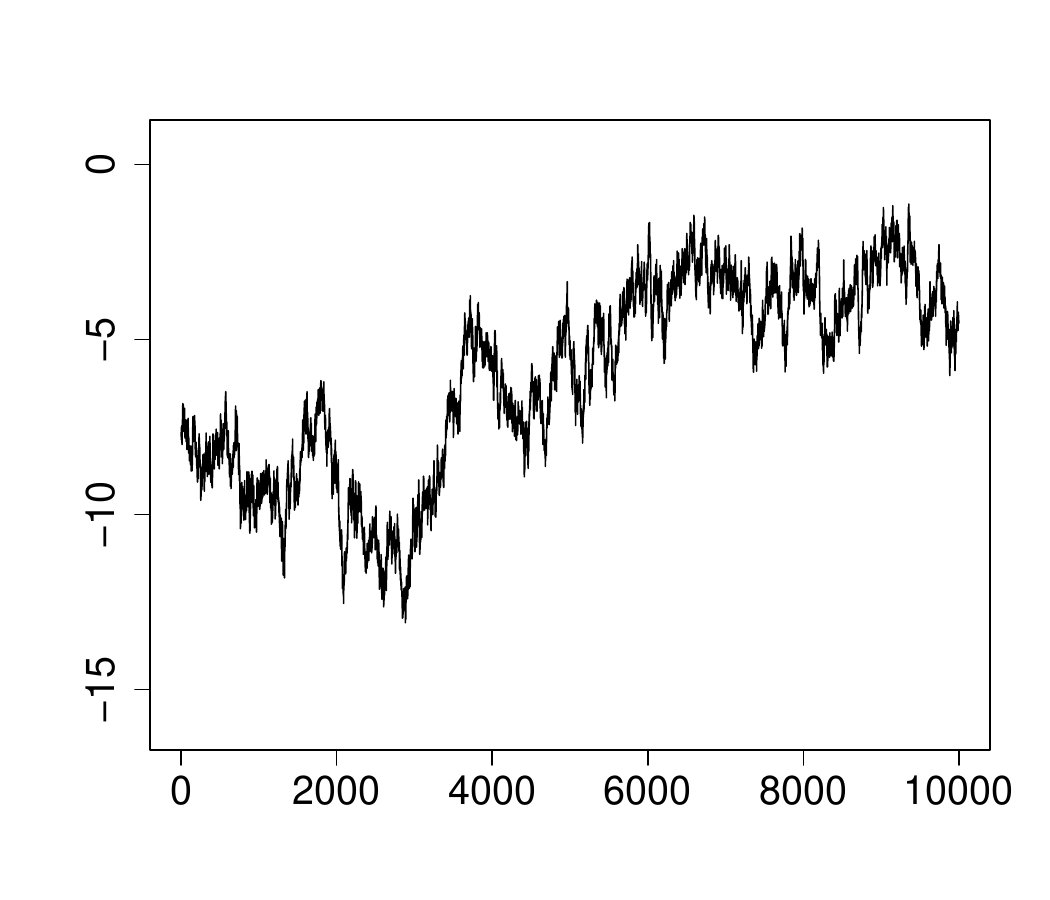}
		\caption{Trace plot from MCMC computations with $q_4$ as proposal distribution.}
		\label{fig:trace_q4_gas}
	\end{subfigure}
	\begin{subfigure}[b]{0.2\textwidth}
		\centering\captionsetup{width=.8\linewidth}
		\includegraphics[width=\textwidth]{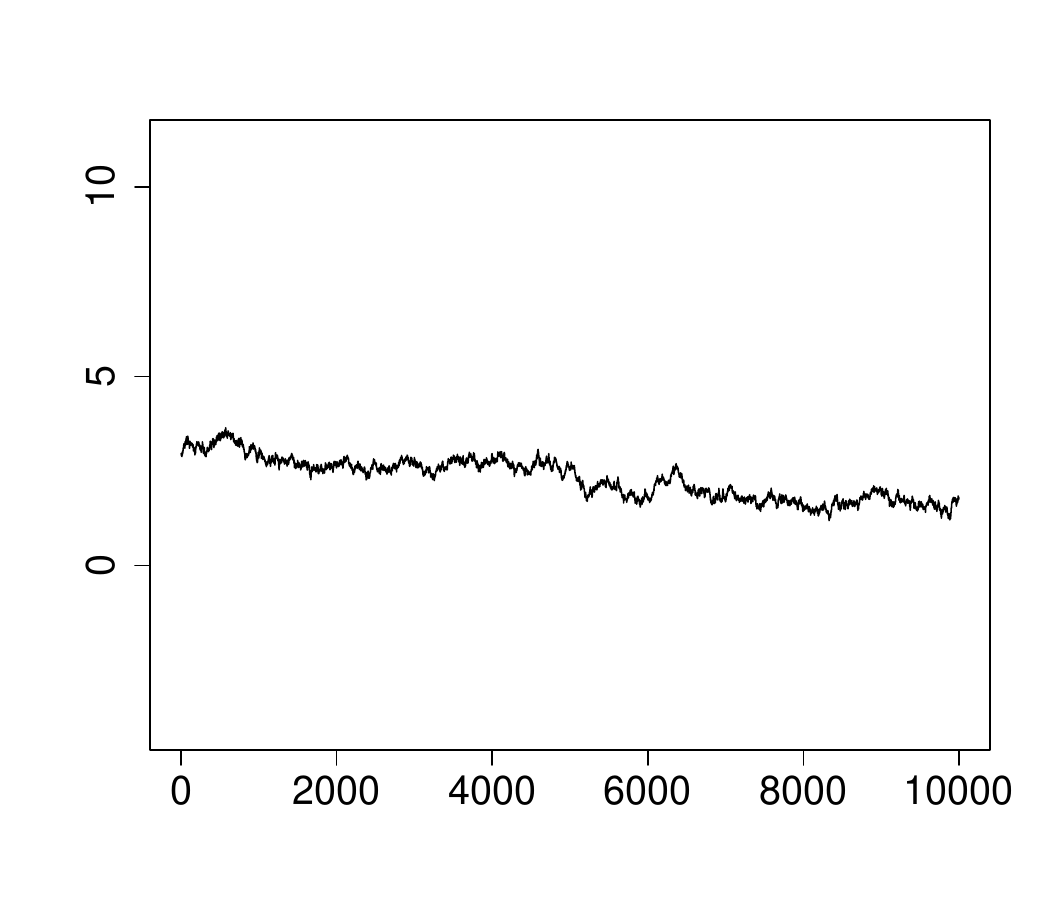}
		\caption{Trace plot from MCMC computations with $q_1$ as proposal distribution.}
		\label{fig:trace_q1_oil}
	\end{subfigure}
	\begin{subfigure}[b]{0.2\textwidth}
		\centering\captionsetup{width=.8\linewidth}
		\includegraphics[width=\textwidth]{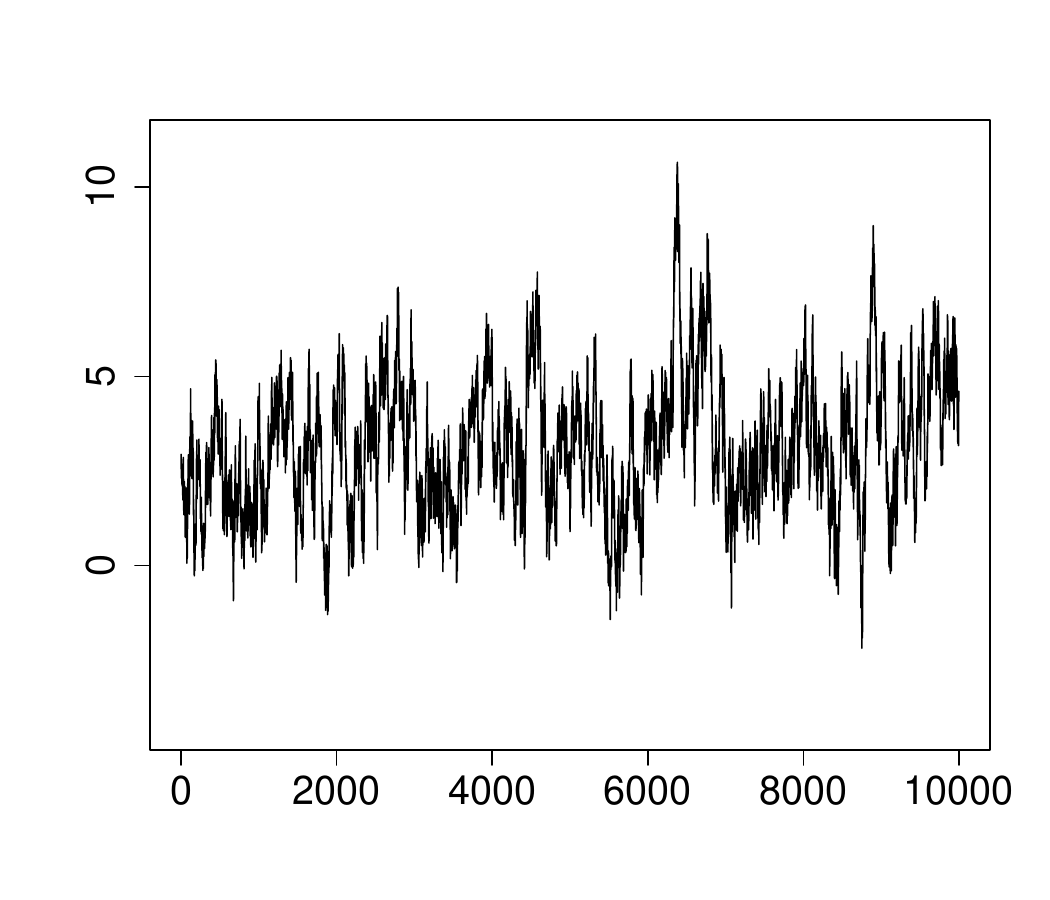}
		\caption{Trace plot from MCMC computations with $q_2$ as proposal distribution.}
		\label{fig:trace_q2_oil}
	\end{subfigure}
	\begin{subfigure}[b]{0.2\textwidth}
		\centering\captionsetup{width=.8\linewidth}
		\includegraphics[width=\textwidth]{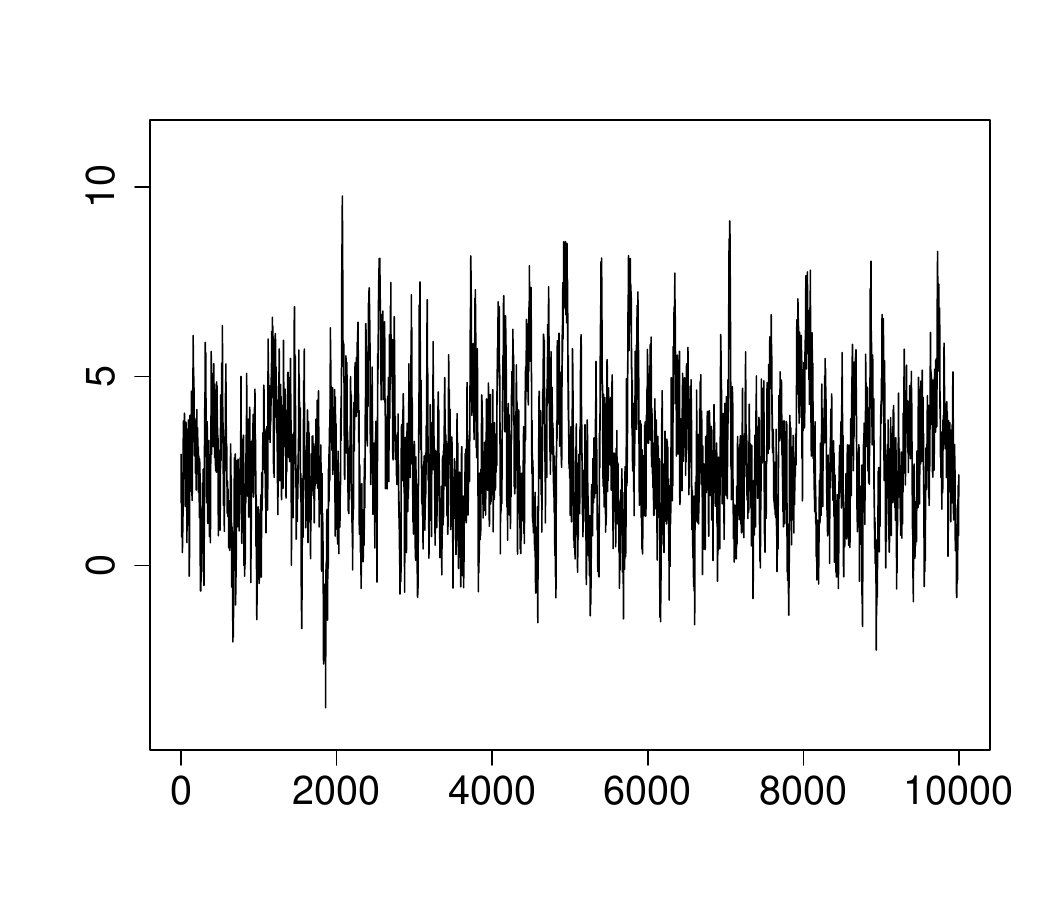}
		\caption{Trace plot from MCMC computations with $q_3$ as proposal distribution.}
		\label{fig:trace_q3_oil}
	\end{subfigure}
	\begin{subfigure}[b]{0.2\textwidth}
		\centering\captionsetup{width=.8\linewidth}
		\includegraphics[width=\textwidth]{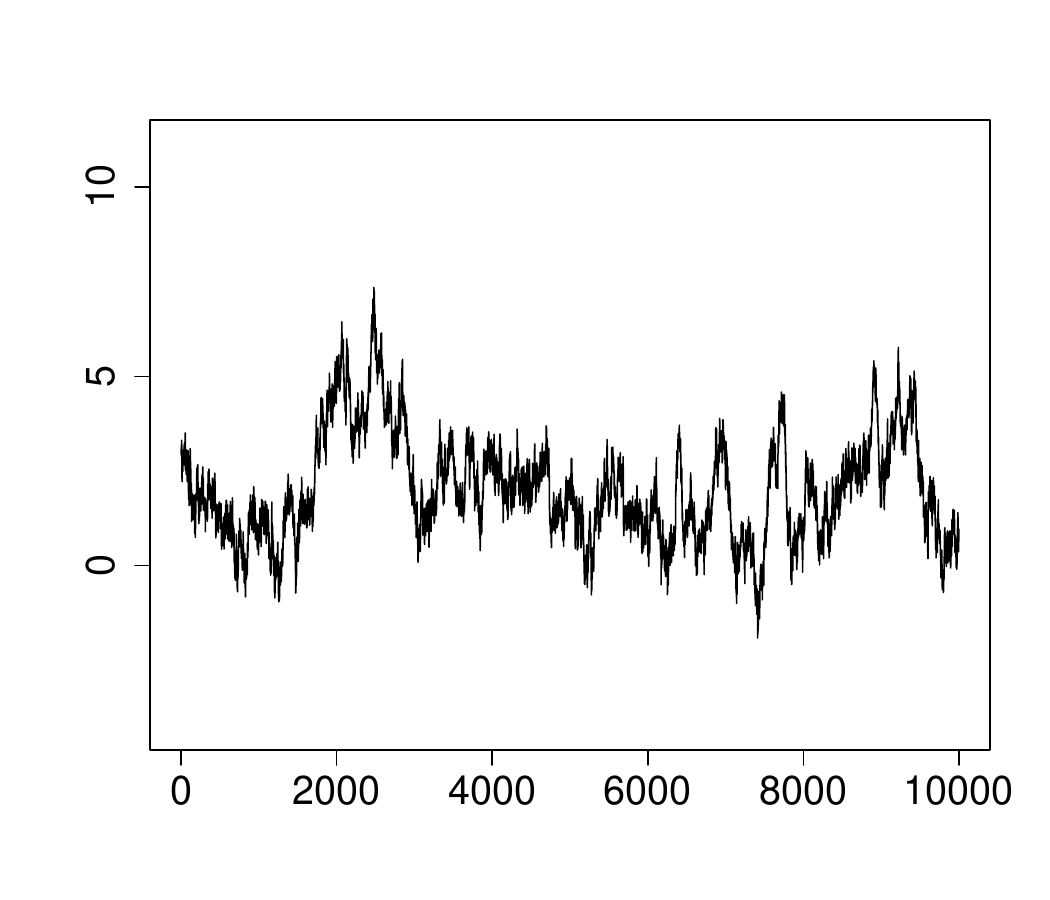}
		\caption{Trace plot from MCMC computations with $q_4$ as proposal distribution.}
		\label{fig:trace_q4_oil}
	\end{subfigure}
	\begin{subfigure}[b]{0.2\textwidth}
		\centering\captionsetup{width=.8\linewidth}
		\includegraphics[width=\textwidth]{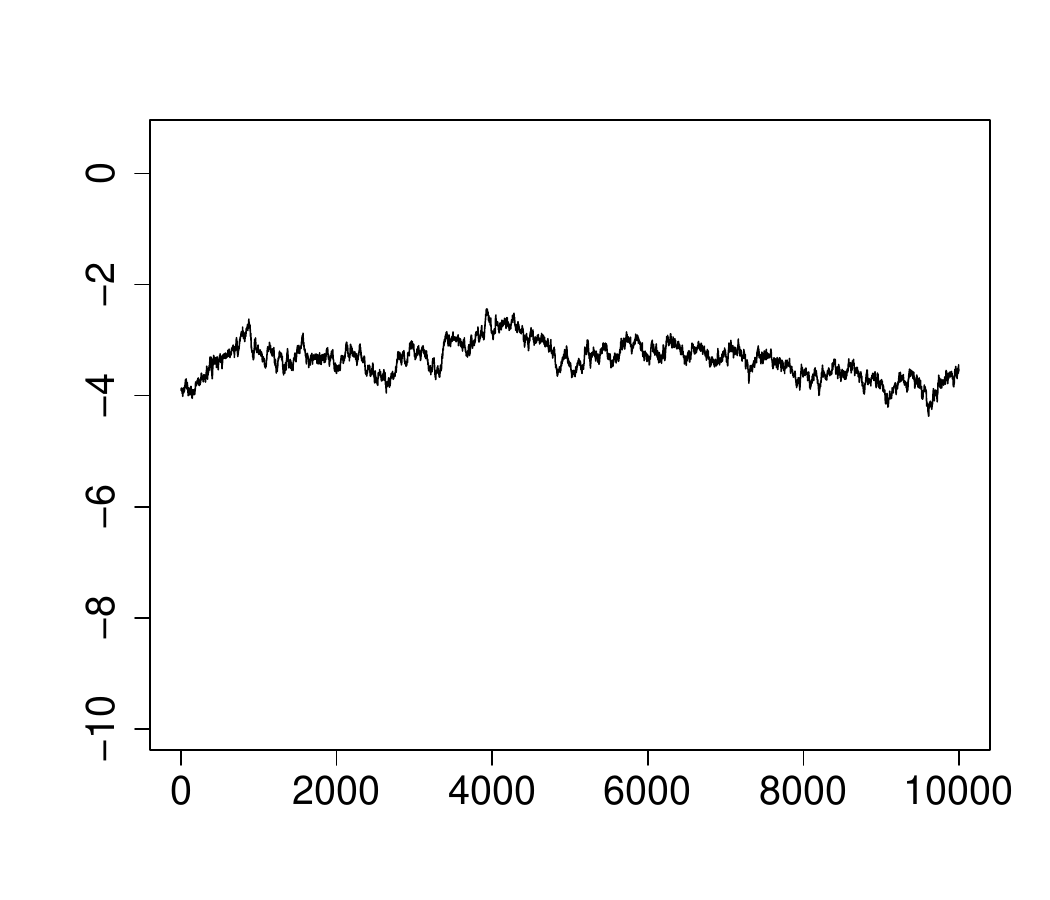}
		\caption{Trace plot from MCMC computations with $q_1$ as proposal distribution.}
		\label{fig:trace_q1_clay}
	\end{subfigure}
	\begin{subfigure}[b]{0.2\textwidth}
		\centering\captionsetup{width=.8\linewidth}
		\includegraphics[width=\textwidth]{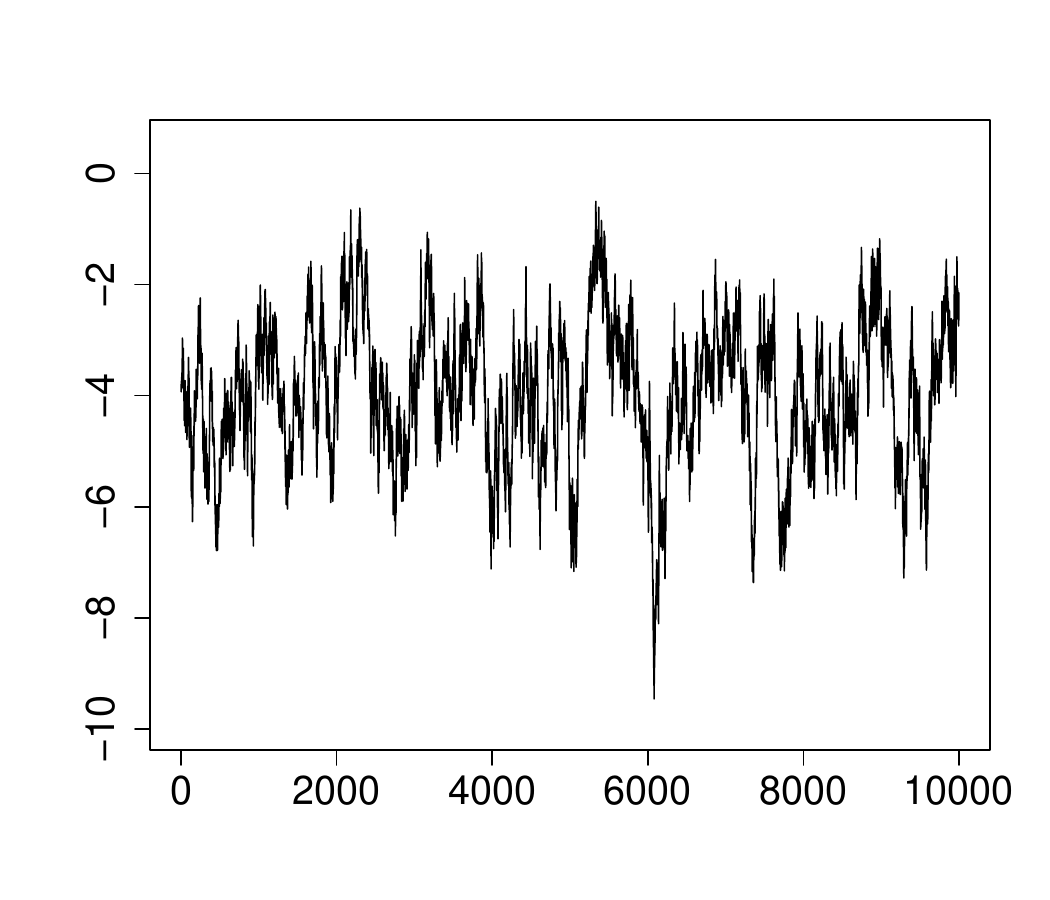}
		\caption{Trace plot from MCMC computations with $q_2$ as proposal distribution.}
		\label{fig:trace_q2_clay}
	\end{subfigure}
	\begin{subfigure}[b]{0.2\textwidth}
		\centering\captionsetup{width=.8\linewidth}
		\includegraphics[width=\textwidth]{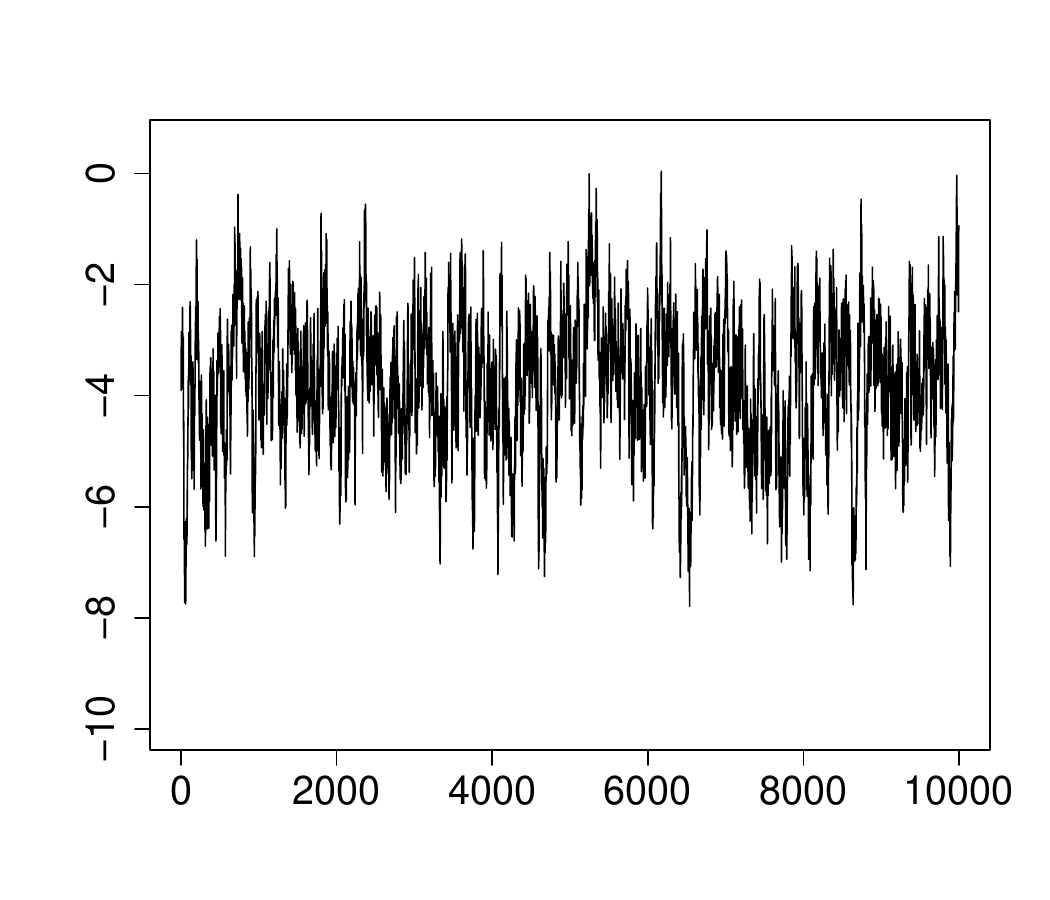}
		\caption{Trace plot from MCMC computations with $q_3$ as proposal distribution.}
		\label{fig:trace_q3_clay}
	\end{subfigure}
	\begin{subfigure}[b]{0.2\textwidth}
		\centering\captionsetup{width=.8\linewidth}
		\includegraphics[width=\textwidth]{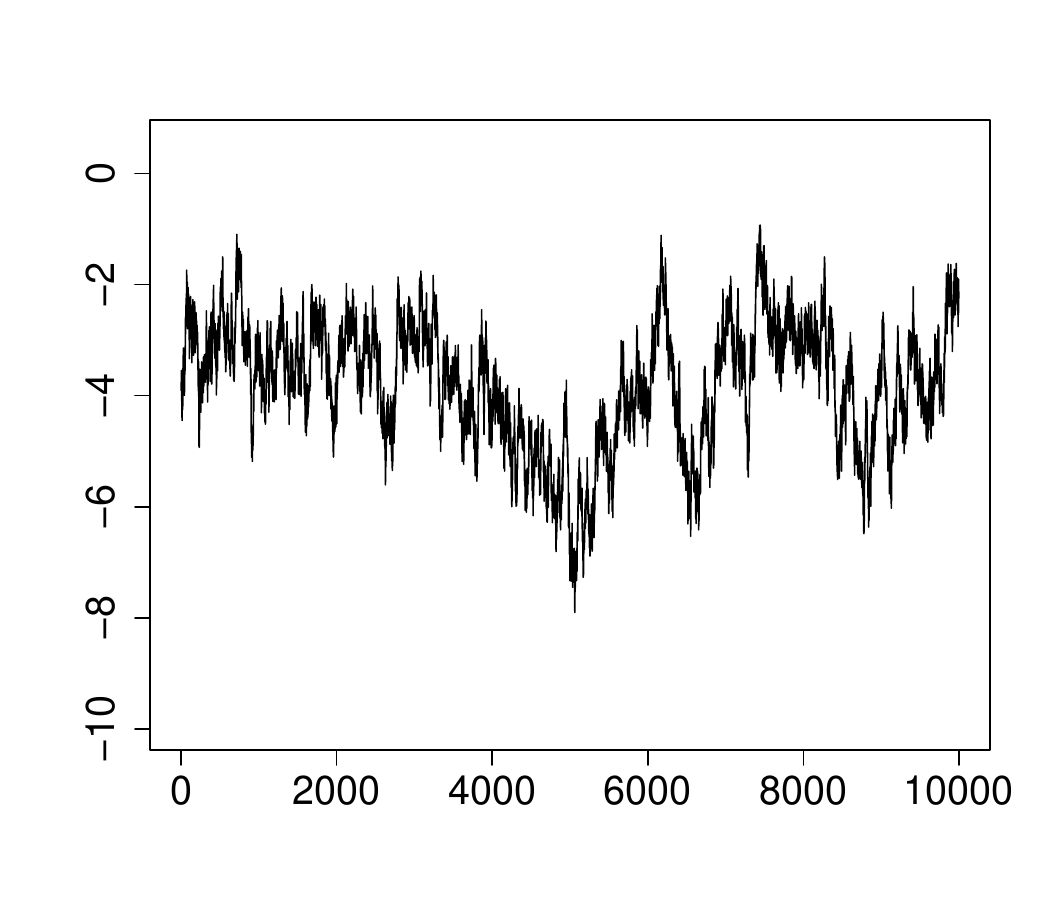}
		\caption{Trace plot from MCMC computations with $q_4$ as proposal distribution.}
		\label{fig:trace_q4_clay}
	\end{subfigure}
	\caption{Trace plots of every $100$th value of $\{x_{g,t}^l\}_{t=0}^{1000000}$ (first row), $\{x_{o,t}^l\}_{t=0}^{1000000}$ (second row) and $\{x_{c,t}^l\}_{t=0}^{1000000}$ (third row) at location $l = 435$ using four MH algorithms. The MH algorithm with proposal distribution $q_1$, $q_2$, $q_3$ and $q_4$ was used to get the results in the first, second, third and fourth columns respectively.}
	\label{fig:trace_post}
\end{figure}

\section{Additional results in simulation studies}		
\label{sc_extra_simulation}

Figure \ref{fig:acf_post} displays ACF plots for every $100$th value of sequences $\{x_{g,t}^l\}_{t=0}^{1000000}$, $\{x_{o,t}^l\}_{t=0}^{1000000}$, and $\{x_{c,t}^l\}_{t=0}^{1000000}$ at location $l = 435$ using four MH algorithms. These plots align with Table \ref{table:ESS_time_q}. The ACF plots in Figure \ref{fig:acf_post} indicate that the MH algorithm with $q_3$ as the proposal distribution produces the least correlated samples. It shows samples approximately $15,000$ time iterations apart, indicating insignificant correlation. Compared to the MH algorithm with $q_1$ as the proposal distribution, the ACF plots for $q_3$ decrease much faster. Additionally, ACF plots for the MH algorithm with proposal distribution $q_2$ also decrease rapidly compared to $q_1$ and $q_4$. While autocorrelation decreases similarly for $q_1$ and $q_4$ in gas plots, MALA decreases noticeably faster in oil and clay plots compared to $q_1$.

The trace plots depicted in Figure \ref{fig:trace_post} are consistent with the findings presented in Table \ref{table:ESS_time_q}. They illustrate that the MH algorithm employing $q_3$ as the proposal distribution exhibits optimal mixing, as evidenced by the trajectory frequently traversing around a central value. Similarly, the trace plots suggest that the MH algorithm utilizing proposal distribution $q_2$ achieves good mixing, albeit with slightly less coverage compared to the most efficient algorithm. Moreover, the trace plots highlight that the MALA mixes notably better than the standard random walk but not as effectively as the MH algorithms employing $q_2$ and $q_3$ as proposal distributions.

However, the trace plots indicate that the MH algorithm with $q_1$ as the proposal distribution has not yet converged, as evidenced by a trend suggesting ongoing exploration of the posterior. Comparison of the trace plots for the standard random walk with those of other algorithms reveals that the former has not reached convergence; for instance, the trace plot in Figure \ref{fig:trace_q1_gas} has not approached values close to $-10$, an area visited by all three other algorithms.

At this particular location, the ACF and trace plots corroborate the results in Table \ref{table:ESS_time_q}. It's noteworthy that mixing behaviors, as reflected in trace and ACF plots, can vary across different locations. Nonetheless, the table's Effective Sample Size (ESS) represents the mean ESS across all locations, indicating that the MH algorithm employing $q_3$ as the proposal distribution exhibited the most favorable overall mixing.

\end{document}